\documentclass[a4paper,fleqn,usenatbib]{mnras}

\usepackage{natbib}
\usepackage[usenames]{color}
\usepackage{multirow}
\usepackage[T1]{fontenc}
\usepackage{ae,aecompl}

\usepackage{graphicx}	% Including figure files
\usepackage{amsmath}	% Advanced maths commands
\usepackage{times}
\usepackage{amssymb}	% Extra maths symbols

\newcommand{\HI}{\ion{H}{i}}

\newcommand{\kms}{\mbox{km~s$^{-1}$}}
\newcommand{\Msol}{\mbox{M$_\odot$}}
\newcommand{\surm}{\mbox{M$_\odot$ pc$^{-2}$}}

\newcommand{\siggas}{\mbox{$\Sigma_{\rm gas}$}}
\newcommand{\sigsfr}{\mbox{$\Sigma_{\rm SFR}$}}

\newcommand{\sighi}{\mbox{$\Sigma_{\rm HI}$}}
\newcommand{\sightwo}{\mbox{$\Sigma_{\rm H_2}$}}

\newcommand{\um}{\mbox{$\micron$}}
\newcommand{\ac}{\mbox{$\arcsec$}}
\newcommand{\Htwo}{\mbox{H$_2$}}

\newcommand{\ropt}{\mbox{$r_{25}$}}

\title[Star Formation and Gas Accretion in Galaxies]{Star Formation and Gas Accretion in Nearby Galaxies}

\author[Kijeong Yim \& J. M. van der Hulst]{
Kijeong Yim$^{1,2}$\thanks{E-mail: kyim@kasi.re.kr} and 
J. M. van der Hulst$^{1}$
\\
$^{1}$Kapteyn Astronomical Institute, University of Groningen, P.O. Box 800, 9700 AV Groningen, The Netherlands\\
$^{2}$Korea Astronomy and Space Science Institute, 776 Daedeok-daero, Yuseong-gu, Daejeon 34055, Korea
}

\date{Accepted XXX. Received YYY; in original form ZZZ}

\pubyear{2016}

\begin{document}
\label{firstpage}
\pagerange{\pageref{firstpage}--\pageref{lastpage}}
\maketitle

\begin{abstract}
In order to quantify the relationship between gas accretion and star formation, we analyse a sample of 29 nearby galaxies from the WHISP survey which contains galaxies with and without evidence for recent gas accretion. We compare combined radial profiles of FUV ($GALEX$) and IR 24 \um\ ({\it Spitzer}) characterizing distributions of recent star formation with radial profiles of CO (IRAM, BIMA, or CARMA) and \HI\ (WSRT) tracing molecular and atomic gas contents to examine star formation efficiencies in symmetric (quiescent), asymmetric (accreting), and interacting (tidally disturbed) galaxies. In addition, we investigate the relationship between star formation rate and \HI\ in the outer discs for the three groups of galaxies. We confirm the general relationship between gas surface density and star formation surface density, but do not find a significant difference between the three groups of galaxies.
\end{abstract}


%%%%%%%%%%%%%%%%%%%%%%%%%%%%%%%%%%%%%%%%%%%%%%%%%%

%%%%%%%%%%%%%%%%% BODY OF PAPER %%%%%%%%%%%%%%%%%%

\section{Introduction}
\label{intro}

It has long been suggested that gas accretion plays an important role in ongoing star formation in galaxies (e.g., \citealt{1980ApJ...237..692L}; \citealt{Sancisi:2008jz}; \citealt{SanchezAlmeida:2014ha}). 
In addition, numerical simulations (e.g., \citealt{2009Natur.457..451D}; \citealt{2009ApJ...694..396B}) predict a positive correlation between gas accretion and star formation. The importance of gas accretion in galaxies has been demonstrated by many studies. For example, the gas consumption problem (\citealt{1980ApJ...237..692L}; \citealt{1983ApJ...272...54K}) can be cured by continuous accretion (\citealt{2012MNRAS.426.2166F}; \citealt{Sancisi:2008jz}). Likewise, the constant star formation rate in the Milky Way (e.g., \citealt{1980ApJ...242..242T}; \citealt{2000MNRAS.318..658B}) can be explained by gas accretion. 
The accreted gas originates from the intergalactic medium (IGM) or is gas removed from galaxies by gravitational interactions with companion galaxies.  The accreted gas could  replenish atomic hydrogen consumed by star formation.
However, the correlation between gas accretion and star formation is not straightforward \citep{Sancisi:2008jz}.   

Indications for gas accretion are considered to be the presence of extra-planar gas (\citealt{Oosterloo:2007kd}; \citealt{2008ApJ...672..298W}),  warped layers \citep{1989MNRAS.237..785O}, and  lopsided discs  (\citealt{2005A&A...438..507B}; \citealt{2011A&A...530A..29V}).  Since \HI\ observations provide possible evidence for gas accretion  (e.g., \citealt{2004IAUS..217..122V}; \citealt{2007A&A...465..787O}),  
 \HI\ observations are expected to be a crucial key revealing the role of accretion. 
Asymmetric structure and kinematics in \HI\ have been investigated by many authors (e.g., \citealt{2002A&A...390..829S}; \citealt{2005A&A...442..137N}; \citealt{2011A&A...530A..29V}). \cite{Sancisi:2008jz} found that about half of the galaxies in the Westerbork \HI\ Survey of Irregular and SPiral galaxies (WHISP; \citealt{1996A&AS..116...15K}; \citealt{2001ASPC..240..451V}) show asymmetries in the  distribution and/or kinematics of the \HI. 

The star formation law  (or Kennicutt-Schmidt law; \citealt{1998ApJ...498..541K}) has been investigated in numerous galaxies (e.g., \citealt{2002ApJ...569..157W}; \citealt{2008AJ....136.2782L}; \citealt{2008AJ....136.2846B}). The power-law correlation between star formation rate (SFR) and total gas (\HI\ + \Htwo) or molecular gas (\Htwo) suggests that star formation is strongly correlated with the gas. 
We quantify star formation (SF) properties in 29 galaxies with resolved \HI, CO, UV and IR data to investigate whether SF characteristics are the same everywhere or depend on interactions and accretion.
We can measure the SFR accurately using resolved $GALEX$ and $Spitzer$ maps and directly show how much the measured SFR is related with the gas properties in different environment such as quiescent, accreting, and tidally disturbed galaxies.  %to look for possible relationships with accretion and interactions. 
The $GALEX$ FUV and  $Spitzer$ 24 \um\ data have proved to be good SFR indicators (\citealt{2007ApJ...666..870C}; \citealt{2008AJ....136.2782L}) and the FUV emission has been detected even in several outer discs \citep{2007ApJS..173..538T} while the 24 \um\ emission is limited  to the optical radius (\ropt). Since \HI\ emission is extended out to  2$\times$\ropt\ or more and the \HI\ gas in the outer regions can be used as a proxy for the accreting gas, the comparison of FUV and \HI\ in these regions will provide the most direct examination for the role of gas accretion in star formation. 
In order to investigate whether gas accretion and interaction affect the star formation rate (SFR), we divide the galaxies into three groups: quiescent, accreting, and interacting. 
We consider kinematical and morphological asymmetries as an indication for accretion and the presence of nearby companions as an indication of interactions. 
The remaining galaxies are symmetric and isolated.

This paper is organized as follows. 
 Section~\ref{data} describes our sample selection and the observations. Section \ref{class} shows how the sample of galaxies is classified into symmetric, asymmetric, and interacting galaxies based on kinematics and morphology. Section \ref{results} presents the results: radial distributions of \sigsfr, \sightwo, \sighi, and \siggas\ in Section \ref{radial}, scaled radial distributions in Section \ref{scaled}, comparisons between the SFR and gas in the inner regions in Section \ref{sflaw} and the outer regions in Section \ref{fuv-hi}, and a comparison of all these properties in relation to the stellar mass of the galaxies in Section \ref{Mstar}.  Section \ref{sum} summarizes and concludes this work.

%in context, cites relevant earlier studies in the field by \citet{Others2013},
%and describes the problem the authors aim to solve \citep[e.g.][]{Author2012}.

\section{Sample and Observational Data}
\label{data}

\subsection{Galaxy Sample}

\begin{table*}
%\centering
\caption{Galaxy Sample}
\label{tab:sample}
\begin{tabular}{lcccccccc}
\hline
\multicolumn{1}{c}{Galaxy} & Distance&\multirow{2}{*}{log$\left(\frac{M_{*}}{\rm M_\odot}\right)$}& \HI\ total flux& r$_{25}$ & Inclination &$V_{\rm sys}$&Galaxy&CO\\
&(Mpc)&& (Jy \kms)& (arcsec)&(\degr)&(\kms)&Class&Telescope\\
\multicolumn{1}{c}{(1)}&(2)&(3)&(4)&(5)&(6)&(7)&(8)&(9)\\
\hline
\verb'UGC 1913 (NGC 925)'&9.3&10.07&326&314&54$^{\rm v}$&554&A/S&IRAM\\
\verb'UGC 2455 (NGC 1156)'&6.5&9.28&64&99&52$^{\rm v}$&375&S \\
\verb'UGC 3334 (NGC 1961)'&56.0&11.61&75&137&47$^{\rm L}$&3934&I&IRAM\\
%UGC 3426 &57.8&2&55&45\tablenotemark{N}&4052&Int\\
\verb'UGC 3851 (NGC 2366)'&3.9&8.44&274&244&68$^{\rm v}$&99&A\\
\verb'UGC 4165 (NGC 2500)'&9.8&9.35&36&87&28$^{\rm ES}$&504&S\\
\verb'UGC 4274 (NGC 2537)'&8.1&9.32&20&52&33$^{\rm M}$&452&S\\
\verb'UGC 4305' &5.0&8.76&253&238&40$^{\rm R}$&142&A\\
\verb'UGC 4862 (NGC 2782)'&39.5&10.84&7&104&30$^{\rm W}$&2543&I\\
\verb'UGC 5079 (NGC 2903)'&7.3&10.55&277&378&64$^{\rm v}$&550&S&IRAM\\
\verb'UGC 5532 (NGC 3147)'&43.0&11.57&32&117&35$^{\rm v}$&2814&S&CARMA\\
\verb'UGC 5557 (NGC 3184)'&10.1&10.28&123&222&21$^{\rm H}$&592&A/S&BIMA\\
\verb'UGC 5789 (NGC 3319)'&13.3&9.67&94&185&62$^{\rm v}$&742&A/S\\
\verb'UGC 5840 (NGC 3344)'&6.9&9.91&186&212&18$^{\rm E}$&588&S\\
\verb'UGC 6537 (NGC 3726)'&17.0&10.70&101&185&49$^{\rm v}$&864&A/S&BIMA\\
\verb'UGC 6856 (NGC 3938)'&15.5&10.44&86&161&24$^{\rm H}$&808&S&BIMA\\
\verb'UGC 6869 (NGC 3949)'&15.8&10.18&45&87&57$^{\rm L}$&800&A/S&CARMA\\
\verb'UGC 7030 (NGC 4051)'&12.9&10.21&44&157&41$^{\rm H}$&704&S&BIMA\\
\verb'UGC 7166 (NGC 4151)'&20.0&10.41&72&189&20$^{\rm N}$&999&S&CARMA\\
\verb'UGC 7256 (NGC 4203)'&22.4&10.89&49&102&51$^{\rm v}$&1083&A\\
\verb'UGC 7278 (NGC 4214)'&3.8&9.10&260&255&44$^{\rm L}$&292&S&IRAM\\
\verb'UGC 7323 (NGC 4242)'&8.8&9.52&49&150&52$^{\rm v}$&517&A/S\\
\verb'UGC 7353 (NGC 4258)'&8.0&10.64&509&559&66$^{\rm v}$&454&S&BIMA\\
\verb'UGC 7524 (NGC 4395)'&3.8&9.02&310&395&47$^{\rm v}$&318&A\\
\verb'UGC 7651 (NGC 4490)'&9.2&10.17&252&189&60$^{\rm H}$&565&I&BIMA\\
\verb'UGC 7766 (NGC 4559)'&9.8&9.95&331&321&67$^{\rm v}$&814&A/S&IRAM\\
\verb'UGC 7831 (NGC 4605)'&4.4&9.37&54&173&56$^{\rm E}$&146&A/S&CARMA\\
\verb'UGC 7853 (NGC 4618)'&8.8&9.21&67&125&36$^{\rm O}$&537&I\\
\verb'UGC 7989 (NGC 4725)'&26.8&11.46&145&321&44$^{\rm v}$&1208&A&IRAM\\
%UGC 8900 (NGC 5395)&56.0&23&87&57\tablenotemark{E}&3468&Int\\
\verb'UGC 12754 (NGC 7741)'&12.5&9.73&53&131&49$^{\rm v}$&752&S\\
\hline
\end{tabular}\\
$Columns:$ (1) Galaxy name; (2) distance adopted from the NED database for $H =$ 73 \kms\ Mpc$^{-1}$; (3) total stellar mass ($M_*$); (4) \HI\ total flux obtained from the masked integrated intensity map using the MIRIAD task HISTO; (5) optical radius r$_{25}$ from RC3 \citep{1991rc3..book.....D}; (6) inclination adopted from $^{\rm v}$\cite{2011A&A...530A..29V}, $^{\rm L}$LEDA, $^{\rm N}$\citet{2006PhDT.........1N},$^{\rm ES}$\citet{1999AJ....117..764E}, $^{\rm M}$\citet{2008AJ....135..291M}, $^{\rm S}$\citet{1999PhDT........27S}, $^{\rm W}$\citet{2013ApJ...777L...4W}, $^{\rm H}$\citet{2003ApJS..145..259H}, $^{\rm E}$\citet{2008MNRAS.390..466E}, $^{\rm O}$\citet{1991AJ....101..829O}; (7) heliocentric systemic velocity adopted from LEDA or NED; (8) galaxy class defined from the \HI\ kinematics and morphology.
\end{table*}

Since spatially resolved data are crucial for comparing star formation and gas properties,  we have selected our sample based on availability of resolved data: \HI\ from the WHISP survey \citep{2001ASPC..240..451V}, CO from the IRAM HERACLES survey  \citep{2009AJ....137.4670L}, the IRAM NUGA survey (\citealt{2003ASPC..290..423G}; \citealt{2009A&A...503...73C}), the BIMA SONG survey \citep{2003ApJS..145..259H}, and the CARMA STING survey (PI: Alberto Bolatto; \citealt{2011ApJ...730...72R}; \citealt{2013ApJ...777L...4W}), FUV from $GALEX$, and IR 24 \um\ from $Spitzer$. 
Additional selection criteria are good S/N in \HI\ and CO and a galaxy inclination of 70$\degr$ or less. This resulted in an initial sample of 16 galaxies. Later, we increased the sample with 13 more galaxies that satisfy all criteria except the availability of CO imaging. 
The final sample of galaxies is listed in Table \ref{tab:sample}. The galaxy class  is determined from the \HI\ kinematics as well as the \HI\ morphology. More details  are given in Section \ref{class}.

\subsection{FUV and IR 24 \um}

Most of the  FUV emission is produced by the young O and B stars. The fraction absorbed by the dust surrounding the stars is re-emitted in the infrared, so the combination (FUV+24 \um) is a good method to estimate the total recent star formation \citep{2008AJ....136.2782L}. 

We have used FUV maps from the $GALEX$ archive. They were taken  from the Nearby Galaxy Survey (NGS; \citealt{2007ApJS..173..185G}), the All-Sky Imaging Survey (AIS), and the Deep Imaging Survey (DIS). 
In the FUV images, foreground sources around galaxies are masked and the sky background mean, obtained from several regions far away from a galaxy, has been subtracted.
The FUV maps have a spatial resolution of 4.3\ac\ and are in units of counts s$^{-1}$ pixel$^{-1}$, where 1 count s$^{-1}$ (cps) is 108 $\mu$Jy (equivalent to $1.4 \times 10^{-15}$ erg s$^{-1}$ cm$^{-2}$ \AA$^{-1}$) according to the $GALEX$ Observer's Guide. %http://galexgi.gsfc.nasa.gov/docs/galex/Documents/ERO_data_description_2.htm#_Toc58822546
These units have been converted to MJy sr$^{-1}$ to match the $Spitzer$ 24 \um\ data. 
For the Galactic extinction correction, we have adopted the extinction $A_{\rm FUV} = 8.24 E(B - V)$  given by \cite{2007ApJS..173..293W}, where the $E(B - V)$ values are obtained from the IDL code and the dust maps provided by \citet{1998ApJ...500..525S}.
We used $Spitzer$ 24 \um\ maps from  the SINGS survey  \citep{2003PASP..115..928K}, Program ID 59 (PI: G. Rieke), Program ID 69 (PI: G. Fazio), Program ID 3124 (PI: D. Alexander), Program ID 3247 (PI: C. Struck), Program ID 30443 (PI: G. Rieke), Program ID 40204 (PI: R. Kennicutt), and Program ID 50639 (PI: C. Danforth). We  downloaded the Basic Calibrated Data (BCD) from the Spitzer Heritage Archive and  used MOPEX (Mosaicking and Point Source Extraction) for background matching and mosaicking the BCD images after removing instrumental artifacts using the Image Reduction and Analysis Facility (IRAF) tasks IMSTAT and IMARITH.  The resolution and units of the maps are 5.9\ac\ and MJy sr$^{-1}$, respectively. 
In the case of galaxies with AGN (UGC 7030, 7166, and 7989), the central regions ($\sim$ 1 kpc) are blanked to reduce the contamination by AGN activity.
%IRAF footnote "IRAF is distributed by the National Optical Astronomy Observatory, which is operated by the Association of Universities for Research in Astronomy, Inc., under cooperative agreement with the National Science Foundation."

\subsection{H$\small\textsc{I}$ and CO}
For the total gas properties, we use \HI\ and CO data with the inclusion of helium (a factor of 1.36).
The \HI\ data are obtained from the WHISP archive \citep{2001ASPC..240..451V}.
There are three kinds of resolutions available in the archive: 12\ac\ $\times$ 12\ac/sin $\delta$ (full resolution), 30\ac\ $\times$ 30\ac\, and 60\ac\ $\times$ 60\ac\ (see \citealt{2002A&A...390..829S} for details of the reduction process). We have downloaded the highest resolution (12\ac\ $\times$ 12\ac/sin $\delta$) images as this is closest to the resolutions of the FUV and IR 24 \um\ images used for determining the SFR.   The downloaded images are masked maps in Westerbork Units (1 W.U. = 5 mJy Beam$^{-1}$) reduced by the WHISP pipeline. We converted them to the integrated intensity maps in units of Jy Beam$^{-1}$ km s$^{-1}$ by multiplying the velocity channel separation using the MIRIAD task MATHS. 

In order to obtain CO data at sufficiently high resolution and sensitivity, we use CO integrated intensity maps from large single dish or interferometric studies: the IRAM HERACLES survey \citep{2009AJ....137.4670L}, the IRAM NUGA survey (\citealt{2003ASPC..290..423G}; \citealt{2009A&A...503...73C}), the BIMA SONG survey \citep{2003ApJS..145..259H}, and the CARMA STING survey (PI: Alberto Bolatto; \citealt{2011ApJ...730...72R}; \citealt{2013ApJ...777L...4W}). All the galaxies with CO data are indicated by their telescope in Table \ref{tab:sample}.

\section{H$\small\textsc{I}$ Classification}
\label{class}

Asymmetries of galaxies in their morphology and kinematics have been studied for many decades (e.g., \citealt{1980MNRAS.193..313B}; \citealt{1994A&A...290L...9R}; \citealt{2011A&A...530A..29V}, \citeyear{2011A&A...530A..30V}). In those studies, they found that  typically about a half of the galaxies shows asymmetries, suggesting that   asymmetric galaxies are very normal (e.g., \citealt{1994A&A...290L...9R}; \citealt{Sancisi:2008jz}). Although the origin of the asymmetries has not been established  so far, some studies suggested that the asymmetric features might be caused by gas accretion (e.g., \citealt{2005A&A...438..507B}; \citealt{2008MNRAS.388..697M}; \citealt{Sancisi:2008jz}) or interactions with companions (\citealt{1997ApJ...488..642J}; \citealt{1997ApJ...477..118Z}). Based on these studies, we assume that mild asymmetries represent  gas accretion, and that strong asymmetries represent interactions. 
 Note, however, that there are also internal mechanisms such as an asymmetric halo potential that can produce asymmetries, which we can not distinguish easily at present. We also looked at the 3.6 \um\ IR distributions for asymmetries and found that in all but two cases both the \HI\ and the IR distributions are asymmetric, the \HI\ in general being more distorted. The two exceptions are UGC 7256 and UGC 7989, which are symmetric in the IR but not the \HI. 

\begin{figure*}
\begin{center}

\includegraphics[width=0.35\textwidth,angle=270]{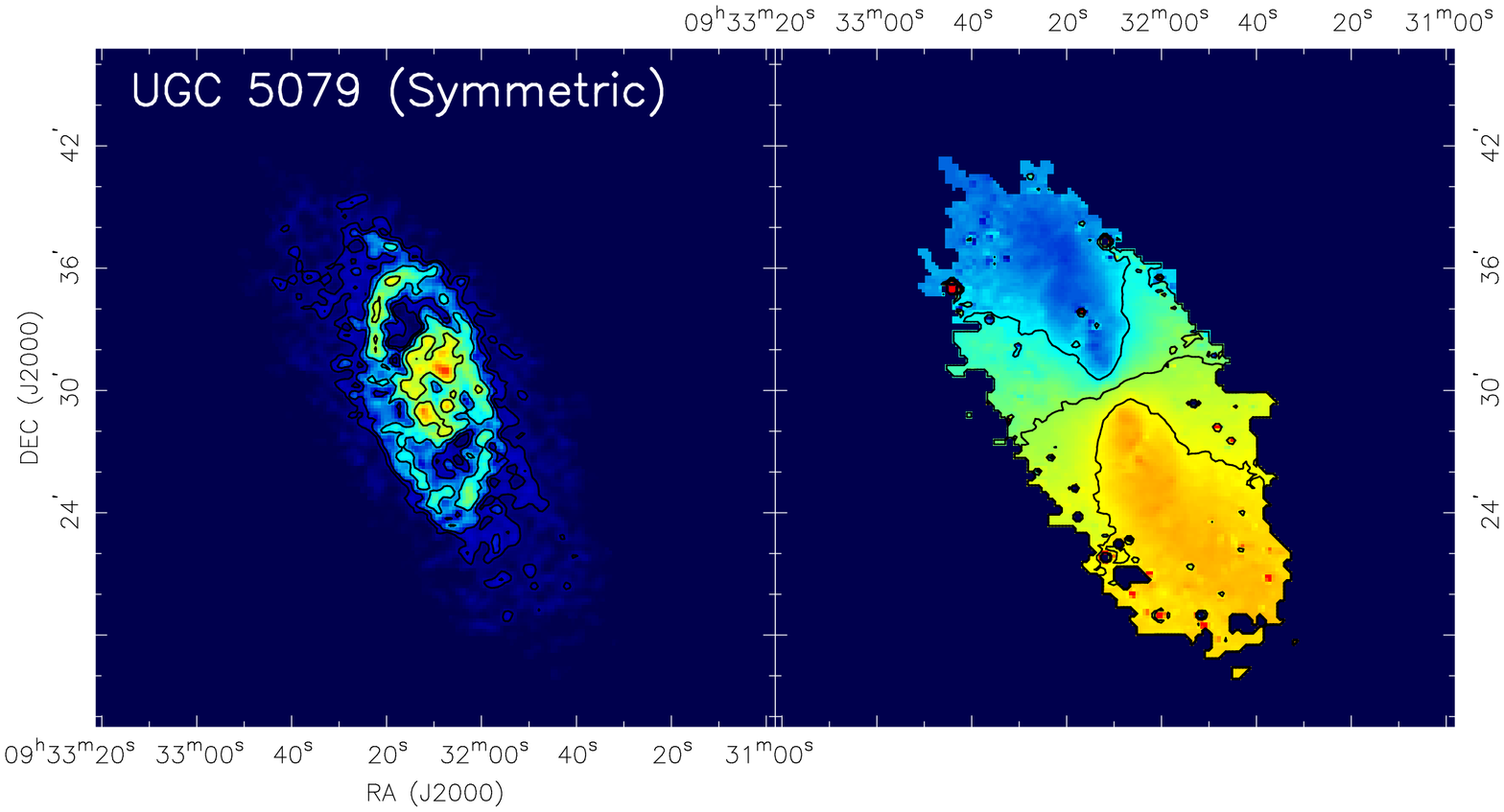}\\
\includegraphics[width=0.35\textwidth,angle=270]{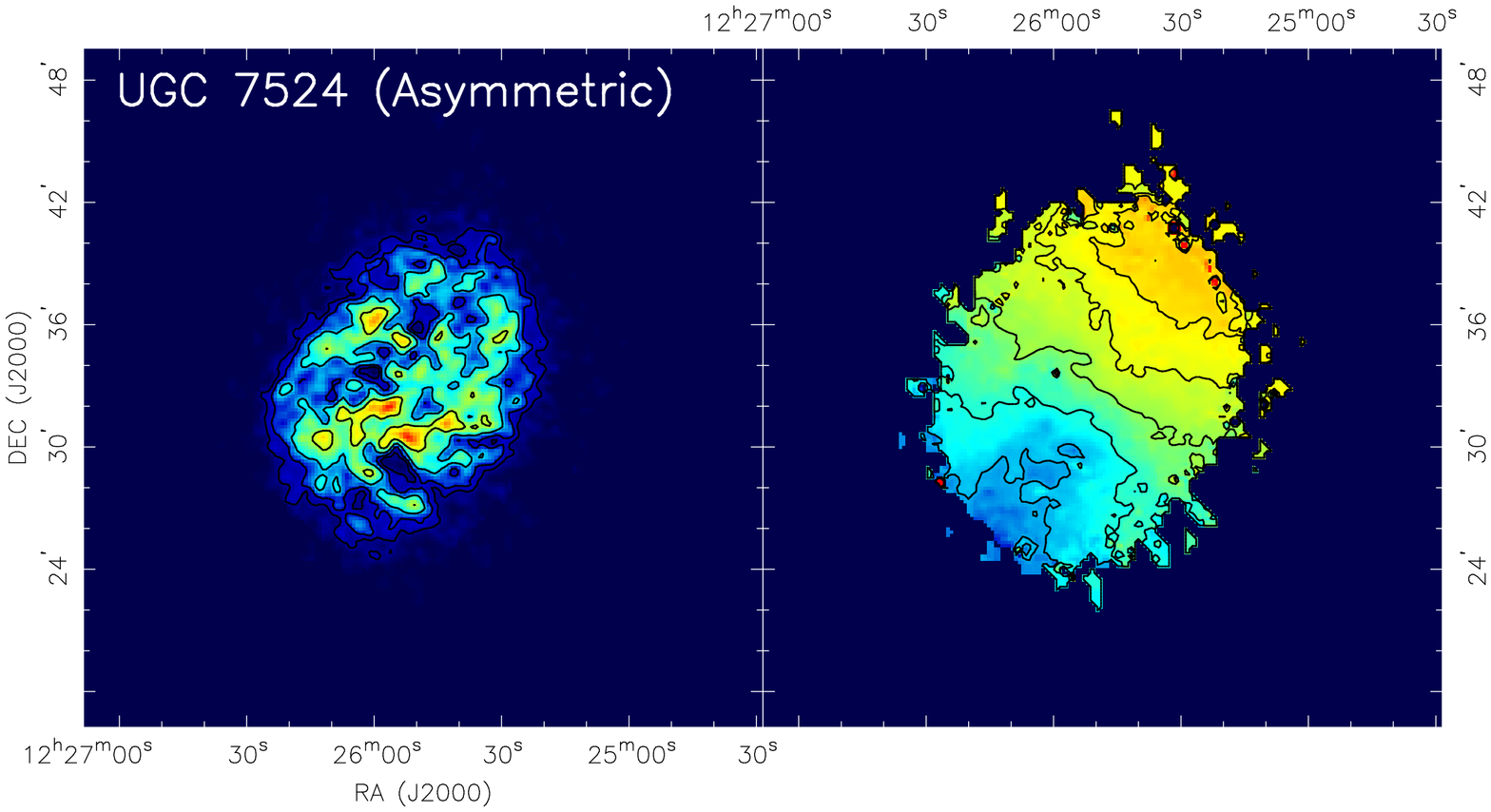}\\
\includegraphics[width=0.35\textwidth,angle=270]{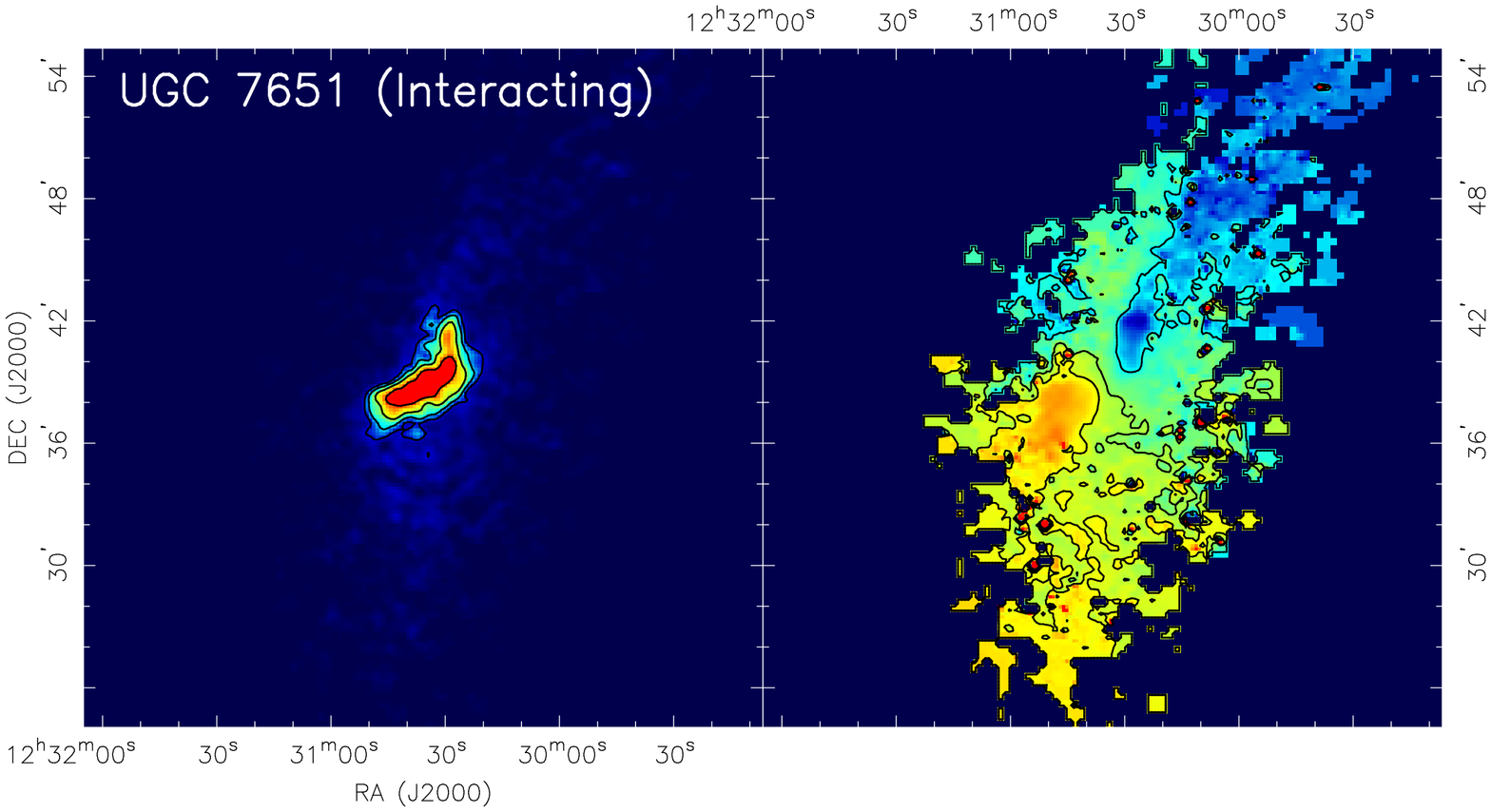}\\ 
\caption{The \HI\ integrated intensity (left panels) and velocity field (right panels) maps for the symmetric (top), asymmetric (middle), and interacting (bottom) galaxies. Left panels: the contours  are 0.18 $\times$ 1.78$^n$ mJy Beam$^{-1}$ km s$^{-1}$, with n=0, 1, 2, 3 for UGC 5079 (NGC 2903), 0.18 $\times$ 1.78$^n$ mJy Beam$^{-1}$ km s$^{-1}$, with n=0, 1, 2, 3 for UGC 7524 (NGC 4395), and 0.84 $\times$ 1.78$^n$ mJy Beam$^{-1}$ km s$^{-1}$, with n=0, 1, 2, 3 for UGC 7651 (NGC 4490). The lowest contour is 10$\%$ of the peak intensity. Right panels: the contours are 450, 550, and 650 km s$^{-1}$ for UGC 5079 (top), 268, 288, 318, 348, and 368 km s$^{-1}$ for UGC 7524 (middle), and 515, 565, and 605 km s$^{-1}$ for UGC 7651 (bottom).
\label{classfig}}
\end{center}
\end{figure*}

We have examined the \HI\ distributions (galaxy morphology),  integrated \HI\  profiles, position-velocity (p-v) diagrams,  and velocity fields of the galaxies and divided our sample into three groups: symmetric, asymmetric, and interacting galaxies based on the criteria mentioned above. Examples of symmetric, asymmetric and interacting galaxies are presented in Figure \ref{classfig}. In order to show the classification more distinctly in the figure, we used the smoothed version (30\ac) instead of the highest resolution images that we use for this study.
During this classification, we found an intermediate class between symmetric and asymmetric features, so we classified the intermediate galaxies as A/S. 
Our classification is presented in Table \ref{tab:sample}. 
When comparing with the previous studies by \citeauthor{2011A&A...530A..29V} (\citeyear{2011A&A...530A..29V}, \citeyear{2011A&A...530A..30V}) for galaxies in common, our classification is generally consistent with their results.

\section{Results}
\label{results}

In order to compare the SFR and the gas surface densities, we first brought all the images to the same resolution using the MIRIAD task CONVOL and then obtained radial profiles of FUV, IR 24 \um, CO, and \HI\ using the Groningen Image Processing System (GIPSY; \citealt{1992ASPC...25..131V}) task ELLINT. In the case of interacting galaxies, the  galaxy merging with UGC 7651 and the tidal tail of UGC 4862 are masked before deriving the  surface densities. All the radial profiles are corrected to face-on by multiplying by cos $i$. The radial sampling is 10\ac\ or 20\ac\ depending on spatial resolution of the object.

\subsection{Radial Distributions}
\label{radial}

The task ELLINT averages the data along tilted rings, so it increases the signal to noise of the measurement. 
Figure \ref{rprof} shows the azimuthally averaged radial profiles of \sigsfr, \sightwo, \sighi, and \siggas\ for the sub-sample of 16 galaxies for which CO data are available. 
Figure \ref{restrp} shows the profiles of \sigsfr\ and \sighi\ for the remaining 13 galaxies in the sample that have no CO data. 
In order to derive the SFR surface density, we combine FUV and 24 \um\ radial profiles using the calibration given by \citet{2008AJ....136.2782L}, which is originally based on previous studies (\citealt{2007ApJ...666..870C}; \citealt{2007ApJ...671..333K};  \citealt{2007ApJS..173..267S}):
\begin{equation}
\sigsfr [\Msol \, {\rm kpc^{-2} \, yr^{-1}}] = 0.081 I_{\rm FUV} \,[{\rm MJy \, sr^{-1}}]\\
+ 0.0032 I_{24 \mu m} \,[\rm MJy \, sr^{-1}].
\label{eq:sfrcombi}
\end{equation}

The gas profiles are converted to units of \surm\ by adopting the standard CO-to-\Htwo\ conversion factor (\citealt{1996A&A...308L..21S}; \citealt{2001ApJ...547..792D}) and assuming optically  thin emission in \HI: 
\begin{eqnarray}
\sightwo [\surm] = 3.2 I_{\rm CO} \,[\rm K \,\kms], \\
\sighi [\surm] = 0.0146 I_{\rm HI} \,[\rm K \, \kms].
\label{eq:coh1}
\end{eqnarray}
The total gas mass is estimated via 1.36(\sightwo\ + \sighi), where the factor of 1.36 is a correction for helium. All the radial profiles for \sighi\ and \sightwo\ are shown in Figure \ref{rprof} (left panels). When we use the HERACLES CO ($J=2\rightarrow1$) data for \sightwo, we adopt a conversion factor of 0.7 \citep{2012AJ....144....3L} for the line ratio CO(2--1)/CO(1--0).
The IRAM PdBI CO data for UGC 3334 given by \citet{2009A&A...503...73C} are not reliable beyond 22\ac\ in radius since the primary beam of PdBI is 43\ac. Instead, for the outside regions beyond 22\ac, we have extracted data points from the FCRAO CO profile  \citep{1995ApJS...98..219Y} using the Dexter tool.  The two different data sets match each other well at the radius of 22\ac\ (see the UGC 3334 in the figure).  
Most galaxies show \sightwo\ $>$ \sighi\ in the central regions as expected, but few cases are different (e.g., UGC 1913, UGC 7278, UGC 7651). The galaxies UGC 1913 (NGC 925) and 7278 (NGC 4214) are known as late-type galaxies with faint emission in CO \citep{2009AJ....137.4670L} and the galaxy UGC 7651 (NGC 4490) is an irregular galaxy interacting with NGC 4485.

\begin{figure*}
\begin{center}
\begin{tabular}{c@{\hspace{0.1in}}c@{\hspace{0.1in}}}
\includegraphics[width=0.35\textwidth]{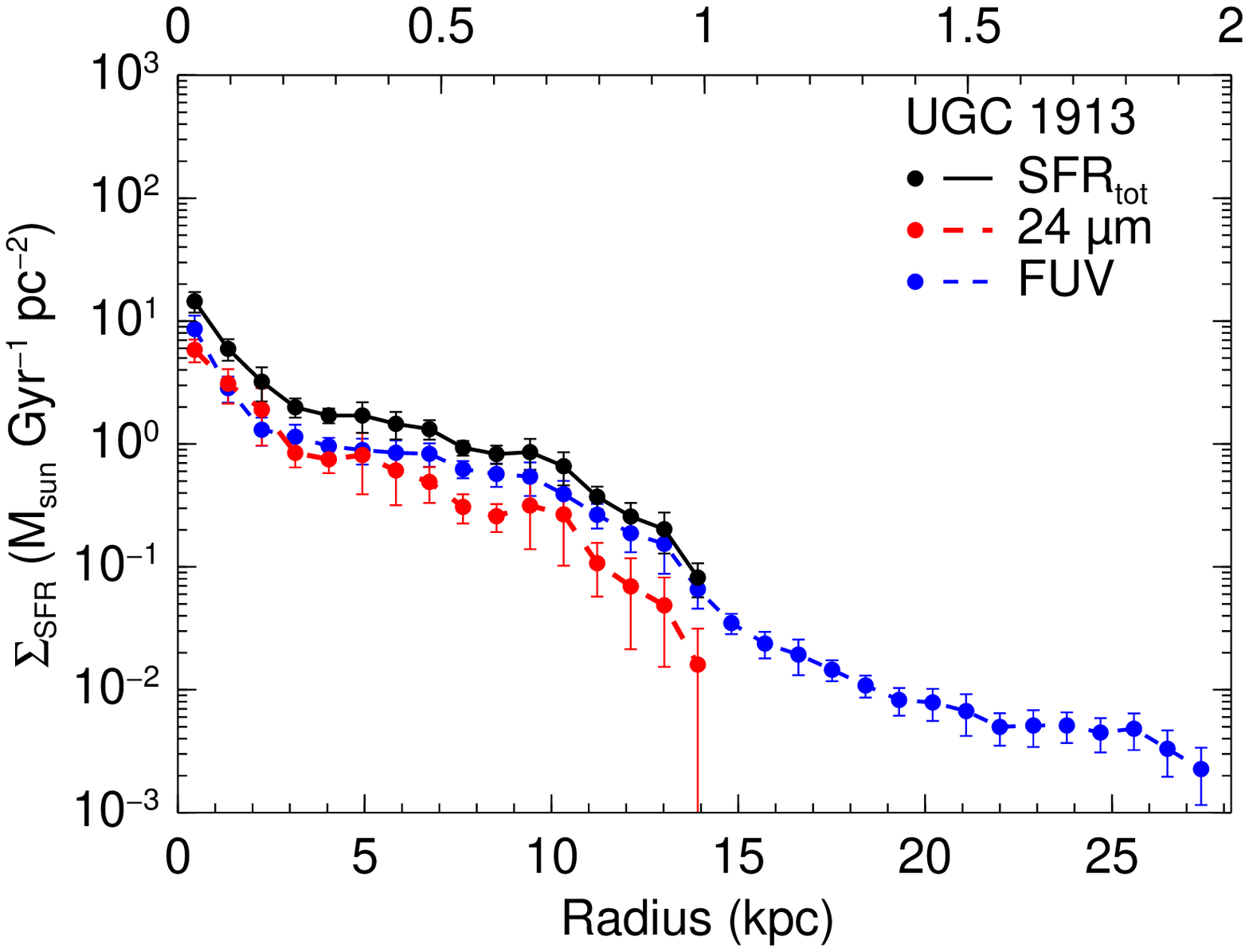}&
\includegraphics[width=0.35\textwidth]{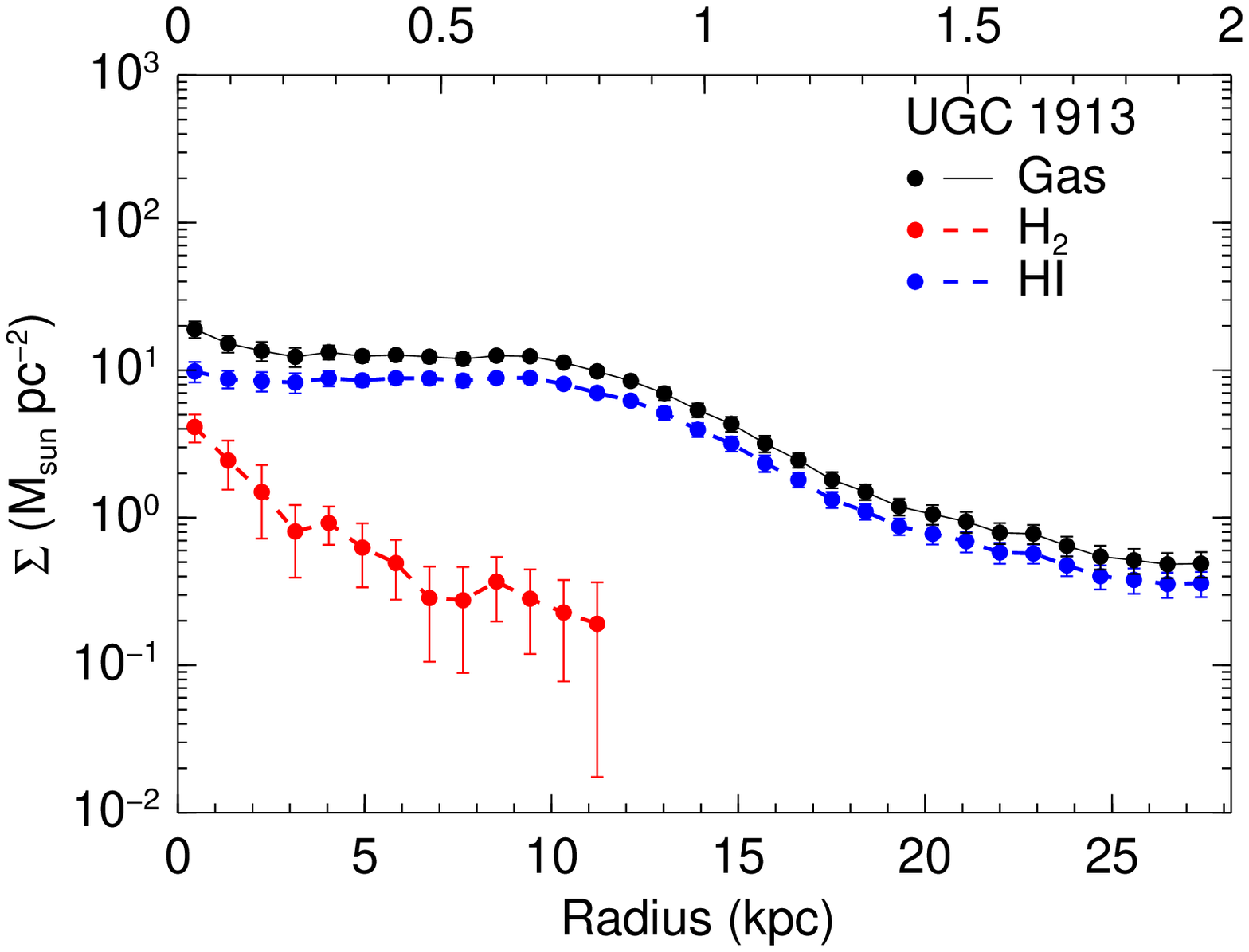}\\
\includegraphics[width=0.35\textwidth]{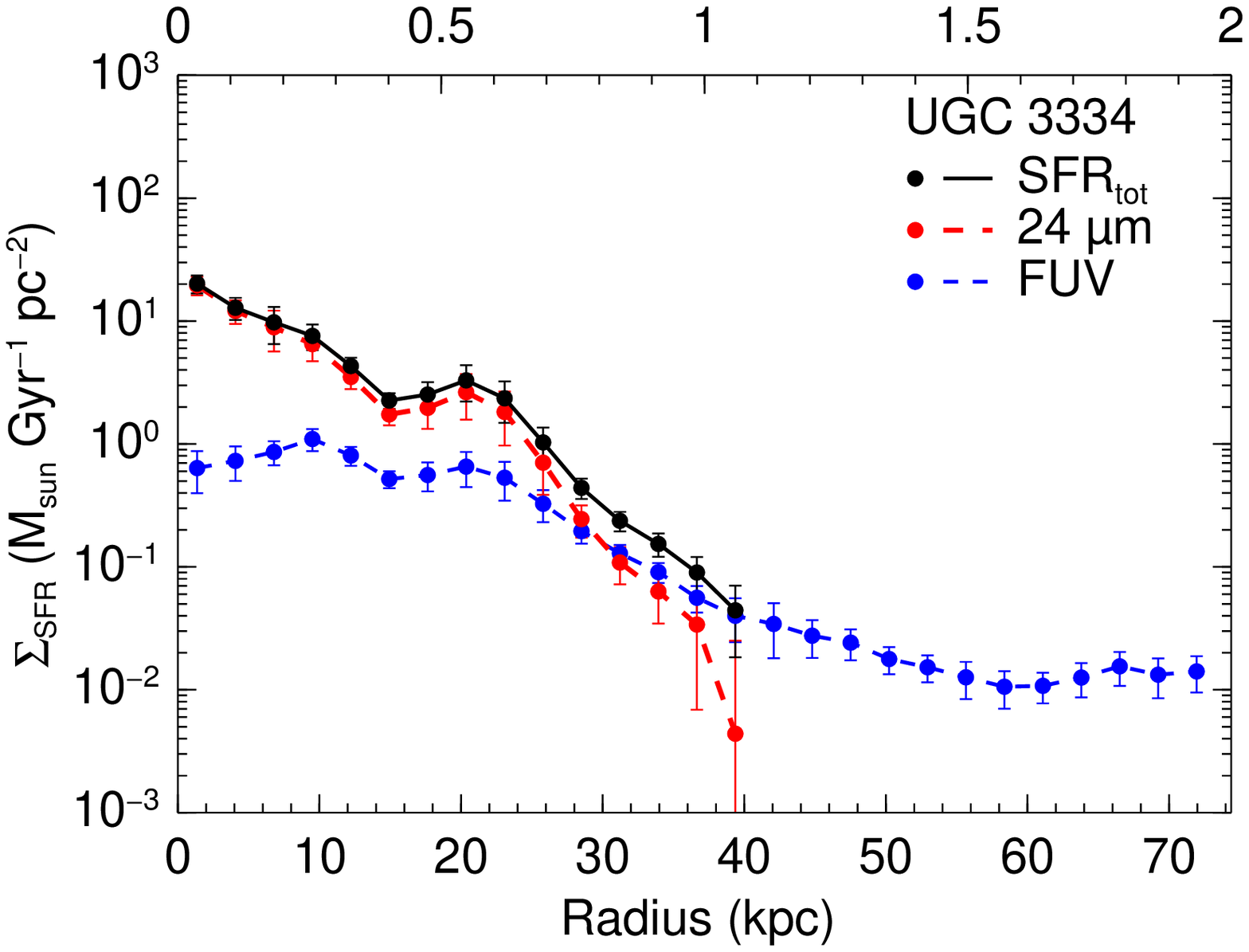}&
\includegraphics[width=0.35\textwidth]{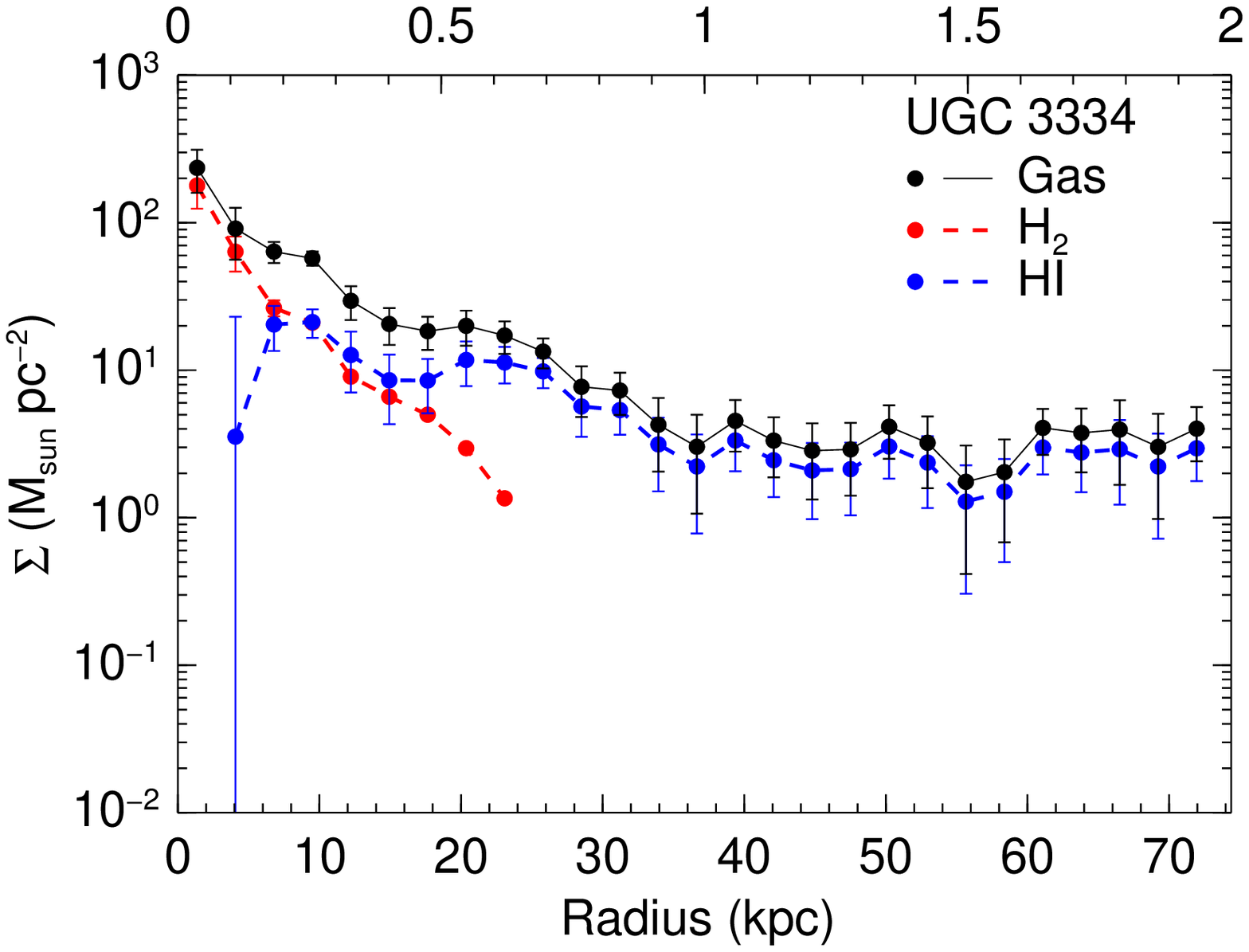}\\
\includegraphics[width=0.35\textwidth]{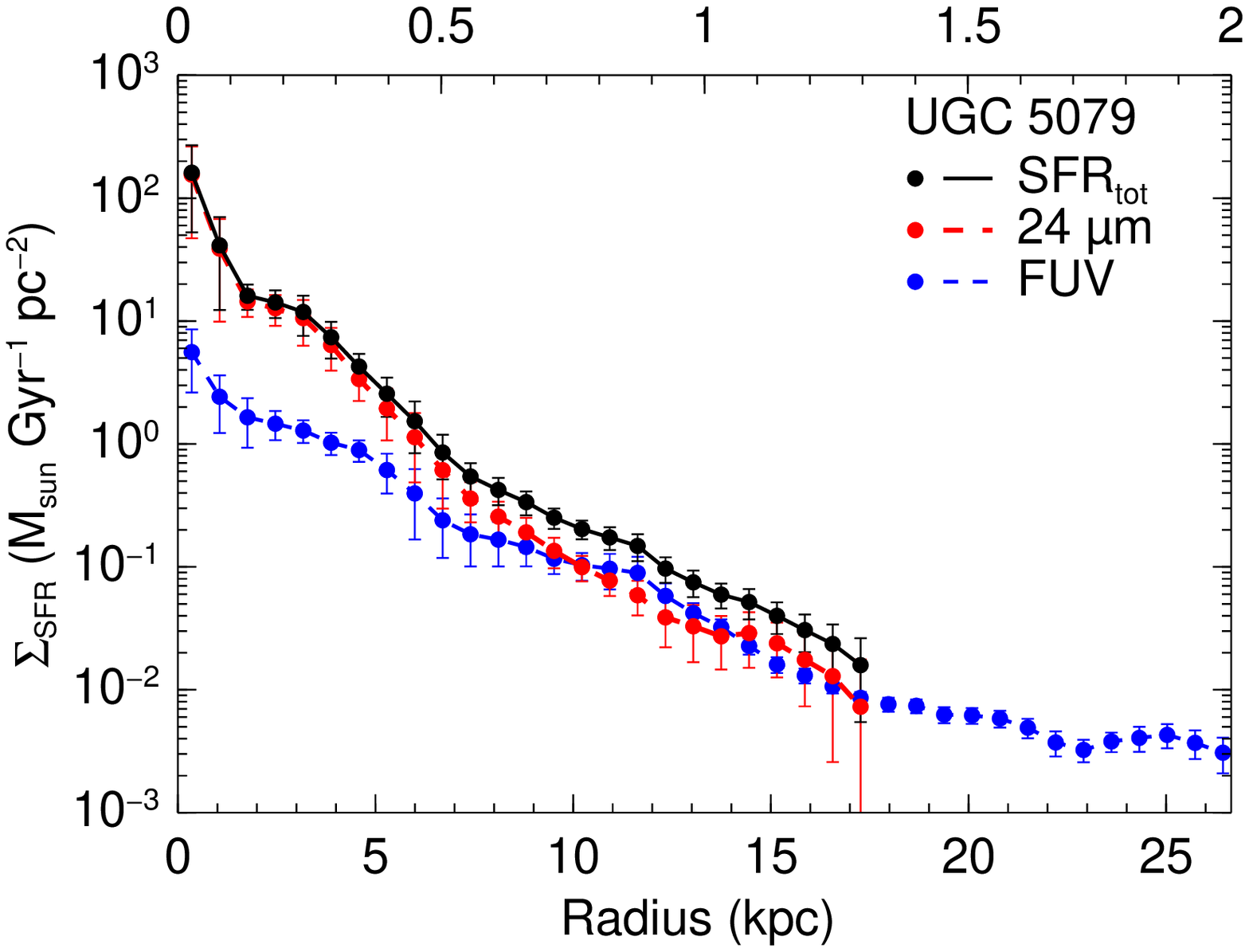}&
\includegraphics[width=0.35\textwidth]{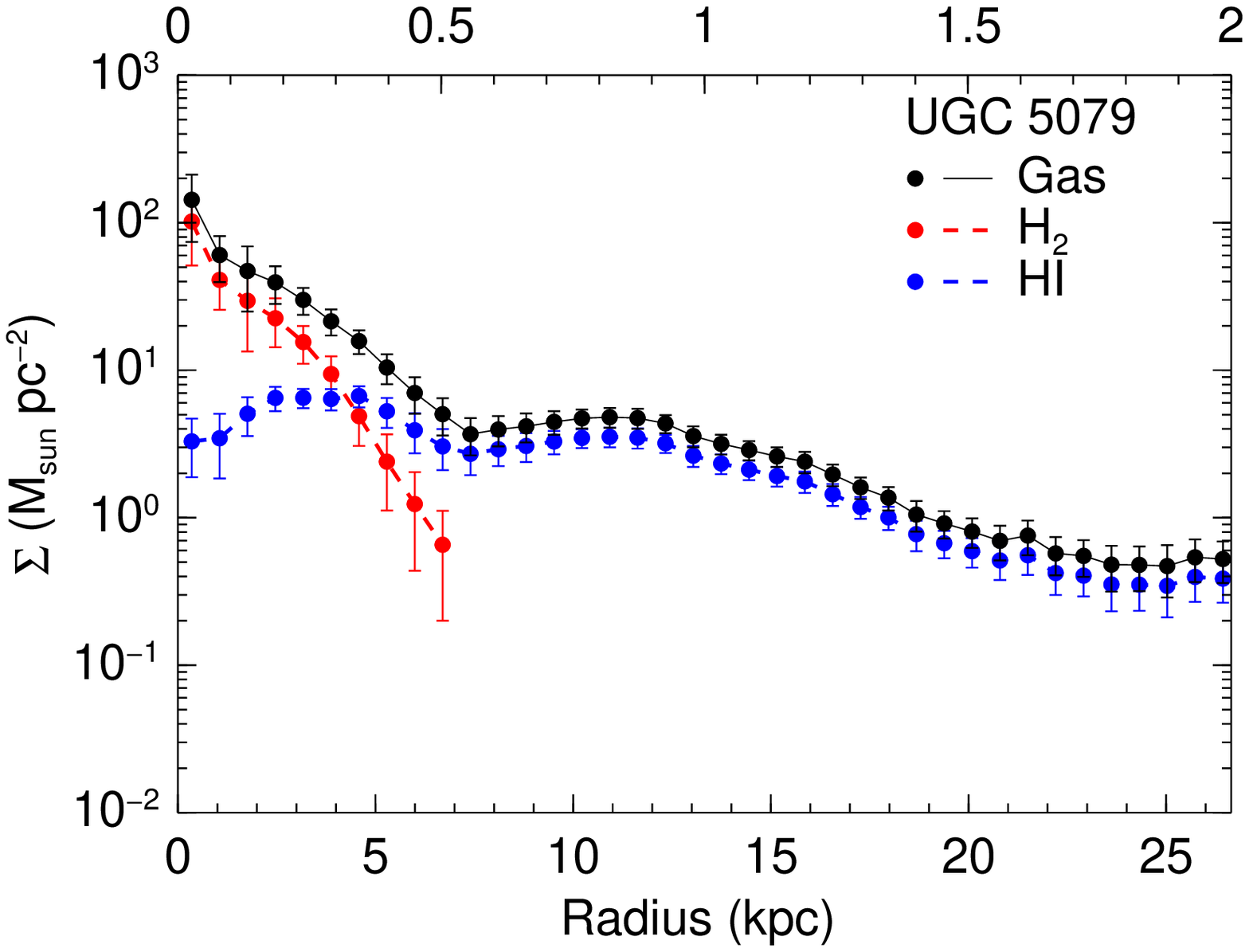}\\
\includegraphics[width=0.35\textwidth]{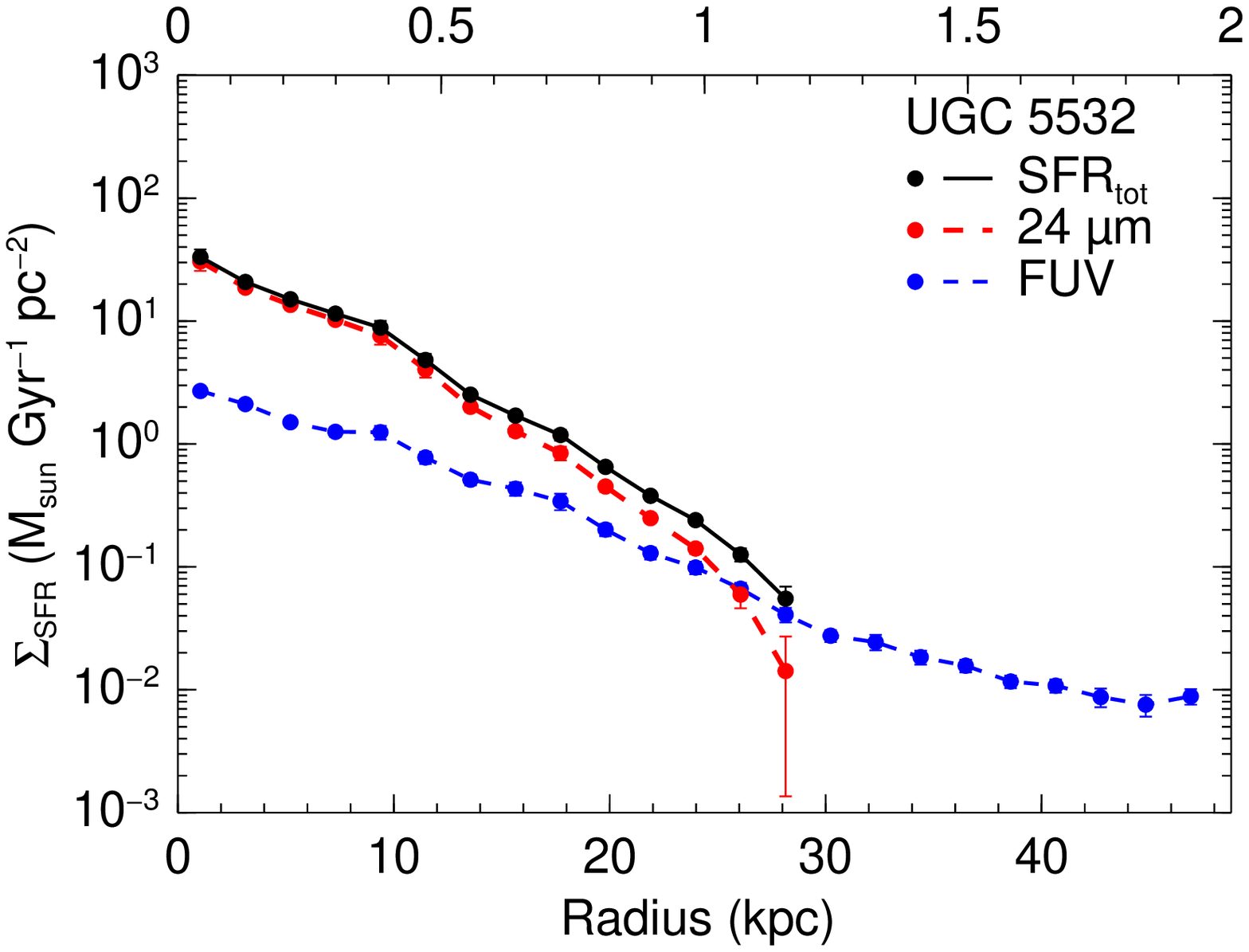}&
\includegraphics[width=0.35\textwidth]{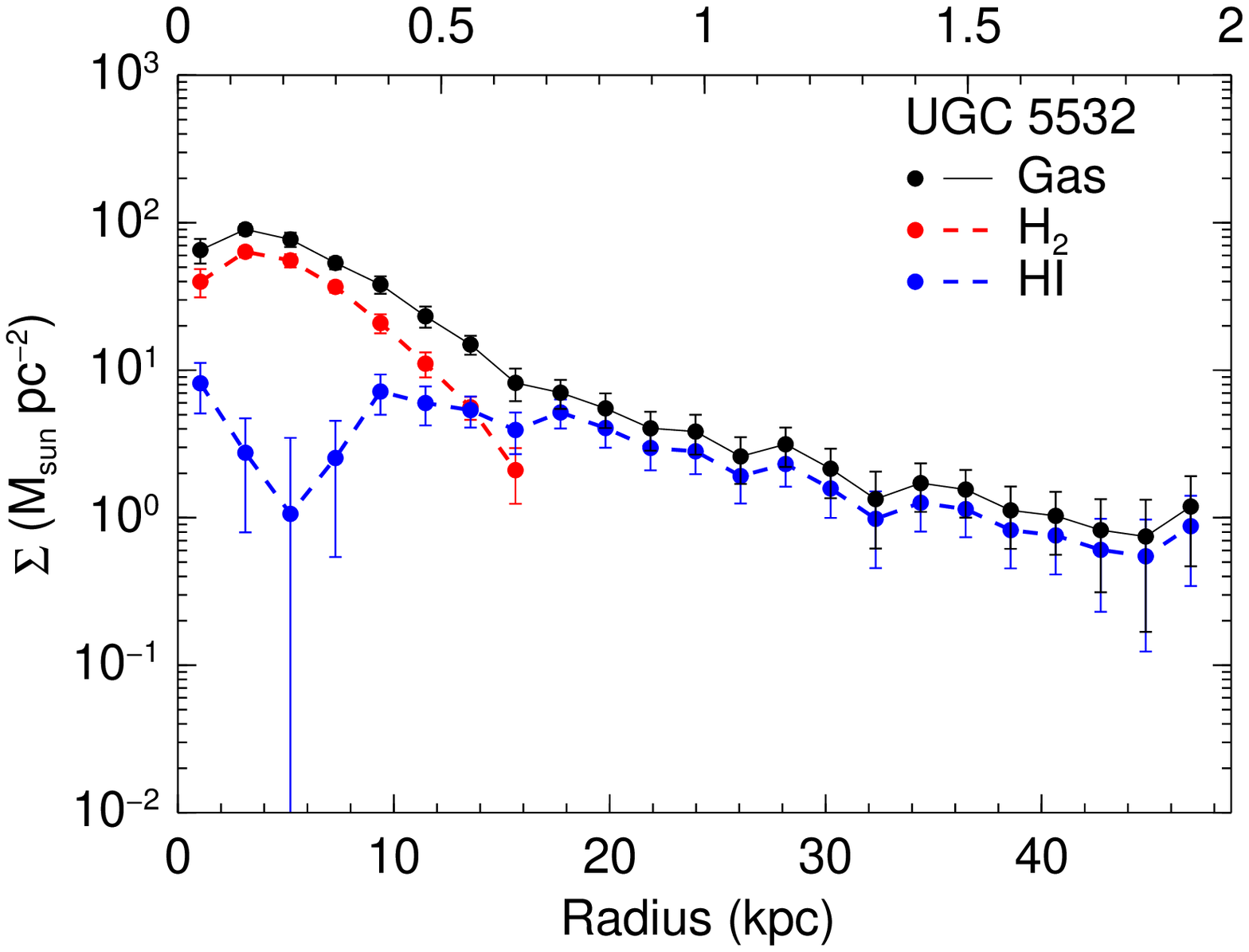}
\end{tabular}
\caption{Left panels: radial profiles of 0.081$I_{\rm FUV}$ (blue), 0.0032$I_{24 \mu m}$ (red), and the combined \sigsfr\ (black) given by Equation \ref{eq:sfrcombi}.  Right  panels:  radial profiles of \sightwo\ (red), \sighi\ (blue), and total gas \siggas\ (black). The helium factor (1.36) is included in the total gas: 1.36(\sightwo\ +\sighi). 
\label{rprof}}
\end{center}
\end{figure*}

\addtocounter{figure}{-1}

\begin{figure*}
%\figurenum{2}
\begin{center}
\begin{tabular}{c@{\hspace{0.1in}}c@{\hspace{0.1in}}}
\includegraphics[width=0.35\textwidth]{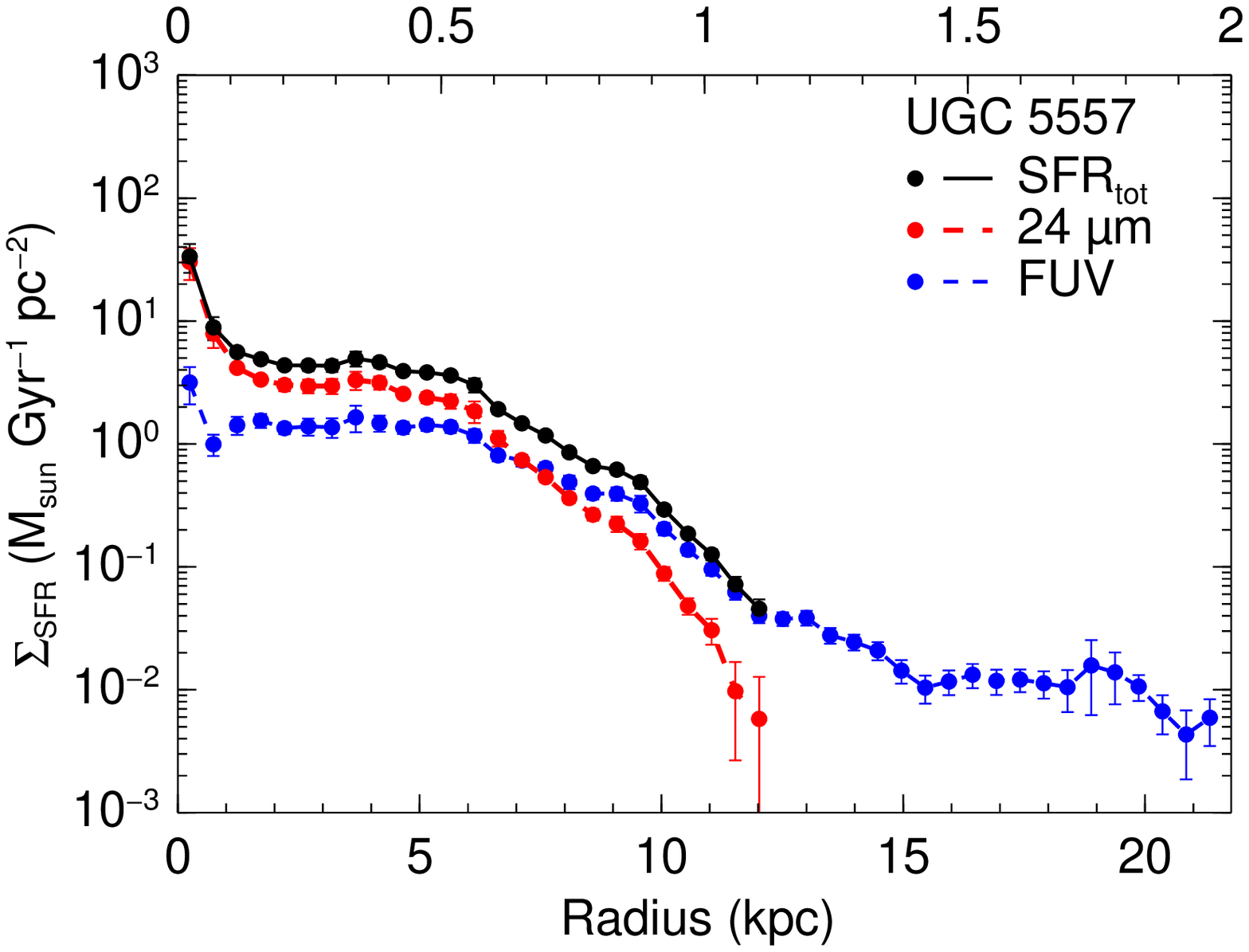}&
\includegraphics[width=0.35\textwidth]{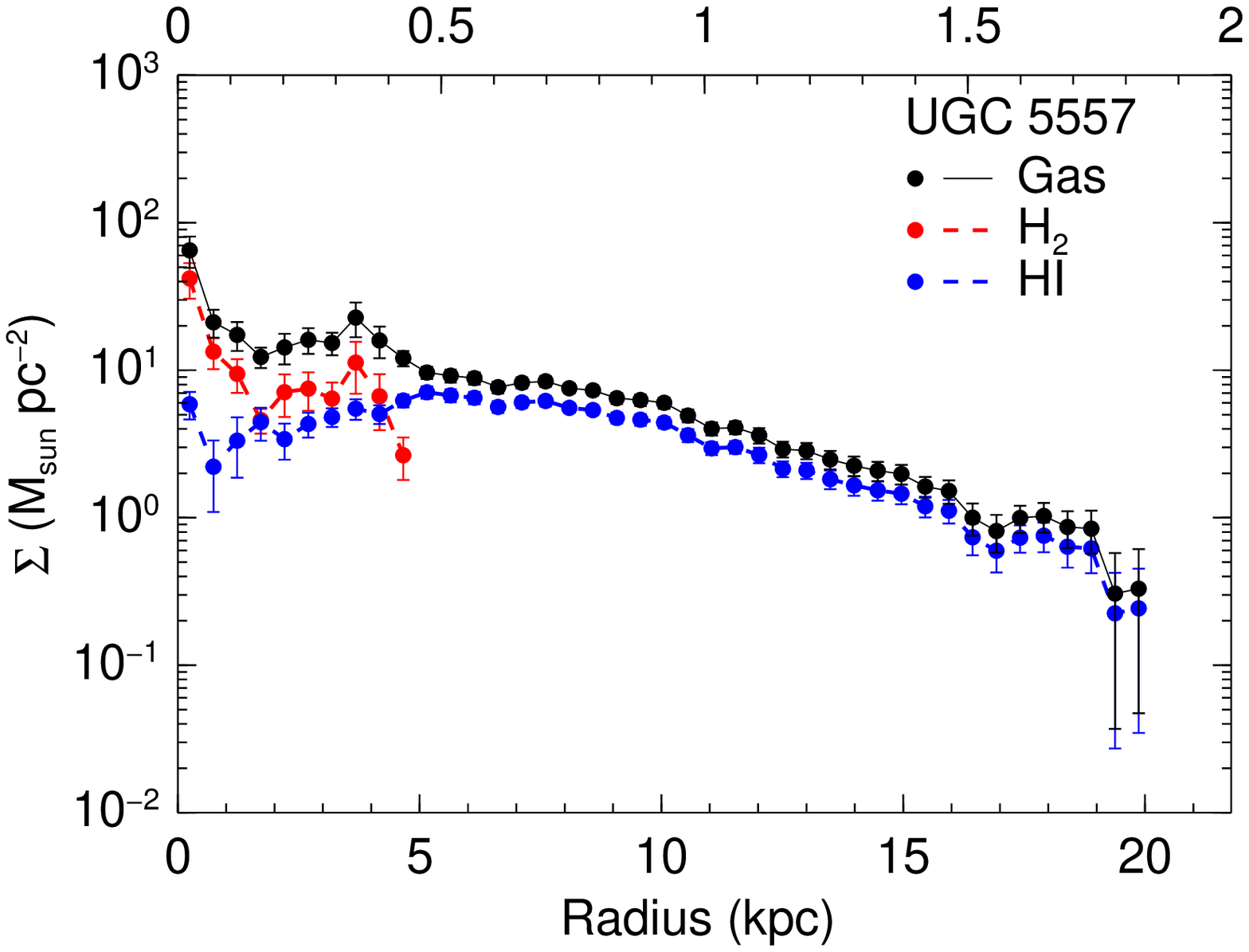}\\
\includegraphics[width=0.35\textwidth]{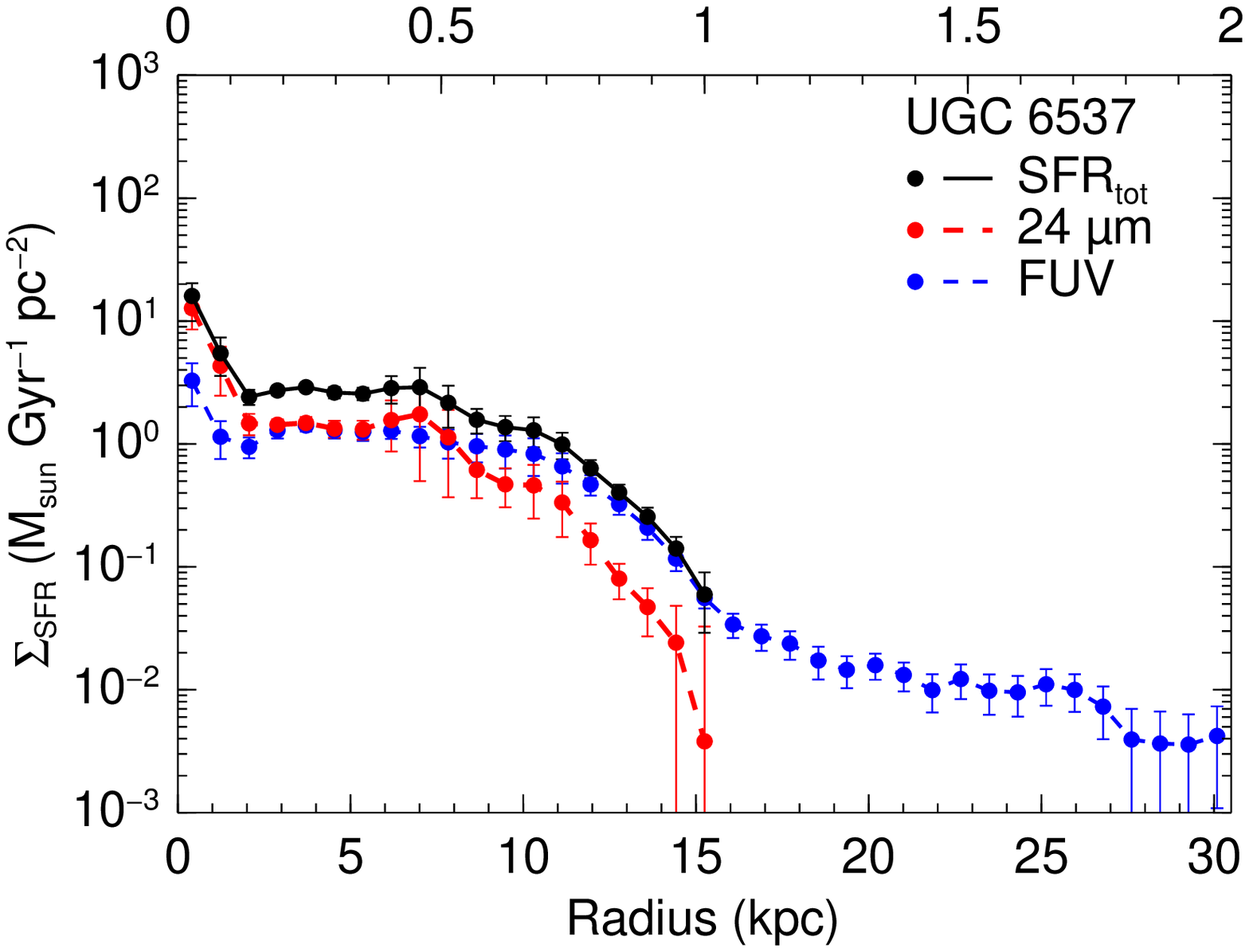}&
\includegraphics[width=0.35\textwidth]{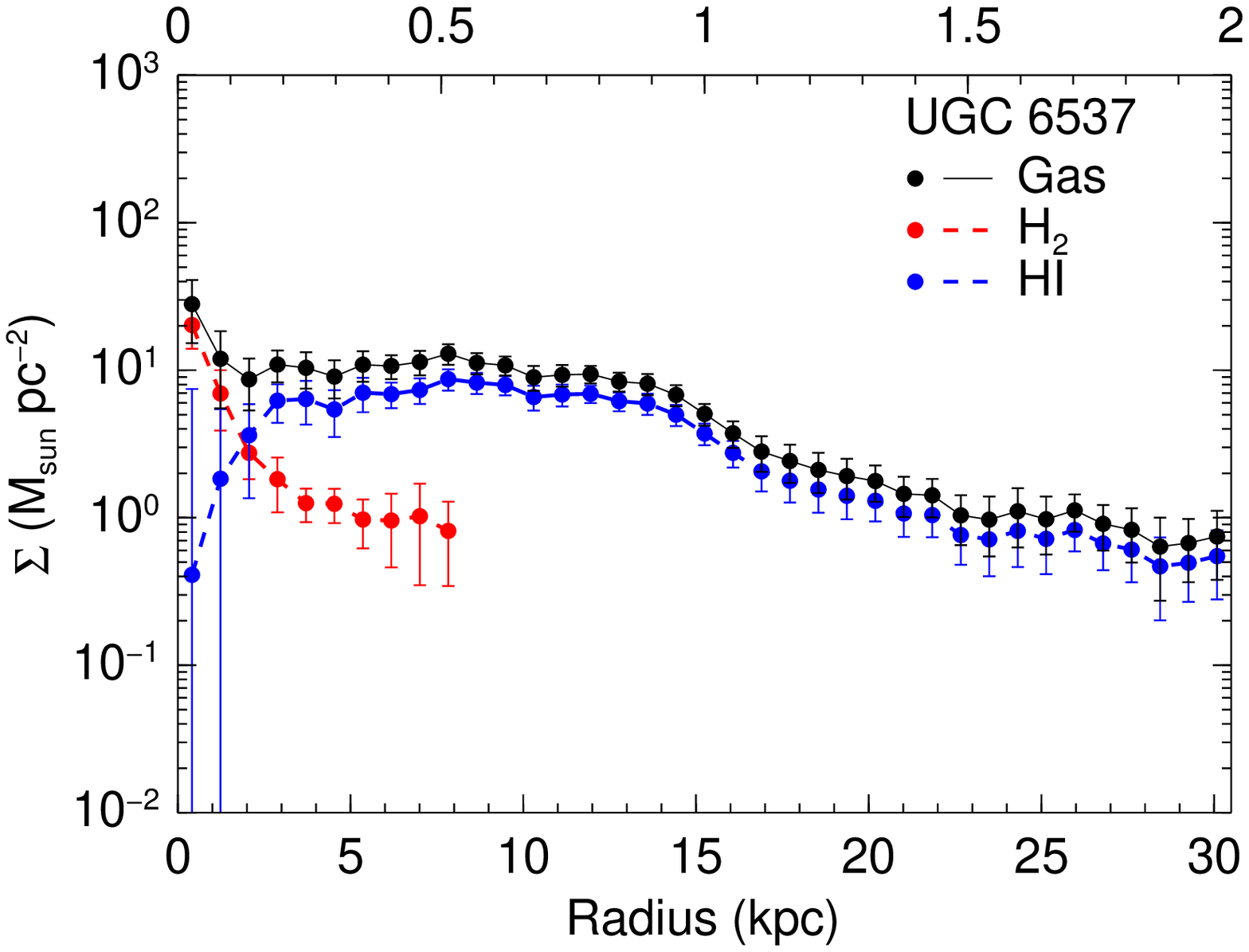}\\
\includegraphics[width=0.35\textwidth]{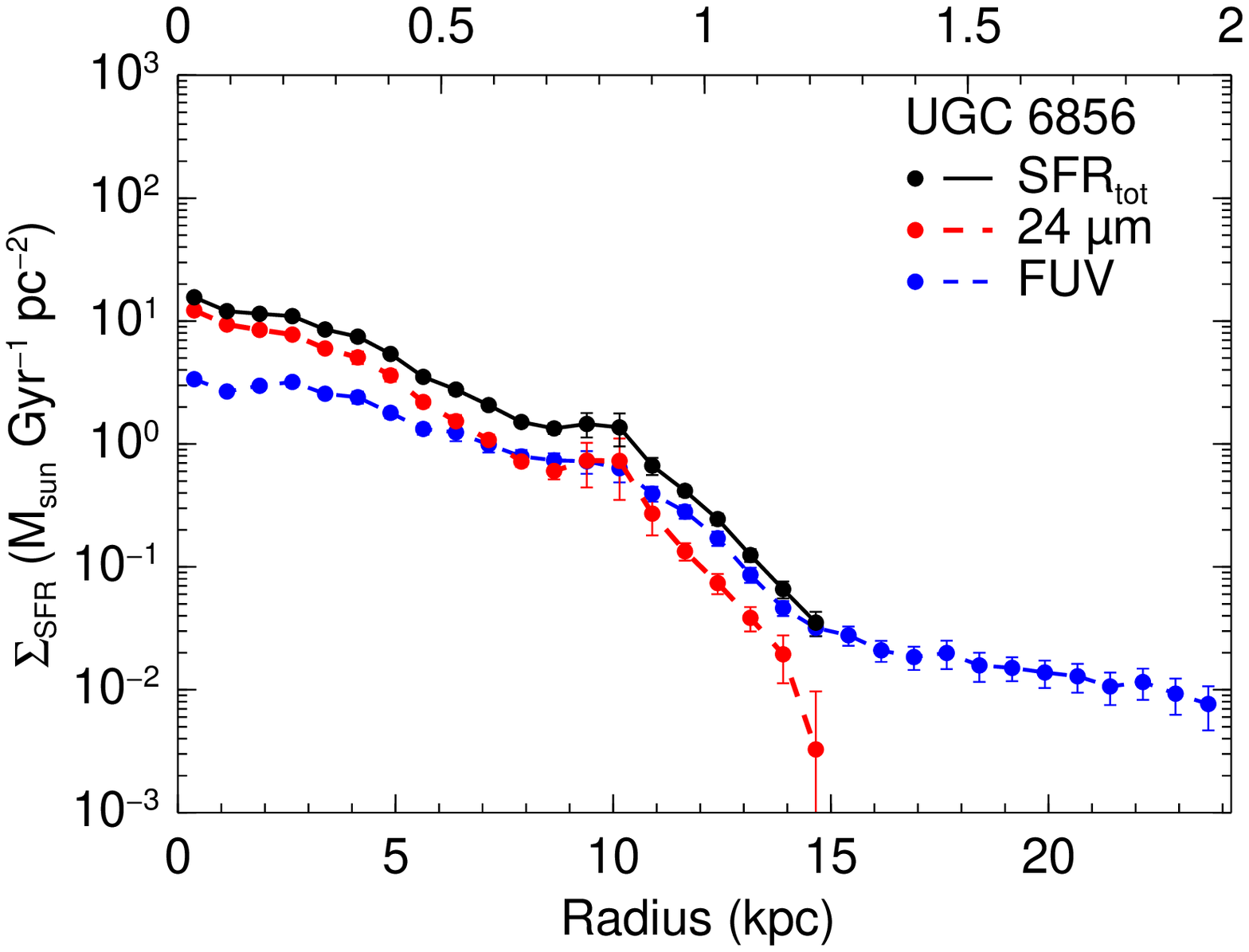}&
\includegraphics[width=0.35\textwidth]{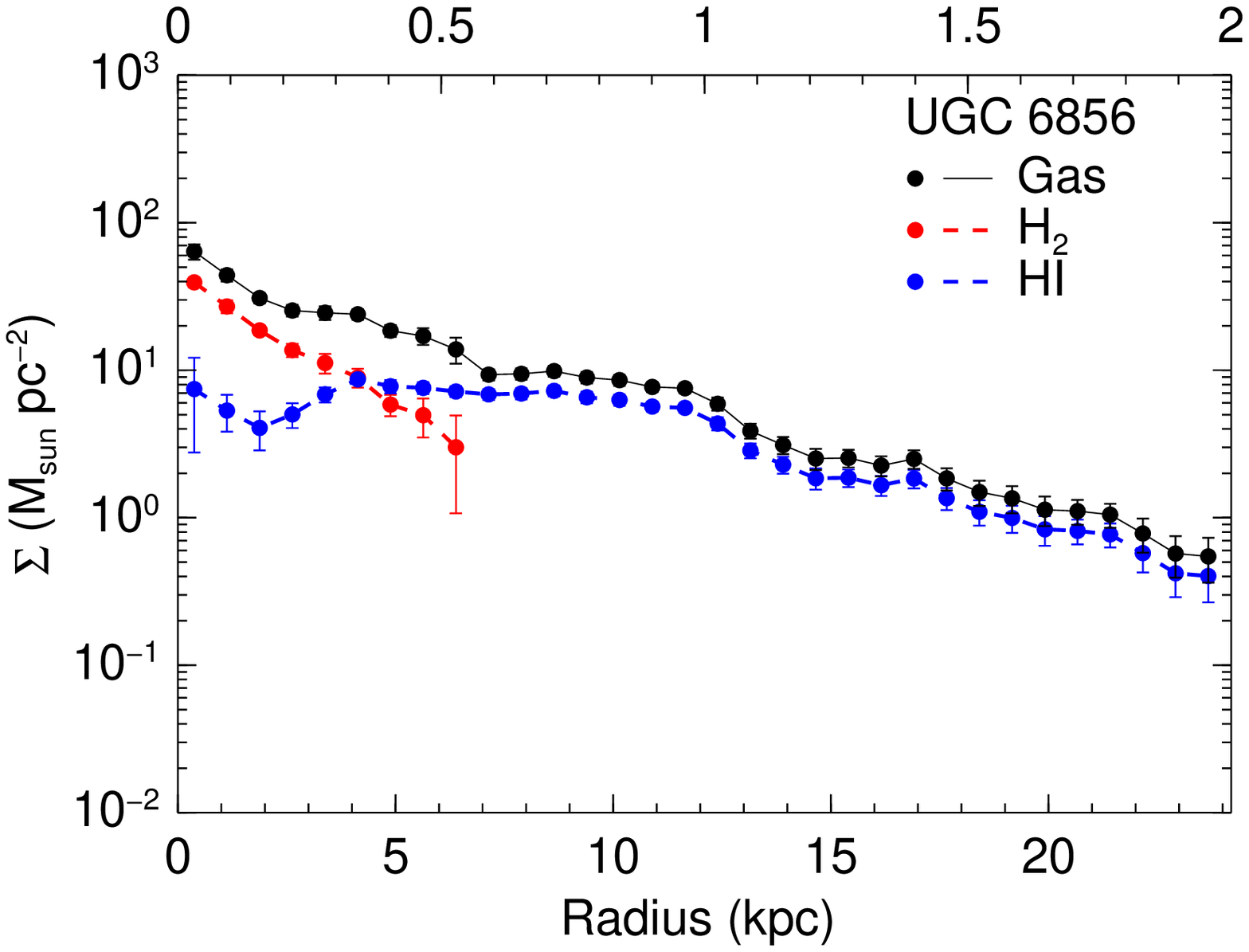}\\
\includegraphics[width=0.35\textwidth]{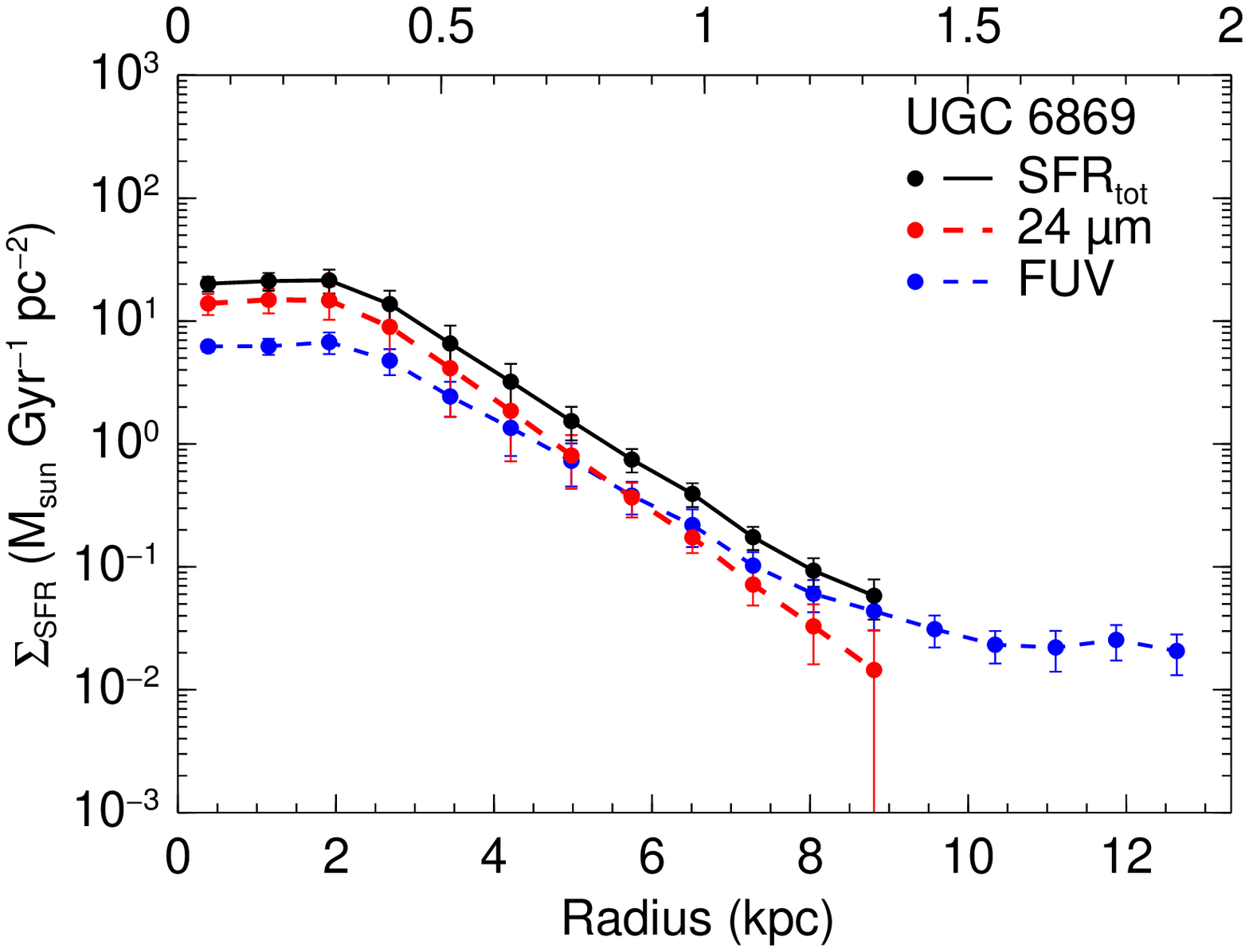}&
\includegraphics[width=0.35\textwidth]{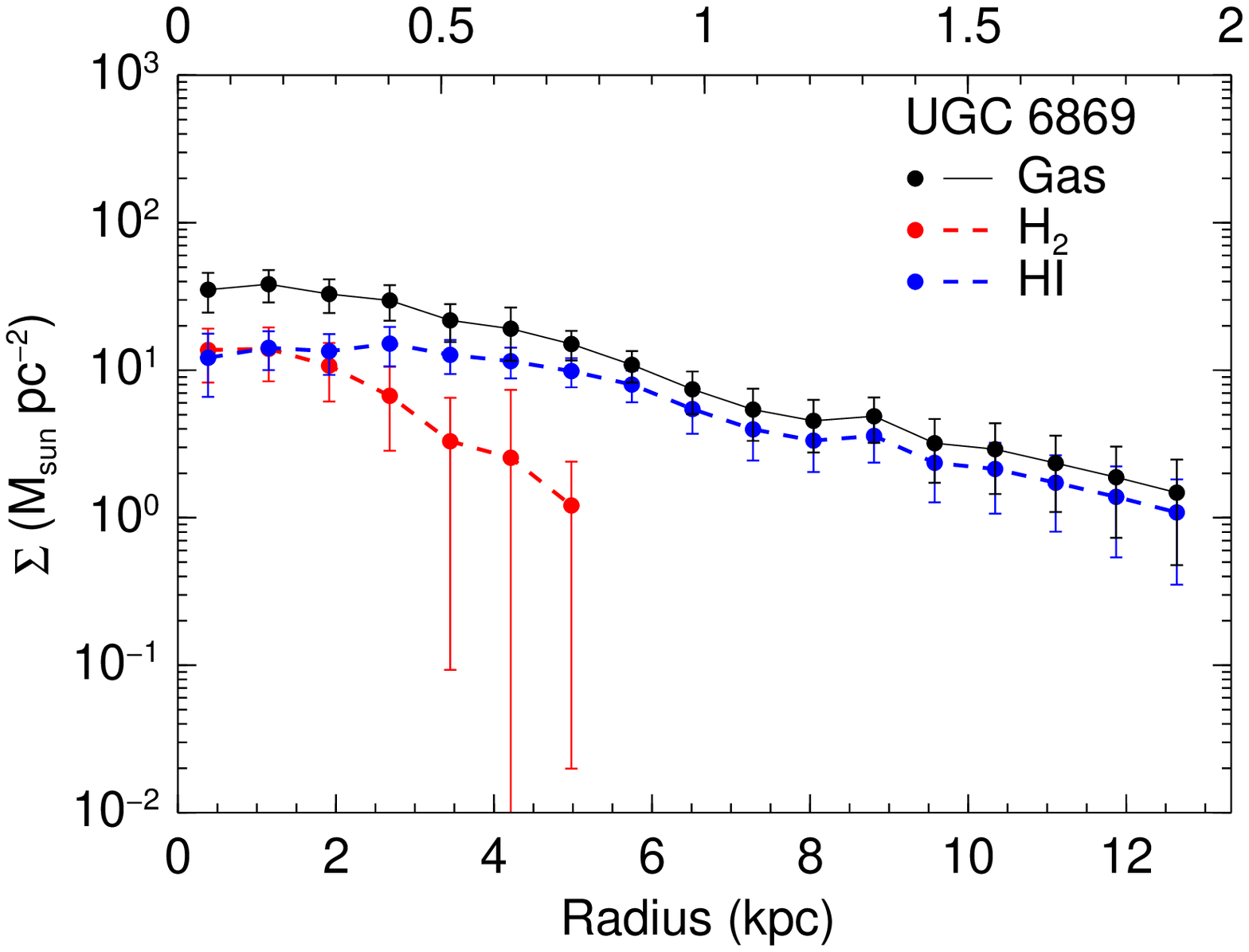}\\
\end{tabular}
\caption{continued}
\label{rprof}
\end{center}
\end{figure*}

\addtocounter{figure}{-1}

\begin{figure*}
%\figurenum{2}
\begin{center}
\begin{tabular}{c@{\hspace{0.1in}}c@{\hspace{0.1in}}c@{\hspace{0.1in}}c}
\includegraphics[width=0.35\textwidth]{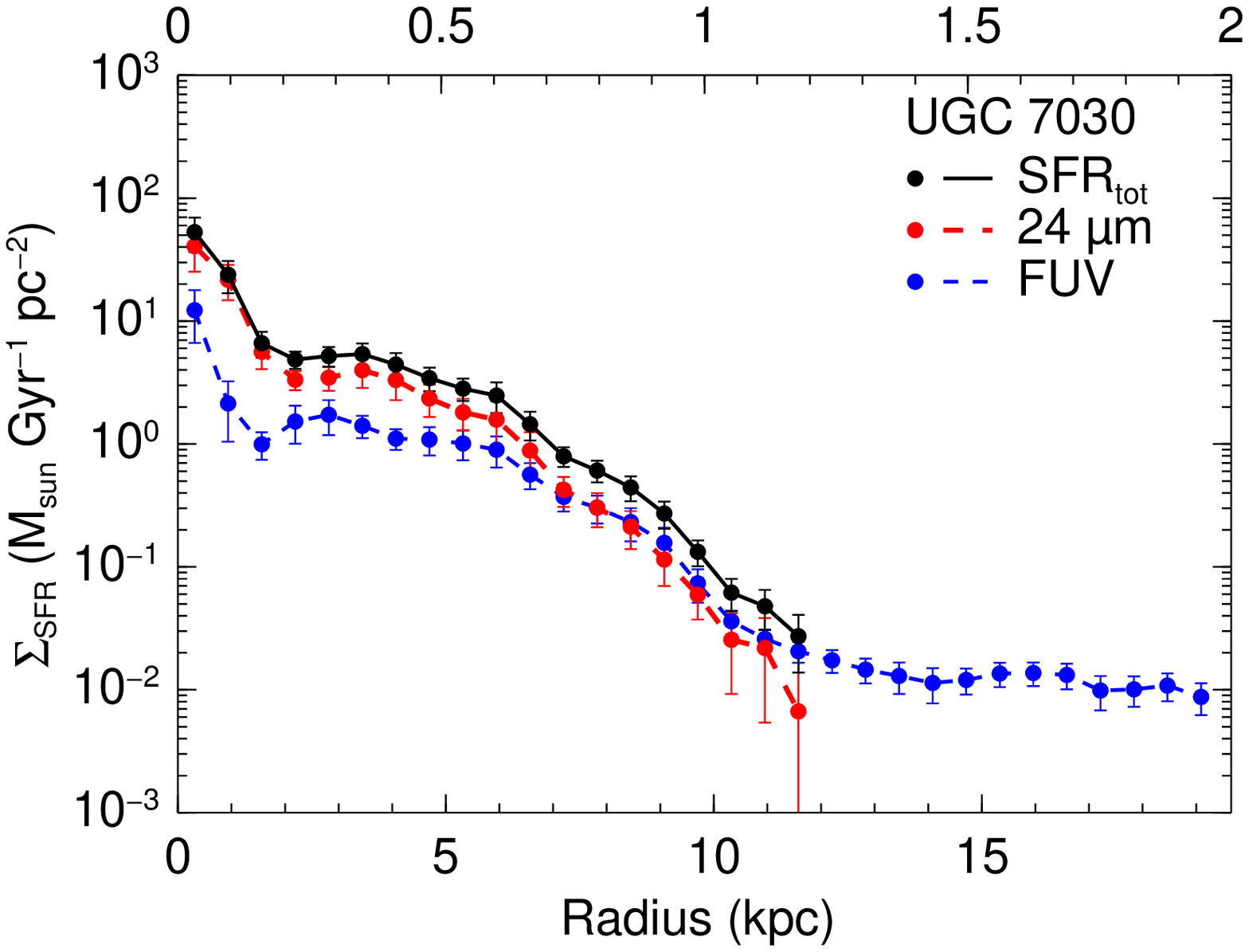}&
\includegraphics[width=0.35\textwidth]{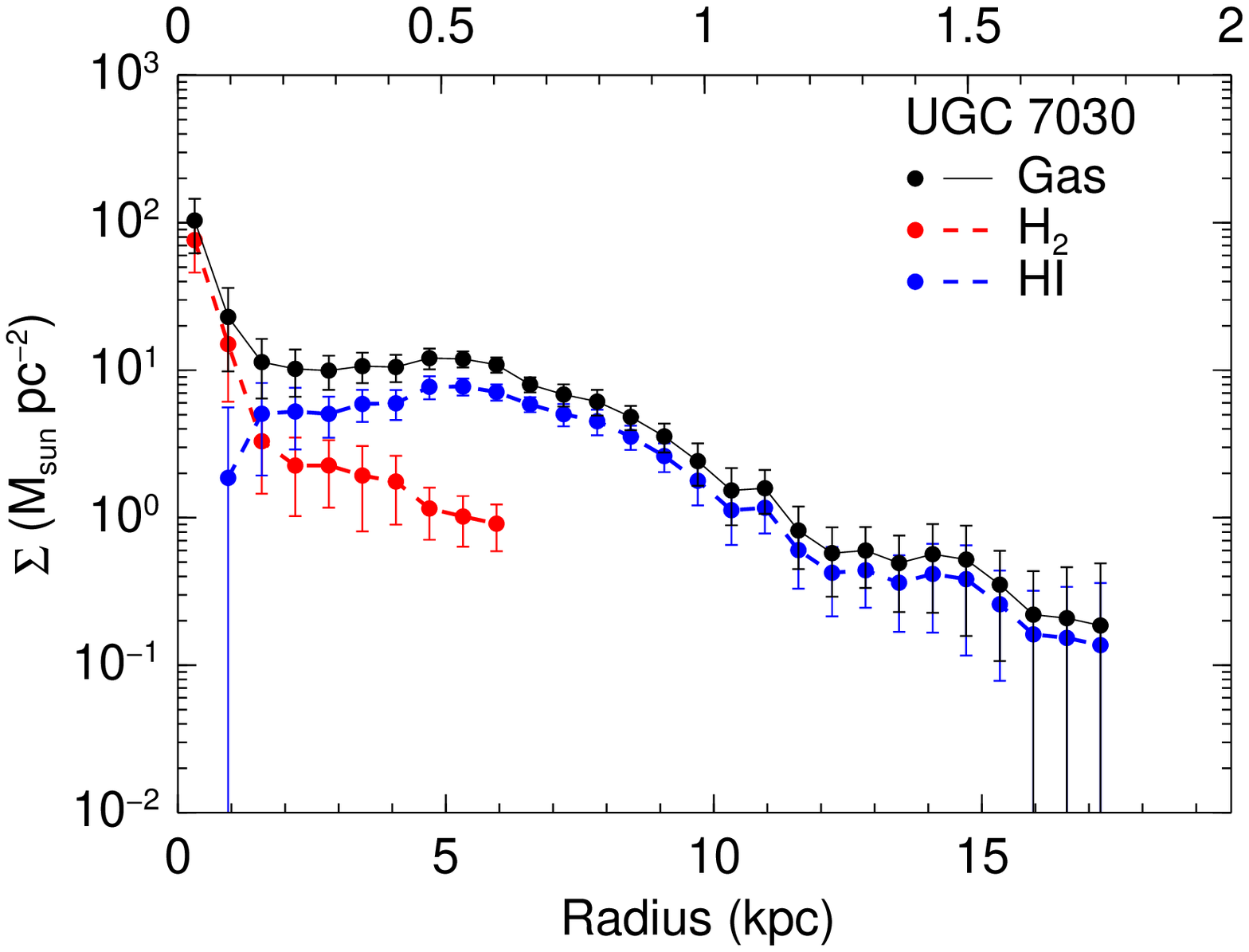}\\
\includegraphics[width=0.35\textwidth]{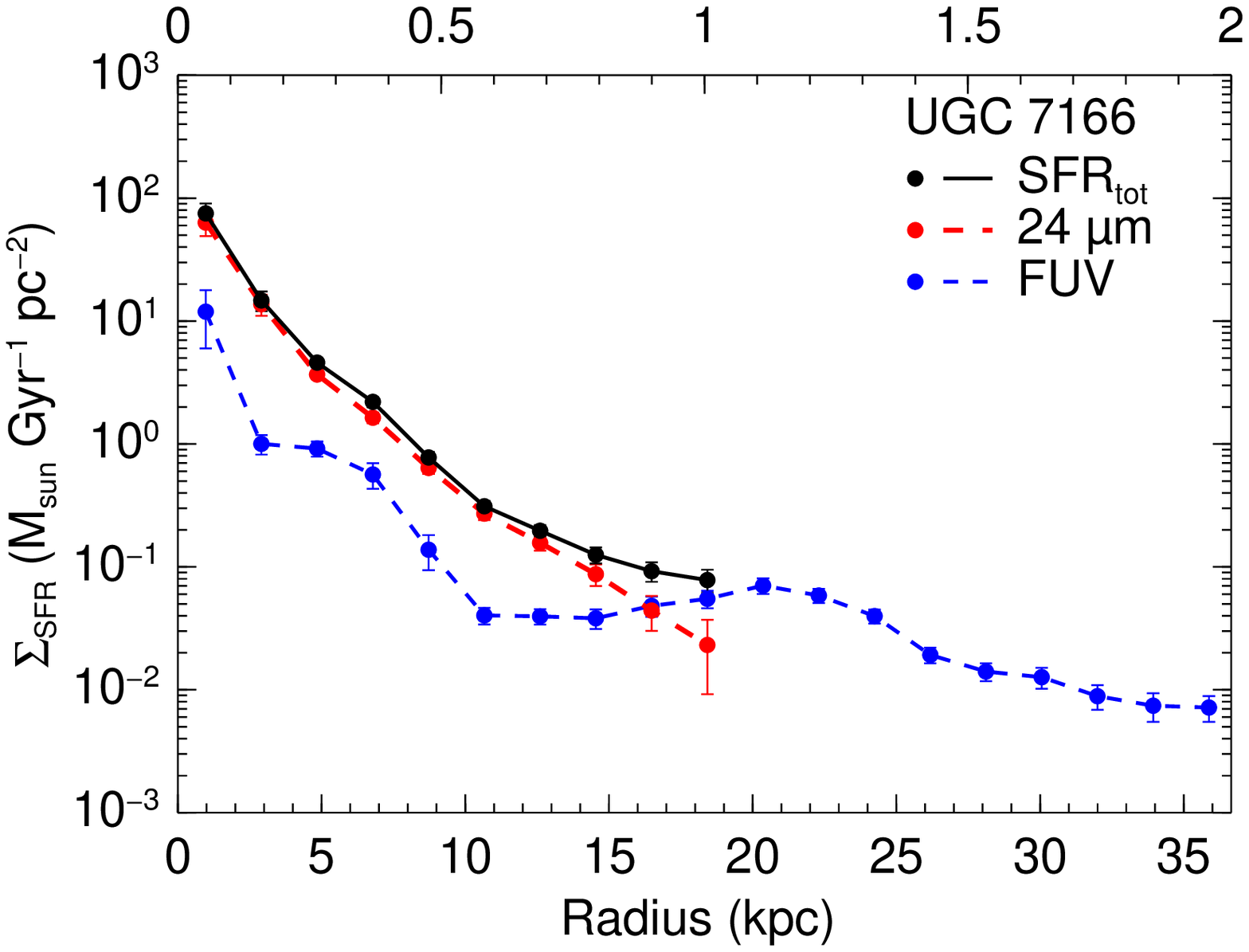}&
\includegraphics[width=0.35\textwidth]{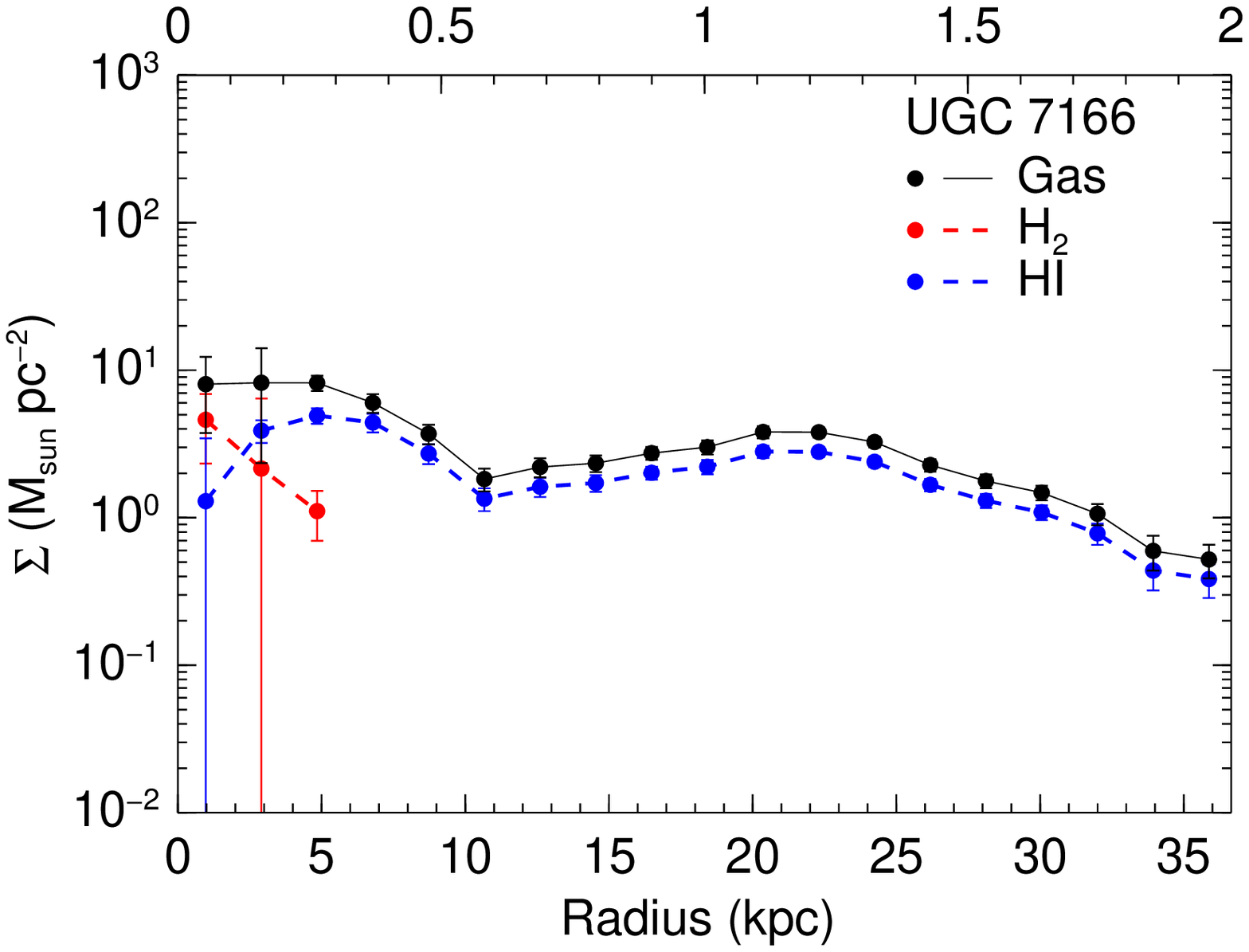}\\
\includegraphics[width=0.35\textwidth]{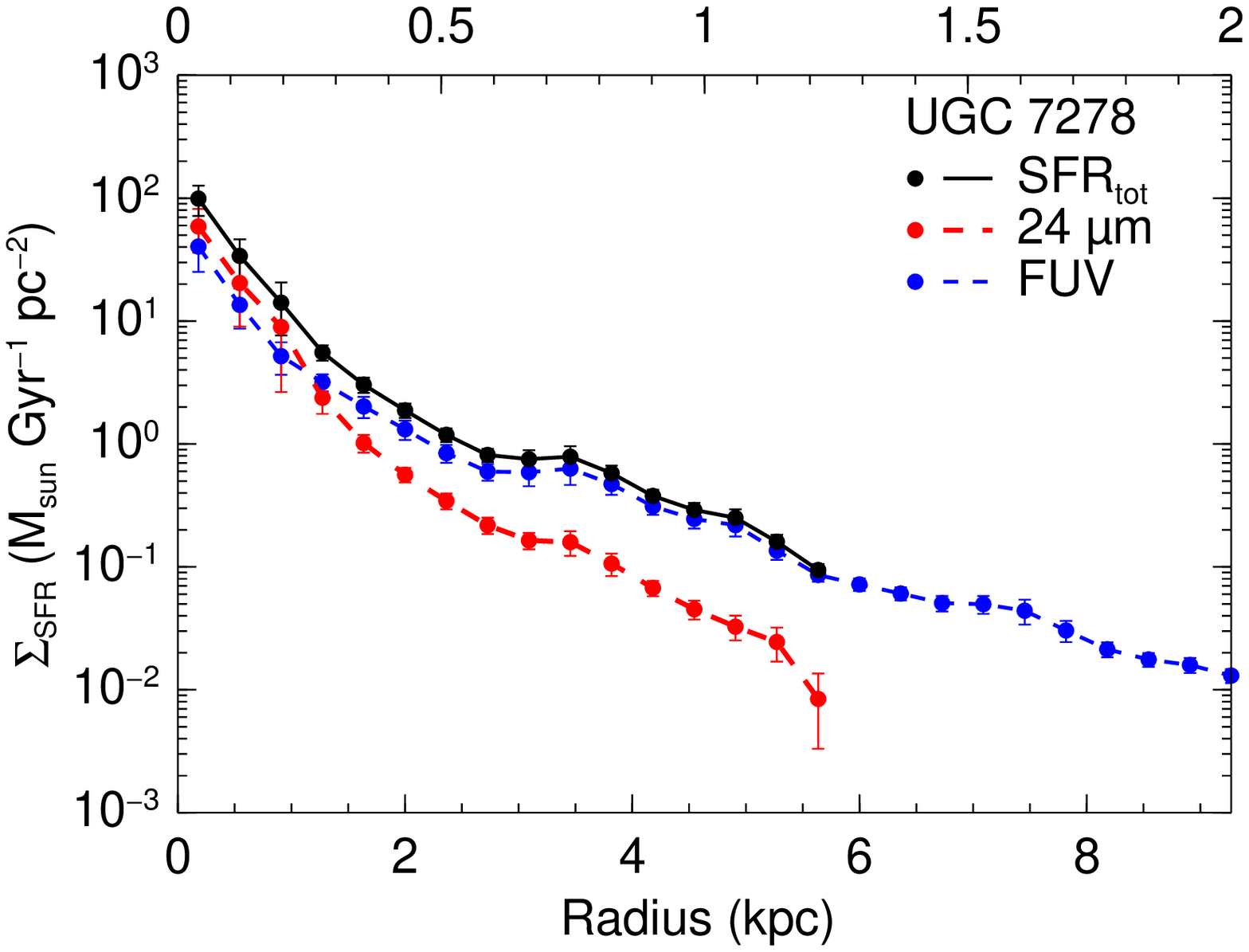}&
\includegraphics[width=0.35\textwidth]{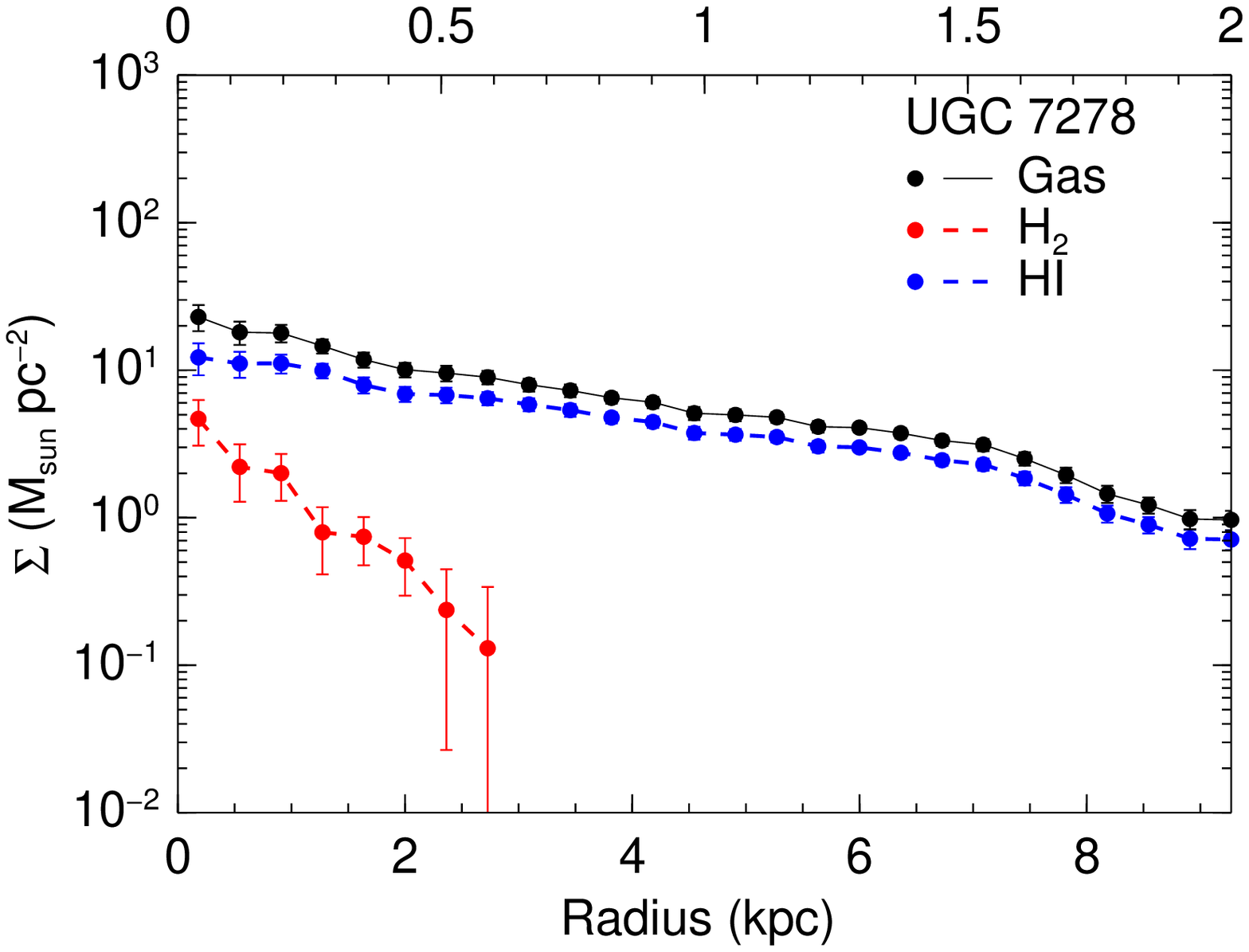}\\
\includegraphics[width=0.35\textwidth]{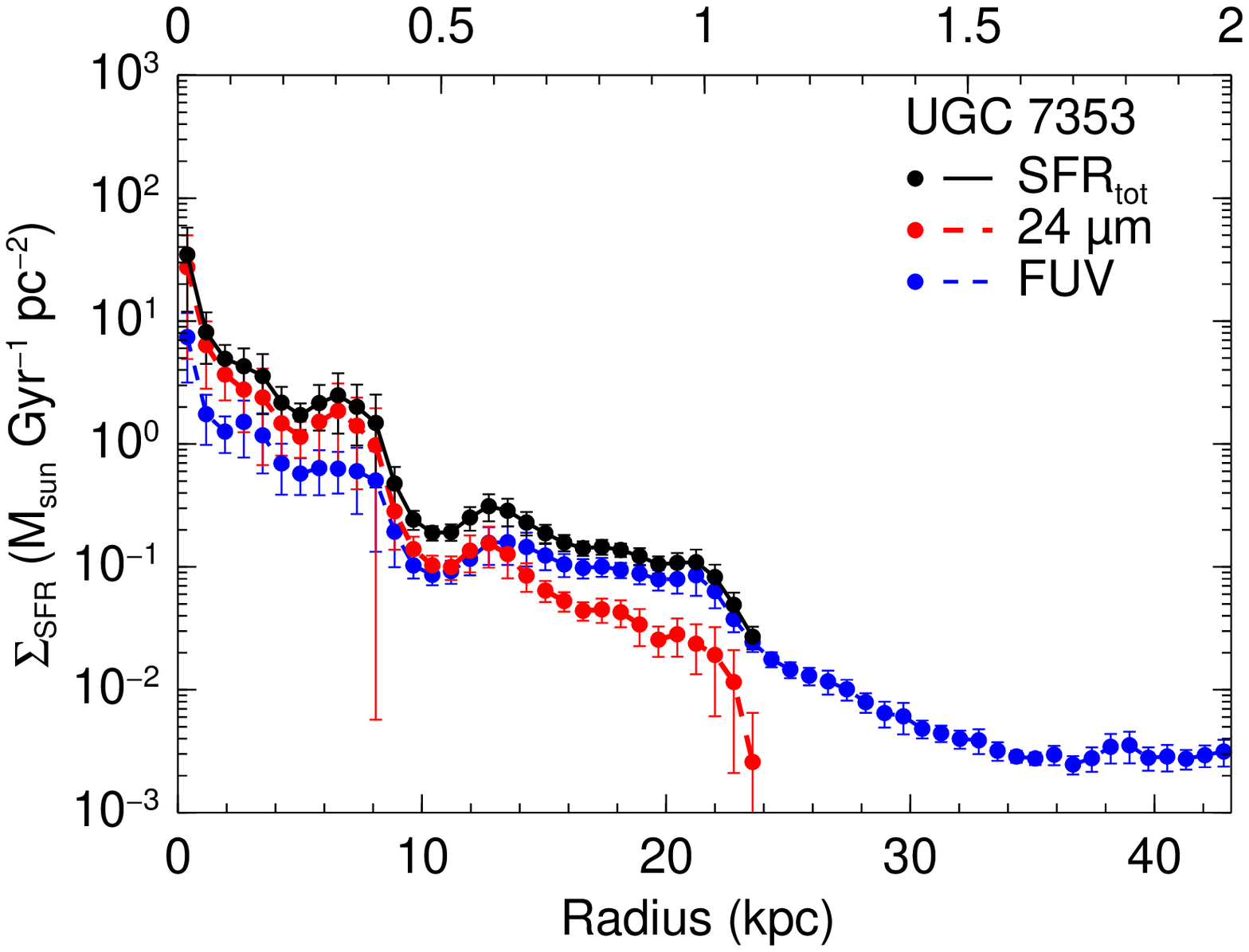}&
\includegraphics[width=0.35\textwidth]{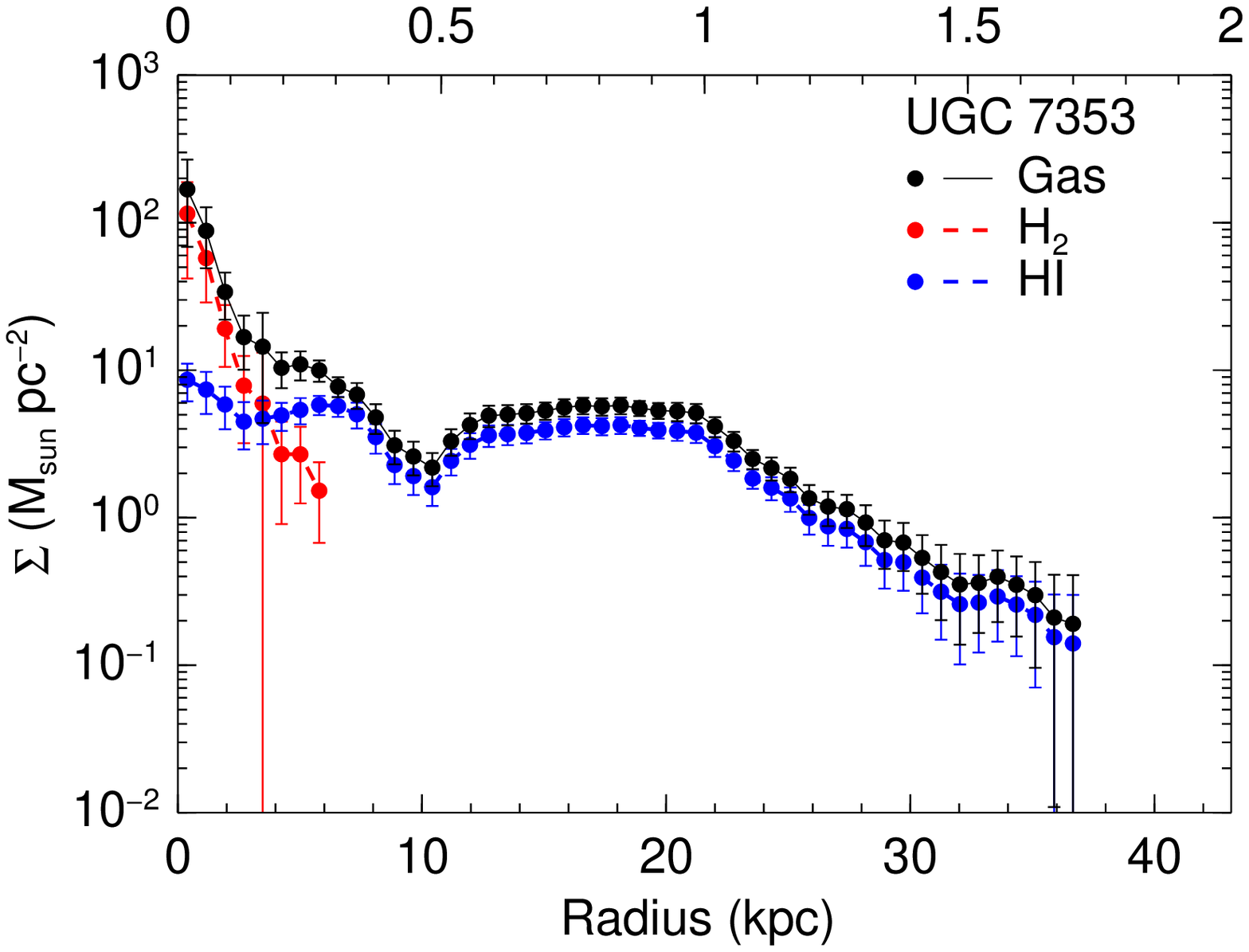}
\end{tabular}
\caption{continued}
\label{rprof}
\end{center}
\end{figure*}

\addtocounter{figure}{-1}

\begin{figure*}
%\figurenum{2}
\begin{center}
\begin{tabular}{c@{\hspace{0.1in}}c@{\hspace{0.1in}}}
\includegraphics[width=0.35\textwidth]{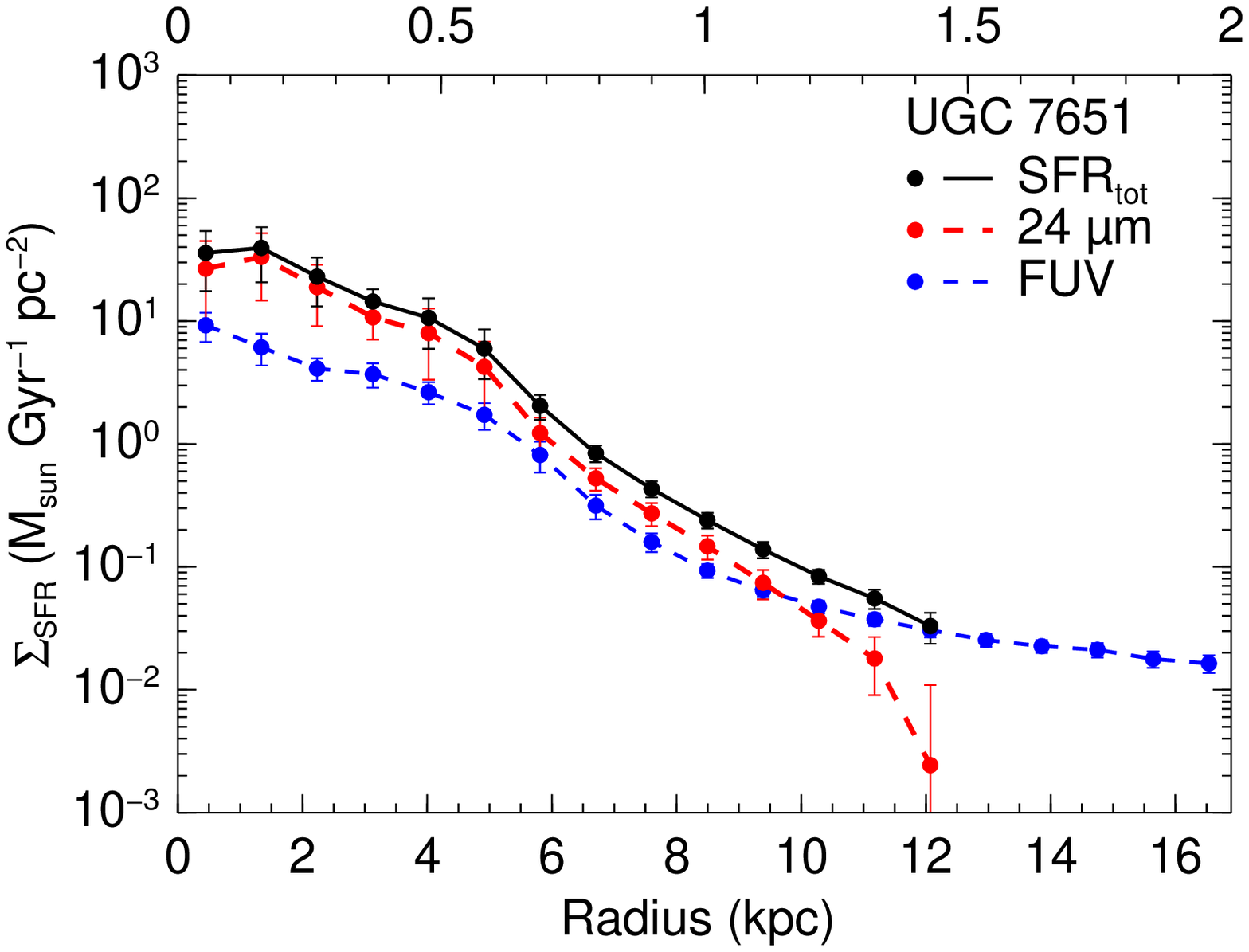}&
\includegraphics[width=0.35\textwidth]{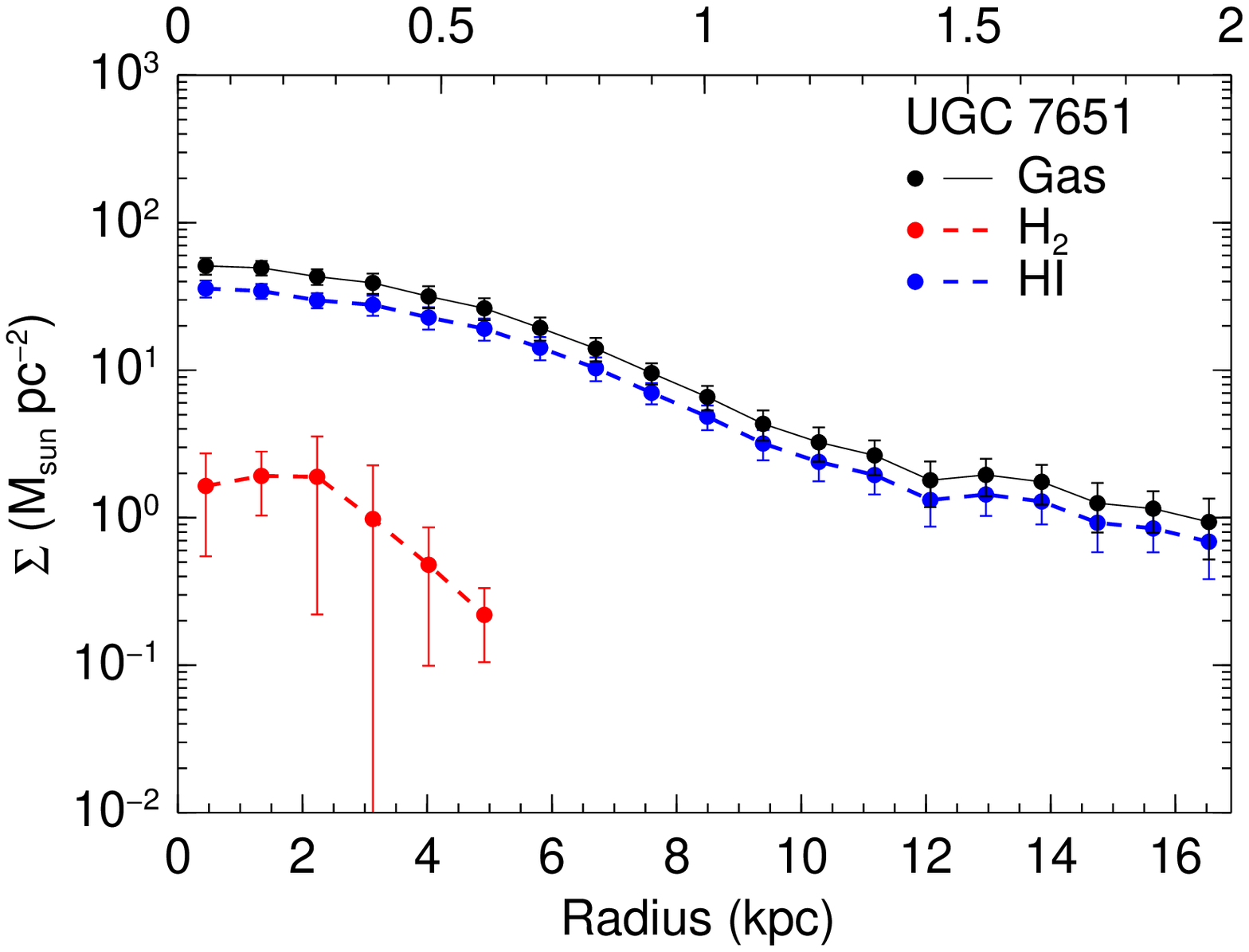}\\
\includegraphics[width=0.35\textwidth]{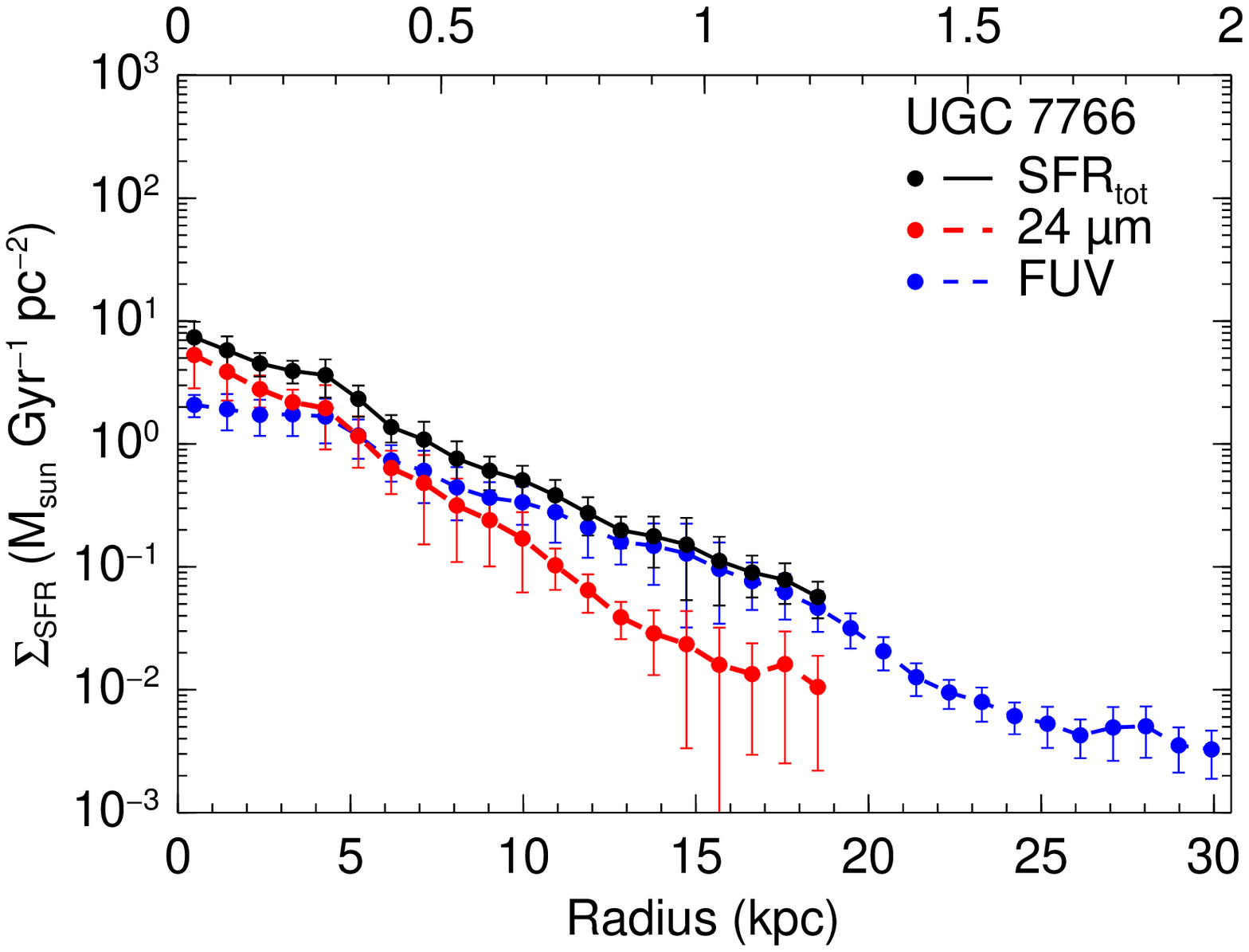}&
\includegraphics[width=0.35\textwidth]{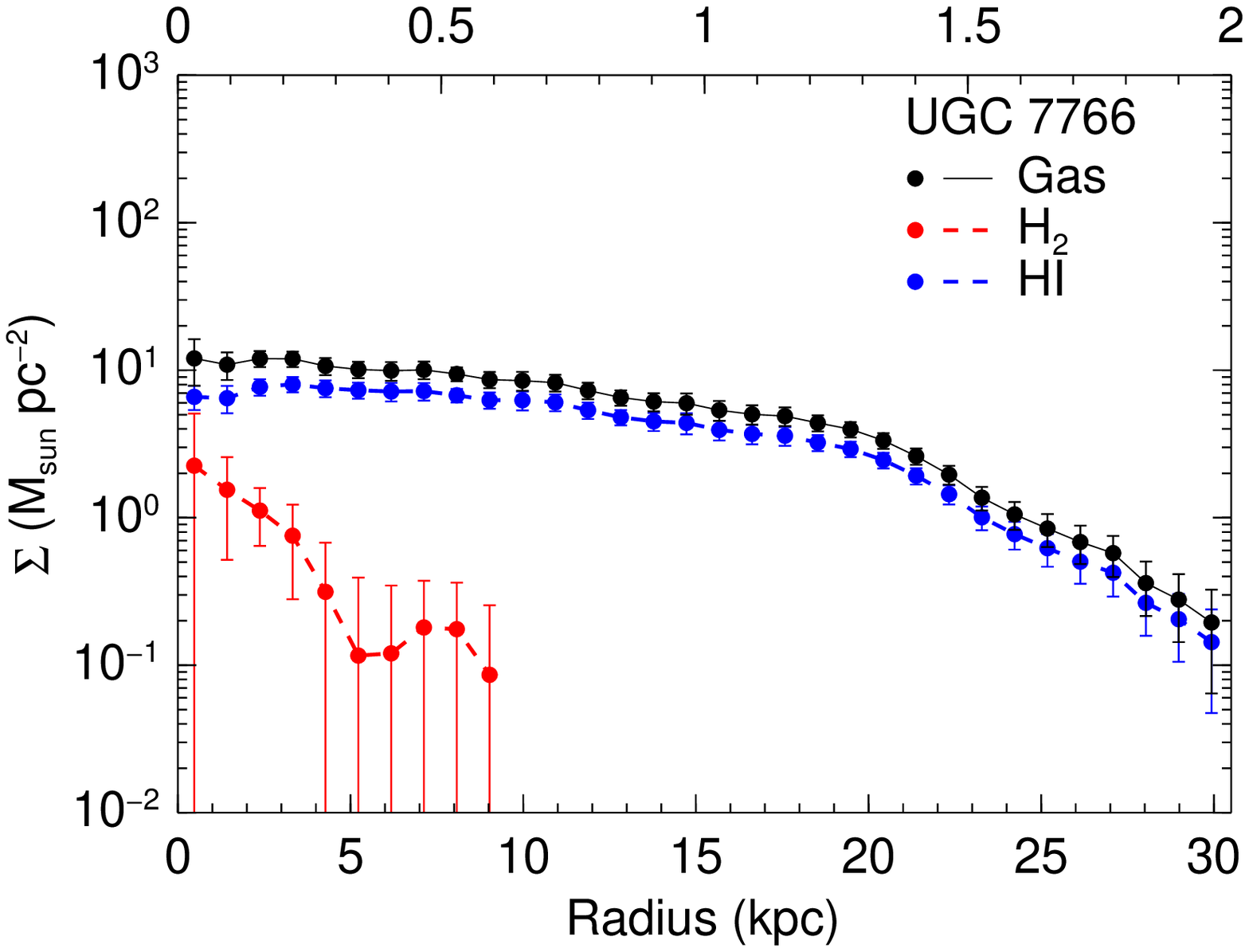}\\
\includegraphics[width=0.35\textwidth]{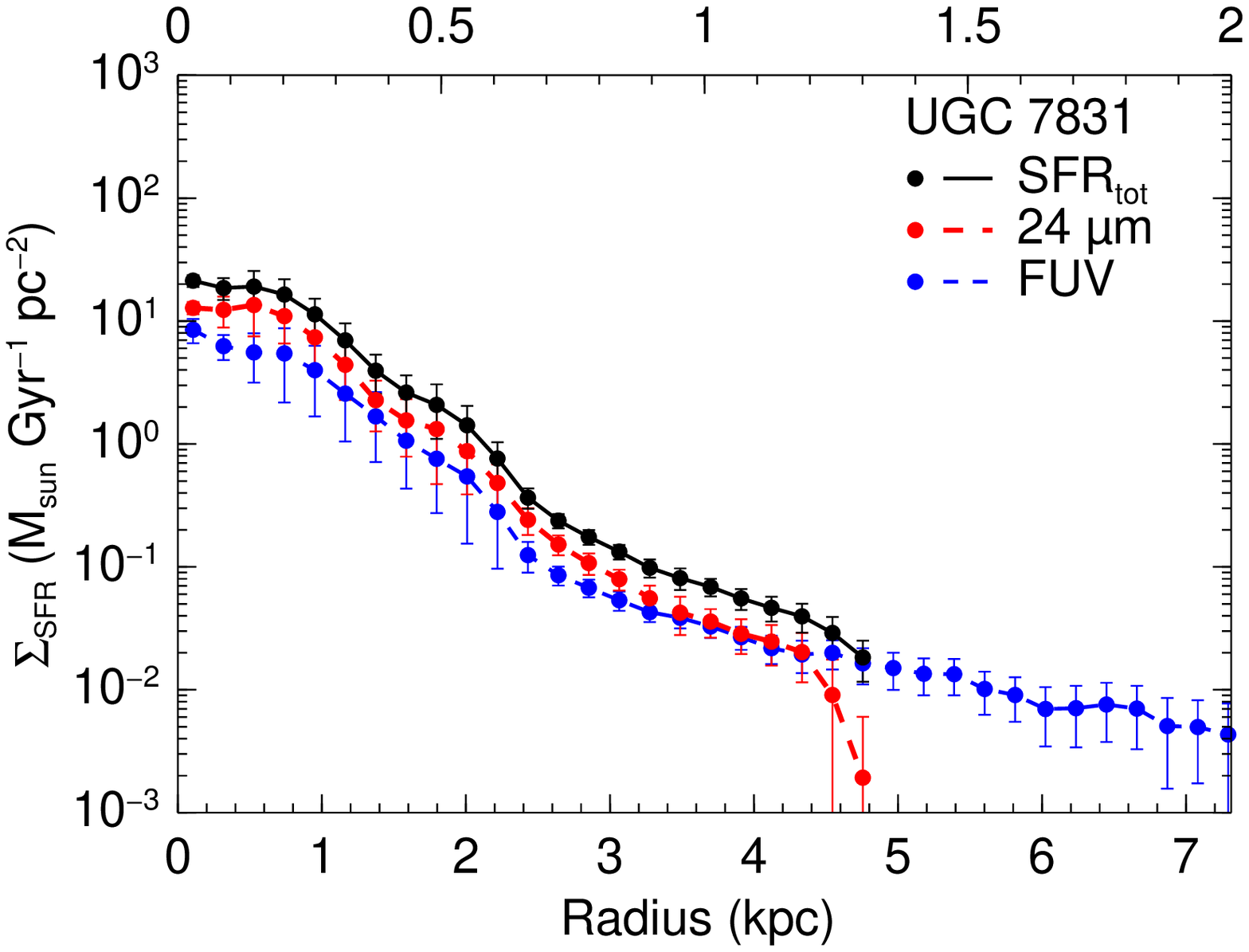}&
\includegraphics[width=0.35\textwidth]{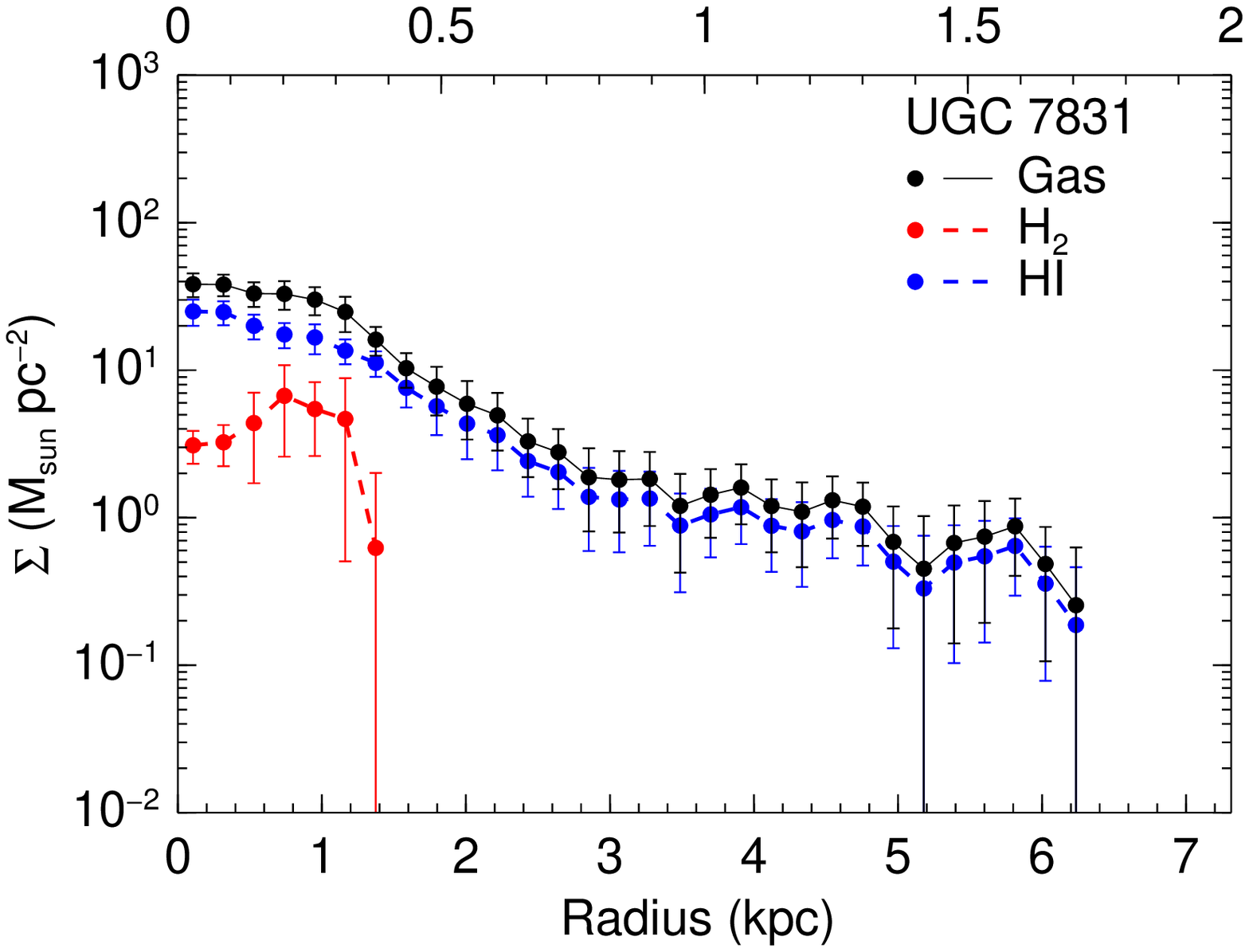}\\
\includegraphics[width=0.35\textwidth]{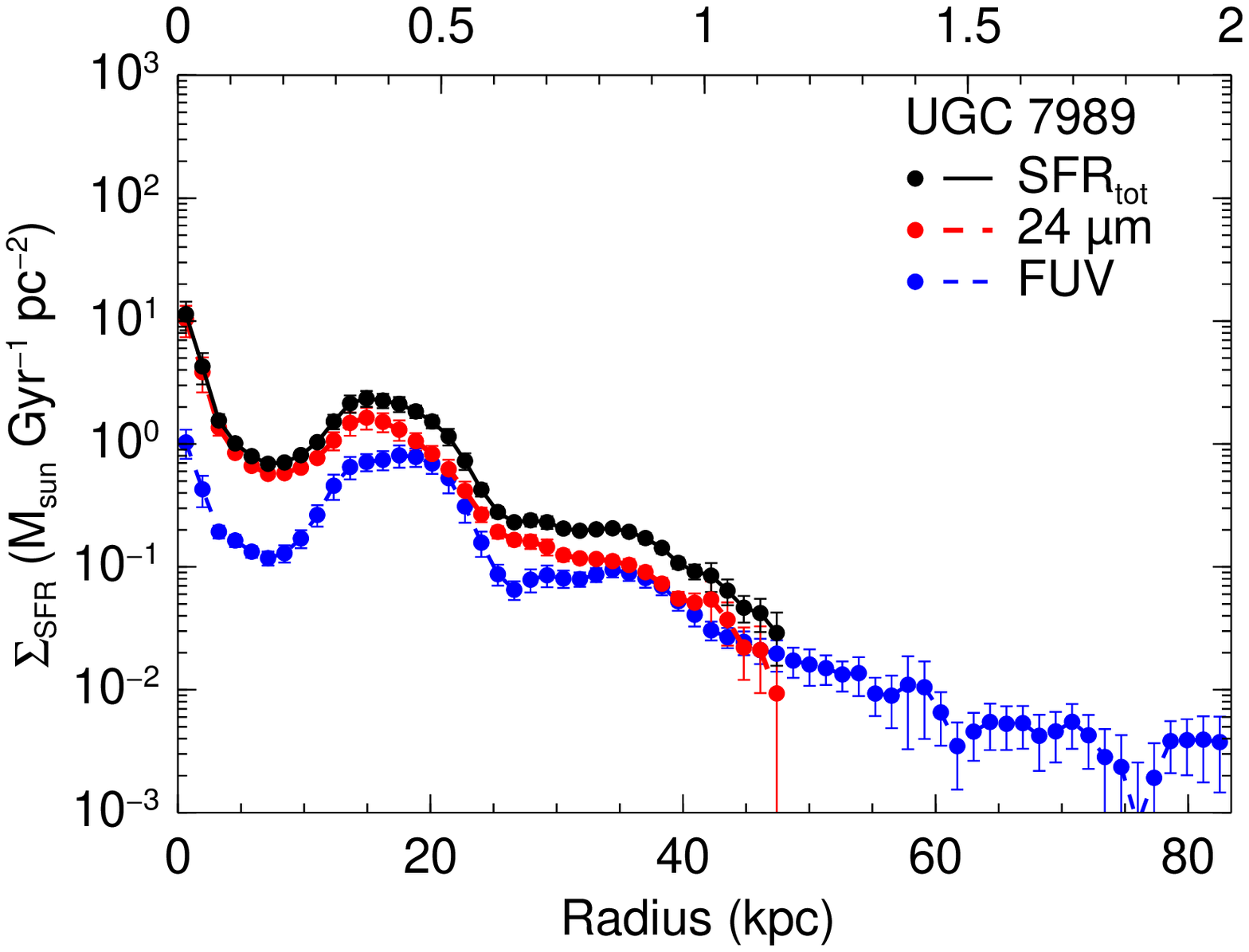}&
\includegraphics[width=0.35\textwidth]{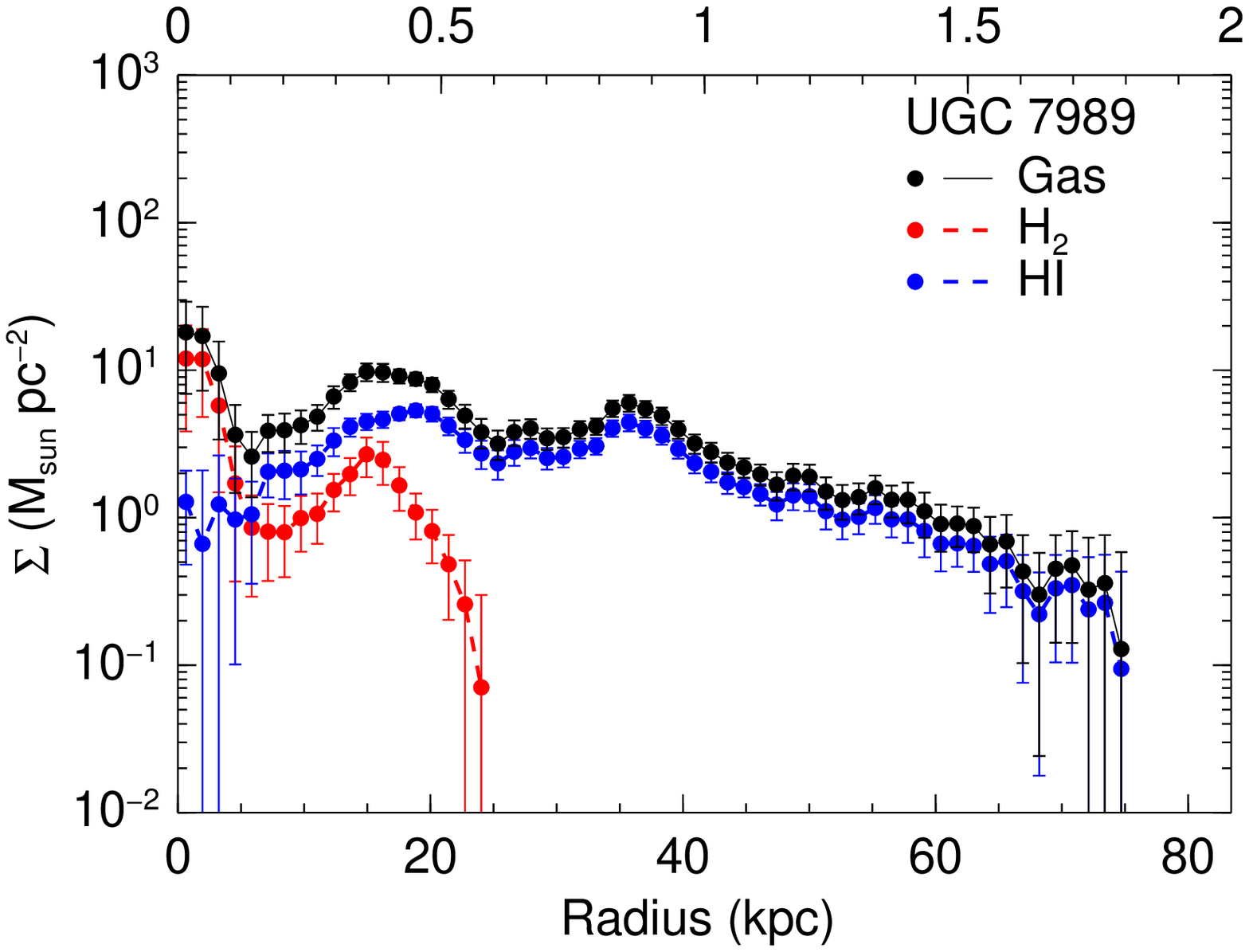}
\end{tabular}
\caption{continued}
\label{rprof}
\end{center}
\end{figure*}

\subsection{Scaled Radial Profiles}
\label{scaled}

\citet{2012ApJ...756..183B} found an exponential scaling relation of the total gas (\sighi\ + \sightwo) within a factor of two uncertainty from scaled radial profiles of \siggas\ for 33 nearby spiral galaxies and the Milky Way. The average profile of the total gas versus normalized radius by \ropt\ is well constrained by the exponential fit.

\begin{figure*}
\begin{center}
\begin{tabular}{c@{\hspace{0.1in}}c@{\hspace{0.1in}}c@{\hspace{0.1in}}c}
\includegraphics[width=0.45\textwidth]{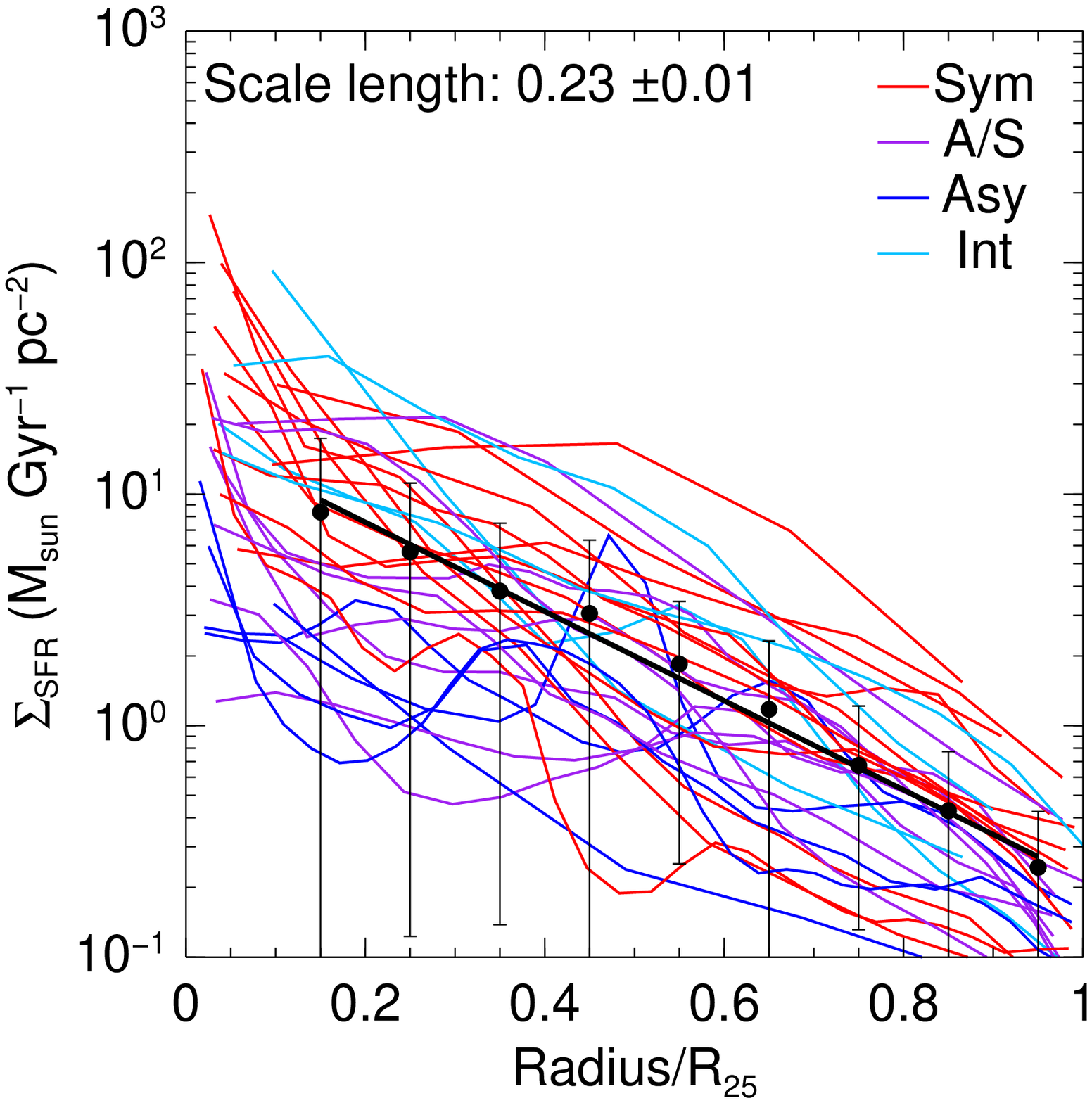}&
\includegraphics[width=0.45\textwidth]{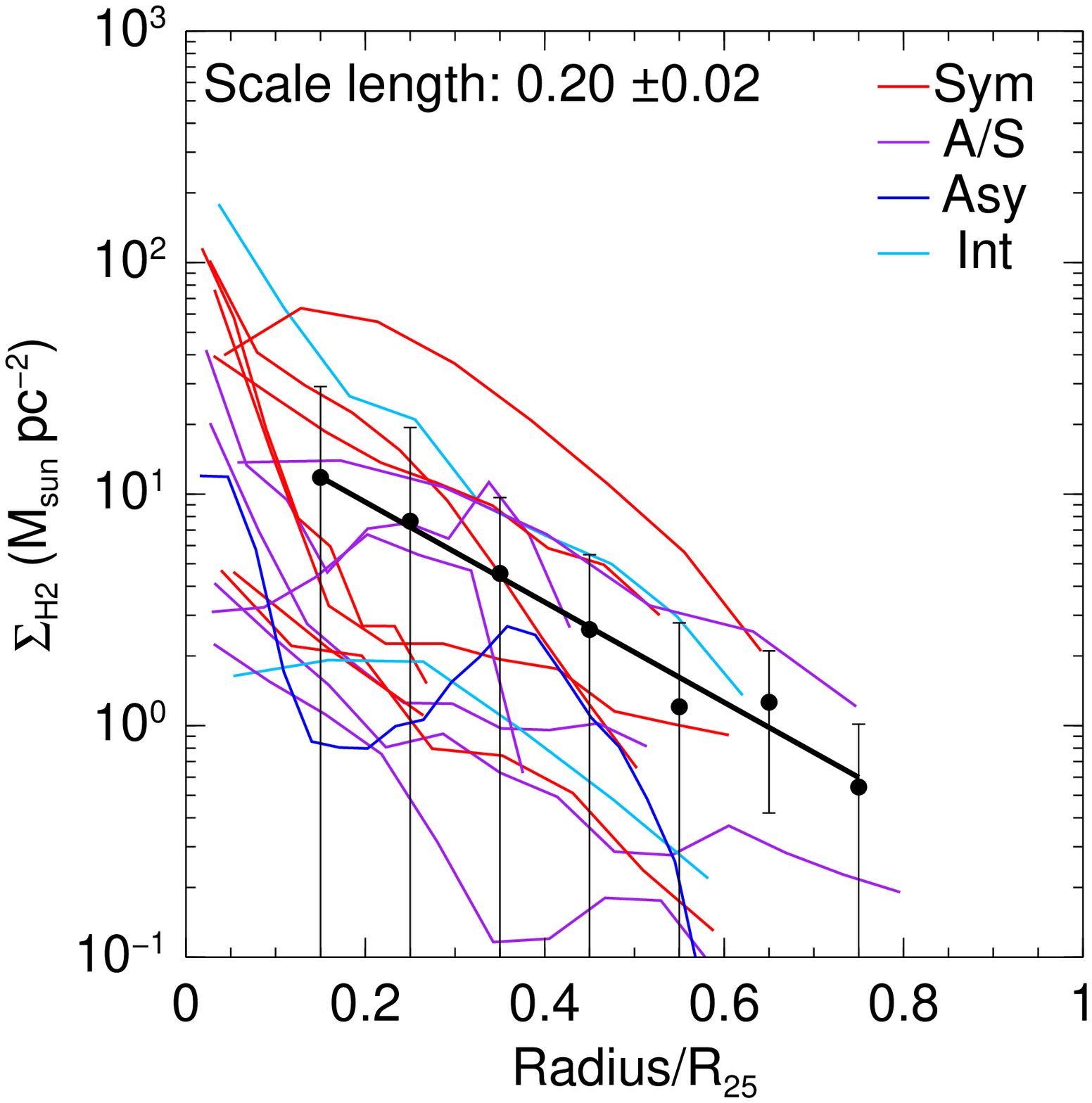}\\
\includegraphics[width=0.45\textwidth]{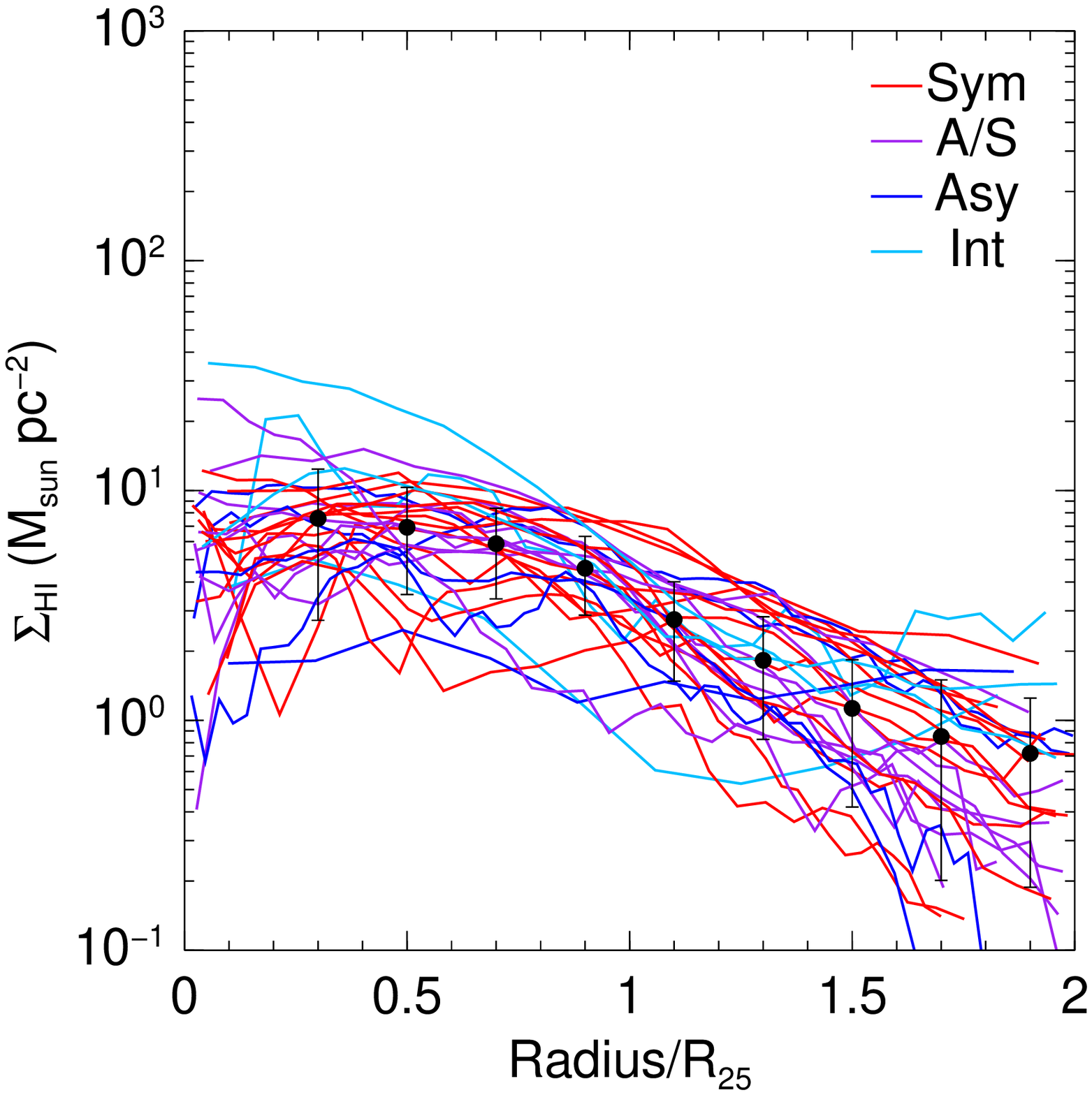}&
\includegraphics[width=0.45\textwidth]{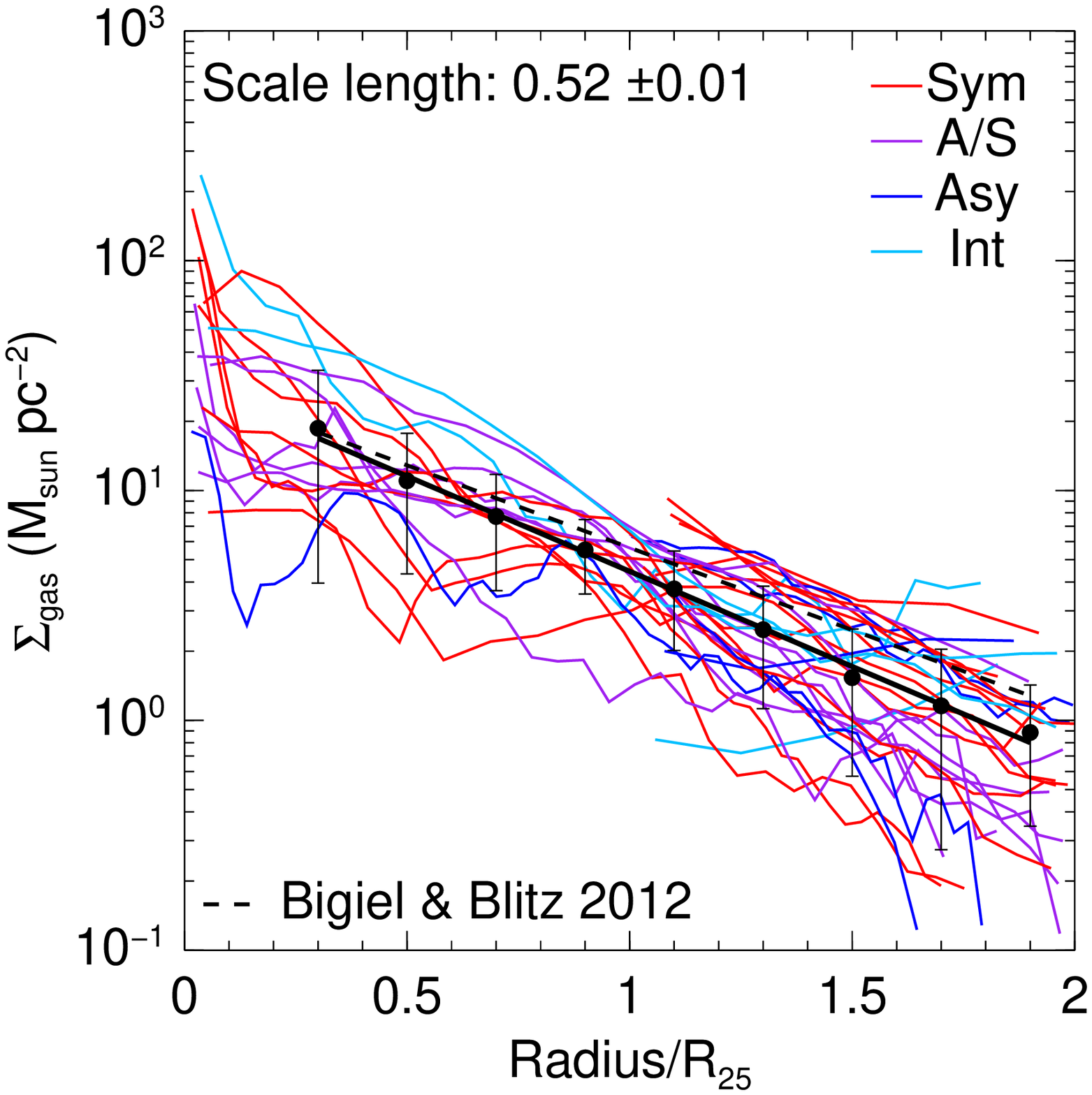}
\end{tabular}
\caption{The scaled radial profiles of \sigsfr\ (top left), \sightwo\ (top right), \sighi\ (bottom left), and \siggas\ (bottom right) for the symmetric (red), A/S (purple), asymmetric (blue), and interacting (light blue) galaxies. The filled circles are the average values in each bin (0.2$\times$\ropt) and the solid line is the exponential fit to the points. The dashed line in the bottom right panel is the fit given by \citet{2012ApJ...756..183B}.
\label{scaledrad}}
\end{center}
\end{figure*}

We plot all the scaled radial profiles of \sigsfr, \sightwo, \sighi, and \siggas\ for the symmetric (red), intermediate A/S (purple), asymmetric (blue), and interacting (light blue) galaxies in Figure \ref{scaledrad}.  From the figures, we noticed that there is no significant difference between the different types in the scaled radial profiles. We fitted an exponential function to the average values (filled circles) in a bin excluding the central regions. The vertical error bars show the standard deviation of the mean. The \sigsfr\ and \sightwo\ profiles are well fitted by the exponential function, but their scatter is too large. On the other hand, the scatter of \sighi\ is relatively small, but it is less well described by the exponential function due to the roughly flat profile within the optical radius. Like the result of \citet{2012ApJ...756..183B}, the exponential fit to the average values of \siggas\ for our sample is tightly constrained within a factor of 2 uncertainty (bottom right panel). In order to increase the sample size in the \siggas\ profiles, the galaxies without CO data are included only for the outer regions (beyond \ropt) where \sighi\ is dominant.  
In the bottom right panel, our exponential fit (solid line) is compared with the universal gas profile (dashed line) given by \citet{2012ApJ...756..183B}:
\begin{equation}
\frac{\siggas}{\Sigma_{\rm tr}} = 2.1 \times e^{-1.65 \times r/r_{25}},
\label{bigiel}
\end{equation}
here $\Sigma_{\rm tr}$ is the surface density at the transition radius where \sightwo\ = \sighi\ and we adopt 14 \surm\ from their study.  Note that this value is not obtained from our sample and we only use it for Equation (\ref{bigiel}) to compare with our scaling relation:
\begin{equation}
\siggas = 30.3 \times e^{-1.92 \times r/r_{25}}.
\end{equation}
In this equation, which is obtained from the fit to our sample, we do not use $\Sigma_{\rm tr}$ to normalize \siggas\ since it is not possible to determine the value for several galaxies that do not have CO data or the transition radius. 
We also fitted the exponential function separately to only symmetric galaxies and  asymmetric (including A/S) galaxies to examine differences in the scaling relation. The scale lengths are 0.56 and 0.52 for the symmetric and asymmetric galaxies, respectively; so not significantly different.

\subsection{Star Formation in the Inner Regions}
\label{sflaw}

\begin{figure*}
\begin{center}
\begin{tabular}{c@{\hspace{0.1in}}c@{\hspace{0.1in}}}
\includegraphics[width=0.35\textwidth]{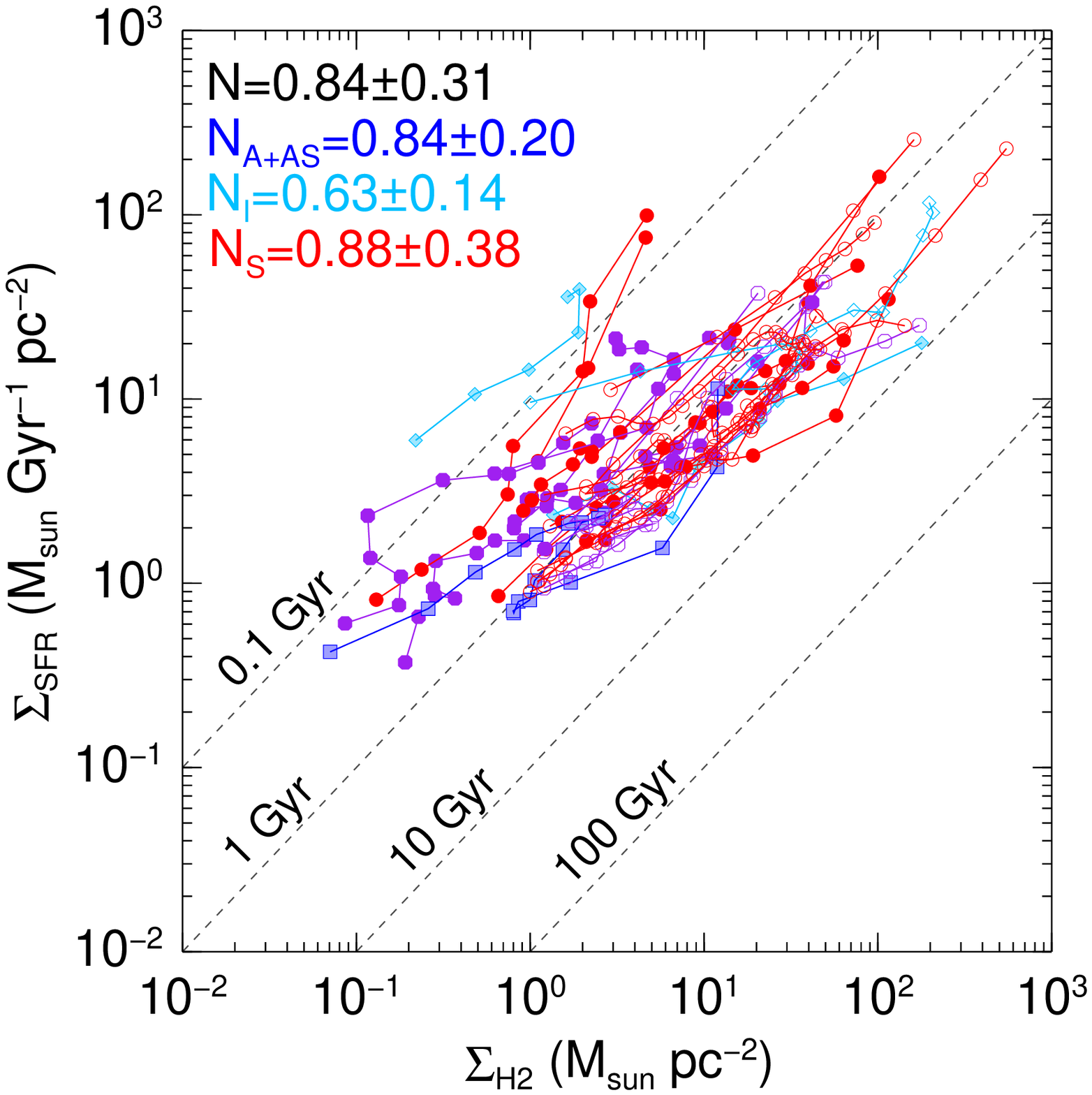}&
\includegraphics[width=0.35\textwidth]{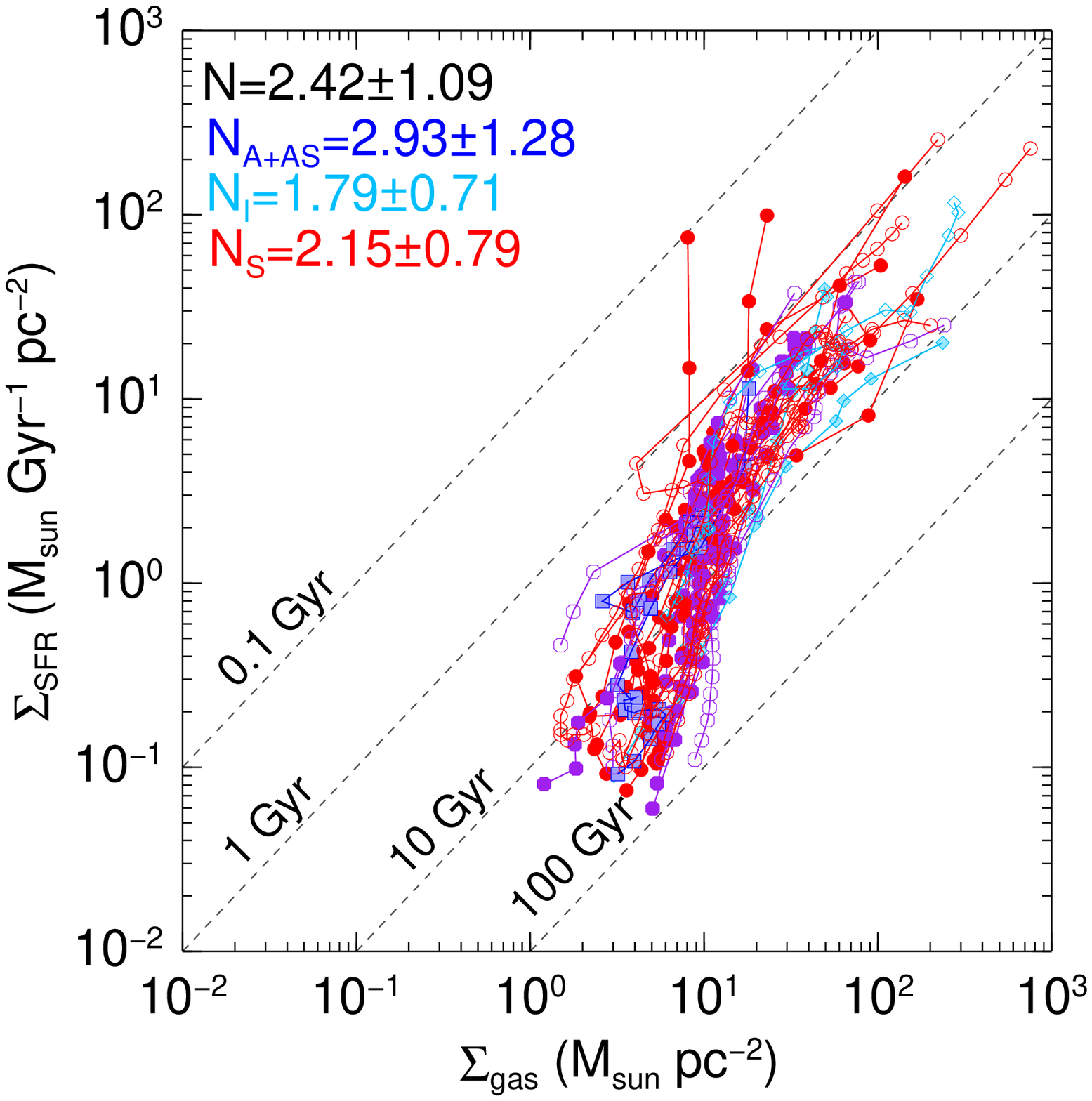}\\
\includegraphics[width=0.35\textwidth]{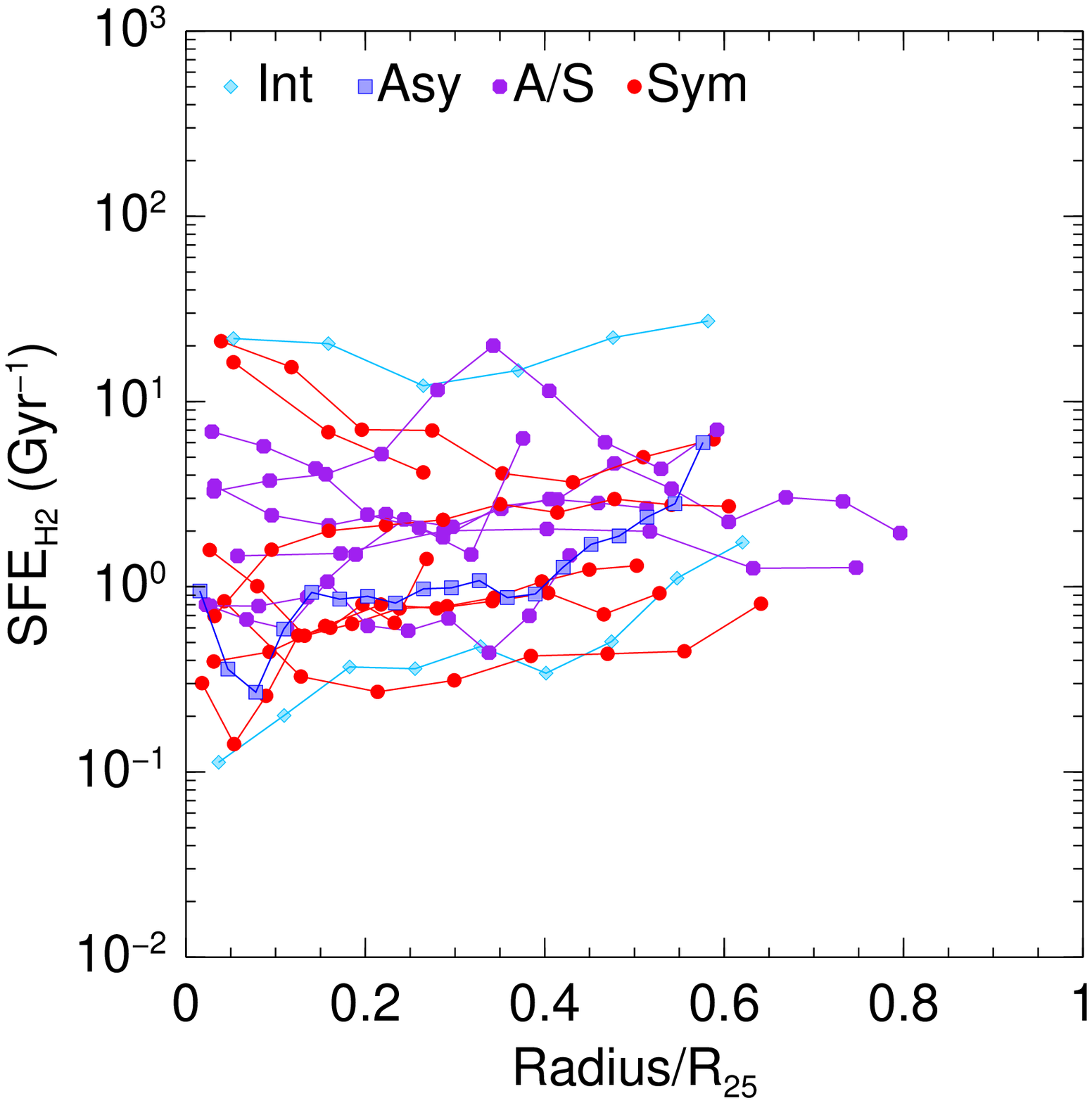}&
\includegraphics[width=0.35\textwidth]{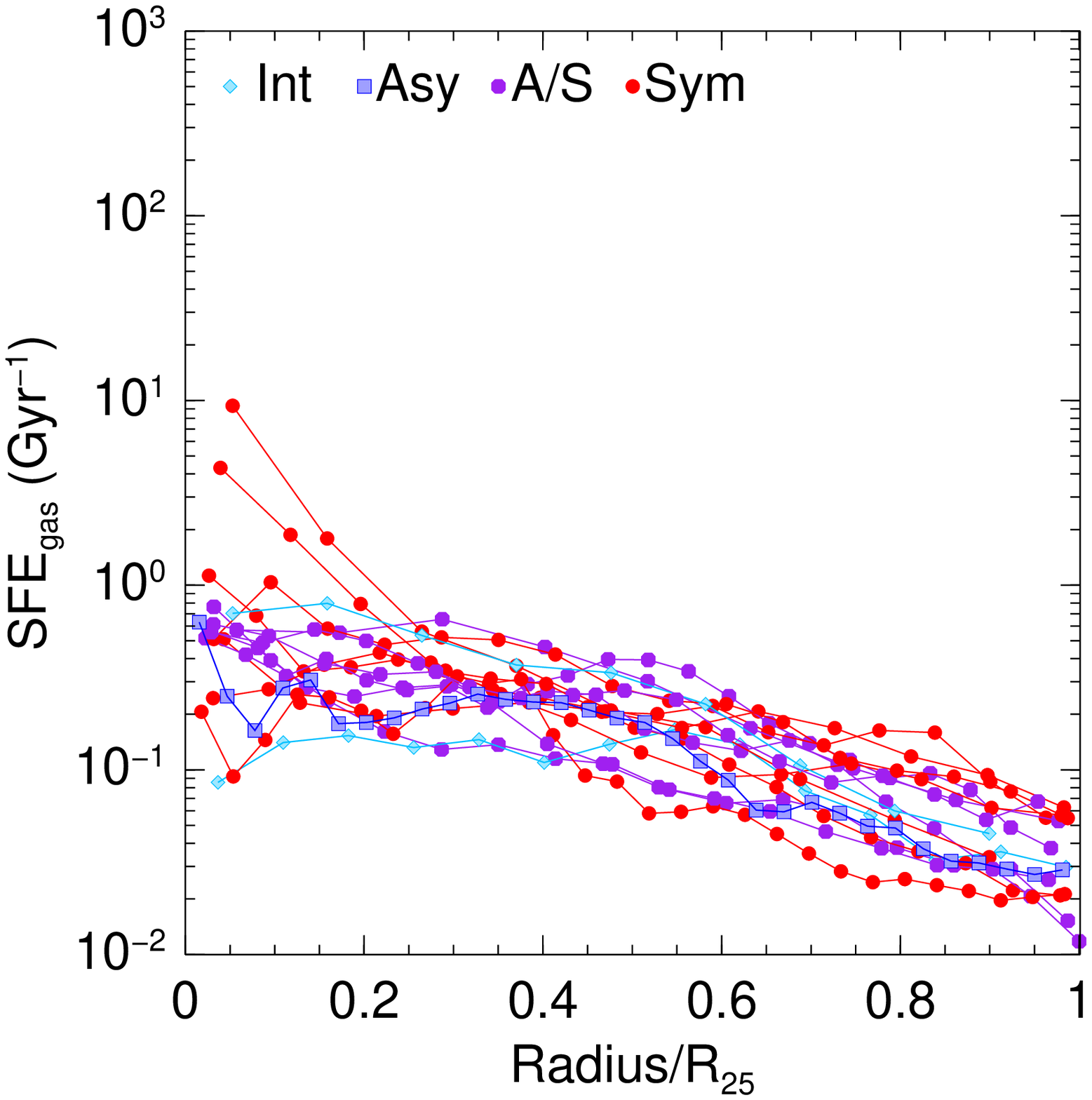}
\end{tabular}
\caption{Top panels: \sigsfr\ versus \sightwo\ (left)  and \siggas (right). 
The average K-S indices are presented as N for all the galaxies, N$_{\rm A+AS}$ for the asymmetric and A/S galaxies,  N$_{\rm I}$ for the interacting galaxies, and N$_{\rm S}$ for the symmetric galaxies.  The different symbols indicate the different galaxy classes; ligt blue diamonds (interacting), blue squares (asymmetric),  purple octagons (A/S), and red circles (symmetric) for the WHISHP galaxies (filled symbols) and the THINGS galaxies (open symbols).
Bottom panels: SFE of the molecular gas (\sigsfr/\sightwo) and  SFE of the total gas (\sigsfr/\siggas) as a function of radius normalized by the optical radius \ropt. 
\label{sfrlaw}}
\end{center}
\end{figure*}

Using the sub-sample with both CO and \HI\ data (16 galaxies), we have examined the K-S law by plotting \sigsfr\ against \sightwo\ and \siggas\ in the inner regions (within \ropt) for the symmetric (red circles), A/S (purple octagons), asymmetric (blue squares), and interacting (light blue diamonds) galaxies in Figure \ref{sfrlaw} (top panels) to investigate whether the correlation is stronger in the asymmetric galaxies (including intermediate galaxies), suggesting a  connection between star formation and gas accretion since we assume that asymmetric galaxies are subject to  gas accretion. 
Note that the asymmetric group includes the intermediate (A/S) galaxies in this section since (1) there is only one galaxy (UGC 7989) in the asymmetric group with CO and (2) the asymmetric and A/S galaxies are not significantly different from each other in their kinematics. However, we indicated each type of  galaxy by different symbols in the figures.   For the purposes of increasing the sample size and comparing with other observations, we included 12 THINGS \citep{2008AJ....136.2563W} galaxies with CO data from \citet{2008AJ....136.2782L}. The THINGS sample consists of 8 symmetric, 3 slightly asymmetric (A/S), and one interacting galaxy. They are plotted as open symbols (but the same symbol shape as the classes of our sample) in the figure. 
The THINGS galaxies show the same overall behaviour as the galaxies from the WHISP sample, and inclusion of these galaxies does not alter the lack of any significant difference between the galaxy classes. 
We used the ordinary least-squares (OLS) bisector \citep{Isobe:1990ft} to fit \sigsfr\ vs. \sightwo\ and \siggas\ in logarithmic space for each galaxy. The power-law index (N) shown in the figure is the average value of the all galaxies, while the indices N$_{\rm A+AS}$,   N$_{\rm I}$, and N$_{\rm S}$ represent the average values of the asymmetric (including A/S), interacting,  and symmetric groups, respectively. Although the average indices are clearly different for the three groups and some symmetric galaxies seem to exhibit a weak correlation in the regions where  \sigsfr\ and \siggas\ values are higher, we do not see  a much tighter correlation among the groups.
In the top right panel of Figure \ref{sfrlaw}, the relation of SFR surface density versus total gas density is slightly steeper for the asymmetric galaxies as compared to the symmetric galaxies, though the difference is hardly significant. 
%It is striking, however, that the scatter among the symmetric galaxies is considerably smaller. 
The individual plots of \sigsfr\ vs. \sighi, \sightwo, and \siggas\  for each galaxy are presented in Figure \ref{restsfl}. 

In the bottom panels of Figure \ref{sfrlaw}, we show the star formation efficiency (SFE) for  molecular gas (SFE$_{\rm H2}$ = \sigsfr/\sightwo) and  total gas (SFE$_{\rm gas}$ = \sigsfr/\siggas) as a function of radius normalized by the optical radius \ropt. Like previous studies (e.g., \citealt{1999AJ....118..670R}; \citealt{2008AJ....136.2782L}), the SFE$_{\rm H2}$ is roughly constant although the scatter is somewhat large.  On the other hand, the SFE$_{\rm gas}$ is decreasing with radius and the scatter is small except the inner region within 0.2$\times$\ropt. 
The scatter in the total gas is reduced by a factor of 10, suggesting that the inclusion of \HI\ in the SFE provides more meaningful results.
Again, there is no significant difference among the groups.

\subsection{Star Formation in the Outer Regions}
\label{fuv-hi}

\begin{figure*}
\begin{center}
\begin{tabular}{c@{\hspace{0.1in}}c@{\hspace{0.1in}}}
\includegraphics[width=0.45\textwidth]{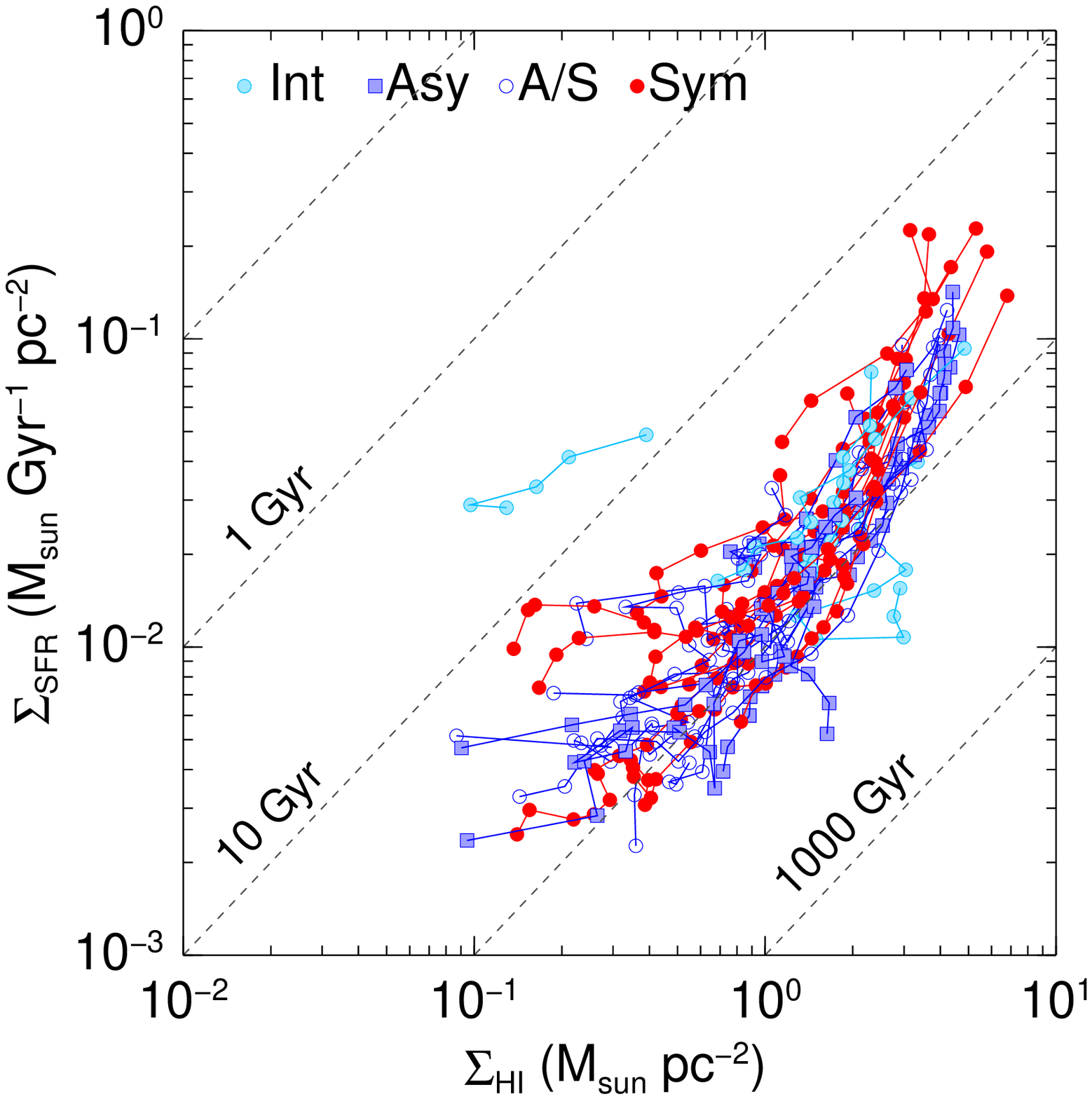}&
\includegraphics[width=0.45\textwidth]{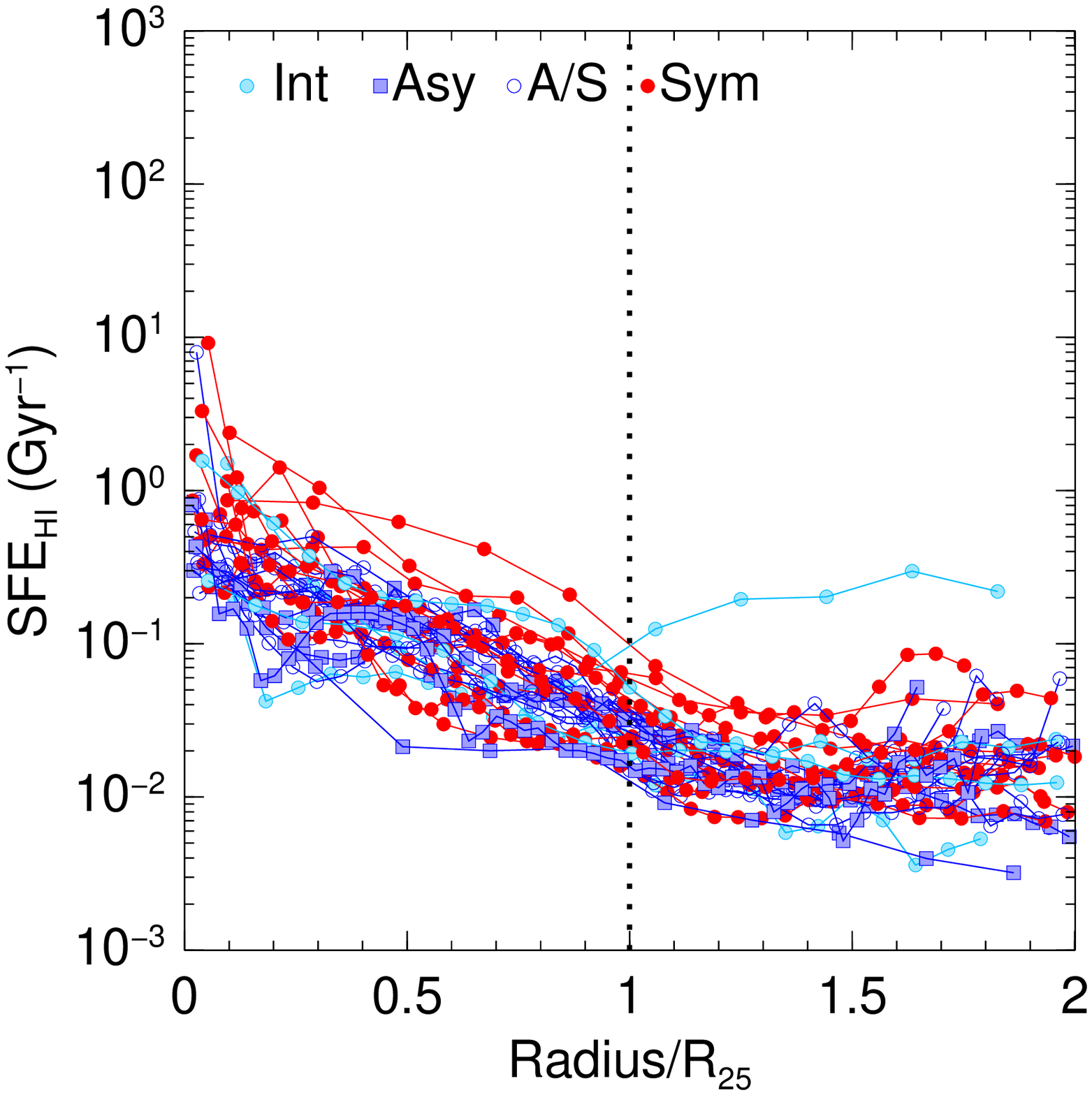}
\end{tabular}
\caption{\sigsfr\ versus \sighi\ (left)  in the outer regions (\ropt\ -- 2$\times$\ropt) and SFE$_{\rm HI}$ with radius normalized by \ropt\ (right). The vertical dotted line represents where $r =$ \ropt. 
\label{outsfr}}
\end{center}
\end{figure*}

Star formation in galaxies does not always stop at \ropt\ or at the edge of the molecular gas disc, as demonstrated by the extended galactic discs in UV (e.g., \citealt{2005ApJ...619L..79T}; \citealt{2005ApJ...627L..29G}). \citet{Bigiel:2010jc} found a strong correlation between \HI\ and FUV in the outer discs of many nearby spiral and dwarf galaxies. Since we are interested in a correlation between gas accretion and star formation, we also investigate a possible relationship between \HI\ and FUV, especially in the outer regions where both \HI\ and FUV extend to 2$\times$\ropt\ or more and where the effects of accretion may be most severe. Here, the outer regions are defined as $r >$ \ropt. Following the suggestion by \citet{Bigiel:2010jc}, we do not consider the internal extinction in FUV since the extinction correction may cause large uncertainties at the outer regions where S/N is low and background sources may contaminate FUV emission. 
In addition, the extinction effect would not be significant in the outer regions. \citet{Bigiel:2010jc} estimated the internal extinction using a typical value of \HI\ column density at the outer discs; the correction factor is about 1.3.

Figure \ref{outsfr} shows \sigsfr\ versus \sighi\ in the outer regions and SFE$_{\rm HI}$ ($=$ \sigsfr/\sighi) versus radius in units of \ropt\ for symmetric (red  circles), intermediate A/S (purple  octagons), asymmetric (blue  squares), and interacting (light blue diamonds) galaxies. 
For estimating \sigsfr\  in the figure, we used the FUV term in Equation (\ref{eq:sfrcombi}) since the 24 \um\ emission is negligible and the \sigsfr\ profile is dominated by FUV in the outer regions. 
Unlike for the inner regions, a tight correlation between \sigsfr\ and \sighi\ has been found in the outer regions for all the groups, although the scatter increases toward lower \sighi.  We obtained the average K-S index of 1.21 using the OLS fit to \sigsfr\ (from only FUV) vs. \sighi\ in the outer regions. 

The very discrepant galaxy in the figures is the interacting galaxy UGC 4862 which has two tidal tails \citep{2012MNRAS.421.3612T}. The most prominent tail in the outer regions is masked in \HI\ and FUV images when driving radial profils. 
The \HI\ profile shows very low density in the outer regions, especially near the optical radius, compared to other galaxies.
The \HI\ gas may be swept out toward the tail during interacting process, causing the low \sighi\ in the regions. On the other hand, the FUV profile shows much higher SFR near the radius of 2$\times$\ropt\ than that of other galaxies. The high SFR in the outer regions might be linked to an extended UV (XUV) disc \citep{2007ApJS..173..538T} of this galaxy. 

The right panel in the figure shows the SFE$_{\rm HI}$ decreasing (roughly exponentially) with radius up to $\sim1.5\times$\ropt\ and flattering (or roughly constant) beyond the radius. All galaxies in our sample appear to behave the same, so apparently the outer discs have very low SFE but also are extremely self regulating, probably because the \HI\ surface density is very close to the critical density. In the end, it tells us more about galaxies than about how they get their gas. 
The roughly constant SFE in the outer regions agrees well with  \citet{Bigiel:2010jc}, reinforcing the notion that the outer disks of spiral and irregular galaxies behave in a similar way, with low but constant star formation efficiencies. Why the outer disks behave this way is not yet very clear.  \citet{2016MNRAS.460.1106W} building on the results of \citet{2013MNRAS.429.2537M} and \citet{2013MNRAS.434.3389Z} show that for a model galaxy with constant Q the SFE declines toward \ropt\ and becomes approximately constant in the outer parts.  As discussed by \citet{Bigiel:2010jc}, this suggests that in the outer parts the disks are overall Q-stable while locally a small faction of the gas clouds become unstable and form stars. 
Several studies have shown that the radial gradient of metallicity abundance is flat in the outer disks (e.g., \citealt{2009ApJ...695..580B}; \citealt{2011MNRAS.412.1246G}). This flat gradient may be related to the constant SFE$_{\rm HI}$ in the outer regions. 
Like in the inner parts, we do not find a significant difference between the three groups of galaxies.

\subsection{Dependence on the total stellar mass}
\label{Mstar}

\begin{figure*}
\begin{center}
\includegraphics[width=0.45\textwidth]{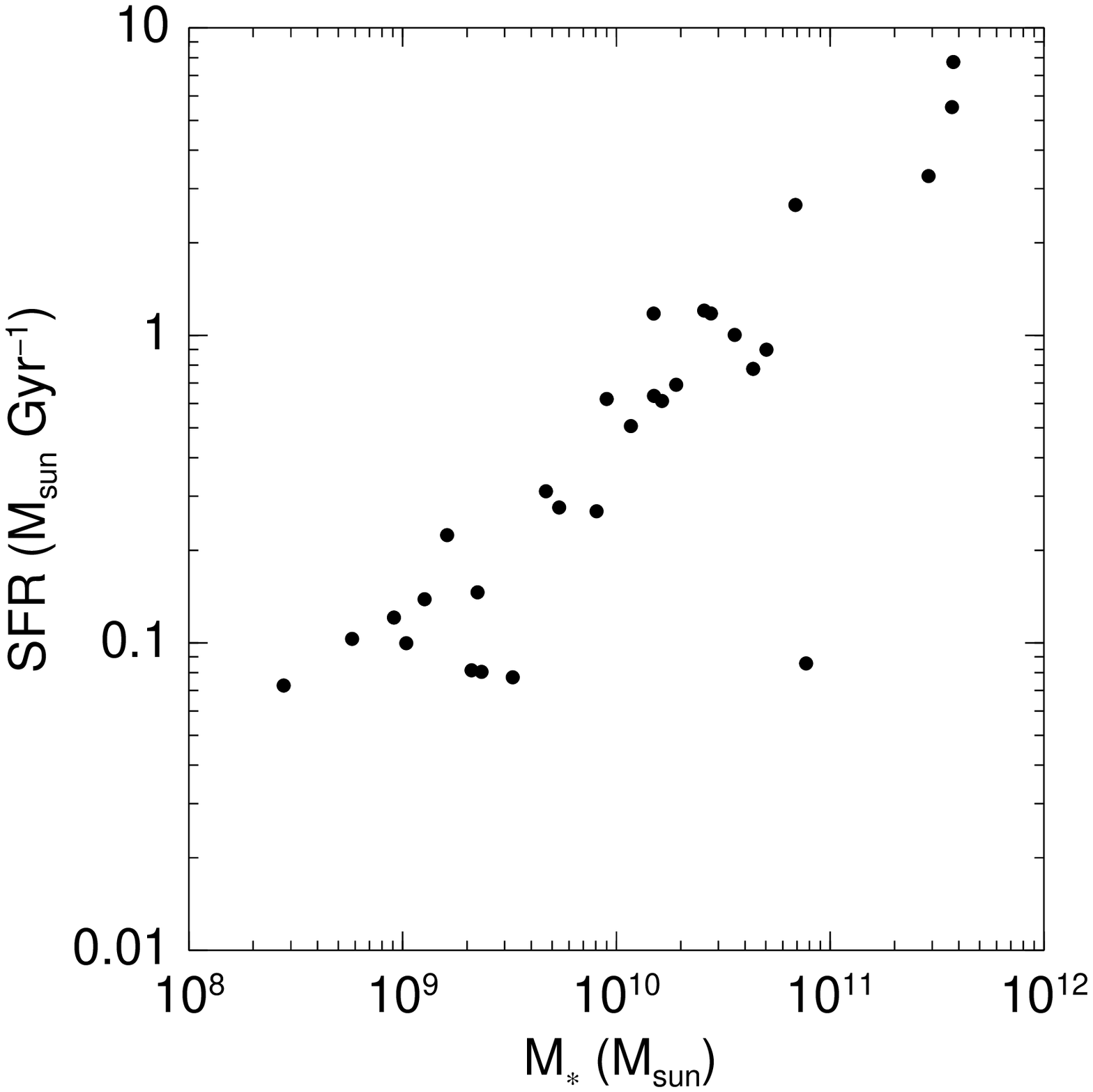}
\caption{ Total SFR versus $M_{*}$ for the sample of galaxies. 
\label{fig:sfr-mstar}}
\end{center}
\end{figure*}

Star formation depends on the total stellar mass (e.g., \citealt{2007ApJ...660L..43N}; \citealt{2008MNRAS.385..147D}). For this reason, there may be differences between the galaxy classes in the SF law (SFL) and SFE for different stellar mass ranges. We therefore re-examine the SFL and SFE, dividing the galaxies into specific stellar mass groups. To investigate this, we first estimated the total stellar mass ($M_*$) using the $Spitzer$ IRAC 3.6 and 4.5 \um\ maps following \citet{2012AJ....143..139E}:
\begin{equation}
M_* = 10^{5.65}F_{3.6}^{2.85}F_{4.5}^{-1.85}\left(\frac{D}{0.05}\right)^2,
\end{equation}
where $F_{3.6}$ and $F_{4.5}$ are fluxes of 3.6 and 4.5 \um\ in units of Jy and $D$ is the distance in units of Mpc. Bright foreground stars in the maps are blanked. we used the MIRIAD tasks CGCURS and HISTO to define  regions of galaxies and to integrate the galaxies, respectively. The estimated stellar masses for our sample are presented in Table \ref{tab:sample}. Figure \ref{fig:sfr-mstar} shows that the galaxies in our sample lie on the relation between star formation and stellar mass. The total SFR values are estimated by integrating the SFR surface density. The discrepant point in the lower right corner is the asymmetric galaxy UGC 7256 (NGC 4203) with low SFR. It is the only early type galaxy (S0) in our sample. 

\begin{figure*}
\begin{center}
\begin{tabular}{c@{\hspace{0.1in}}c@{\hspace{0.1in}}}
\includegraphics[width=0.45\textwidth]{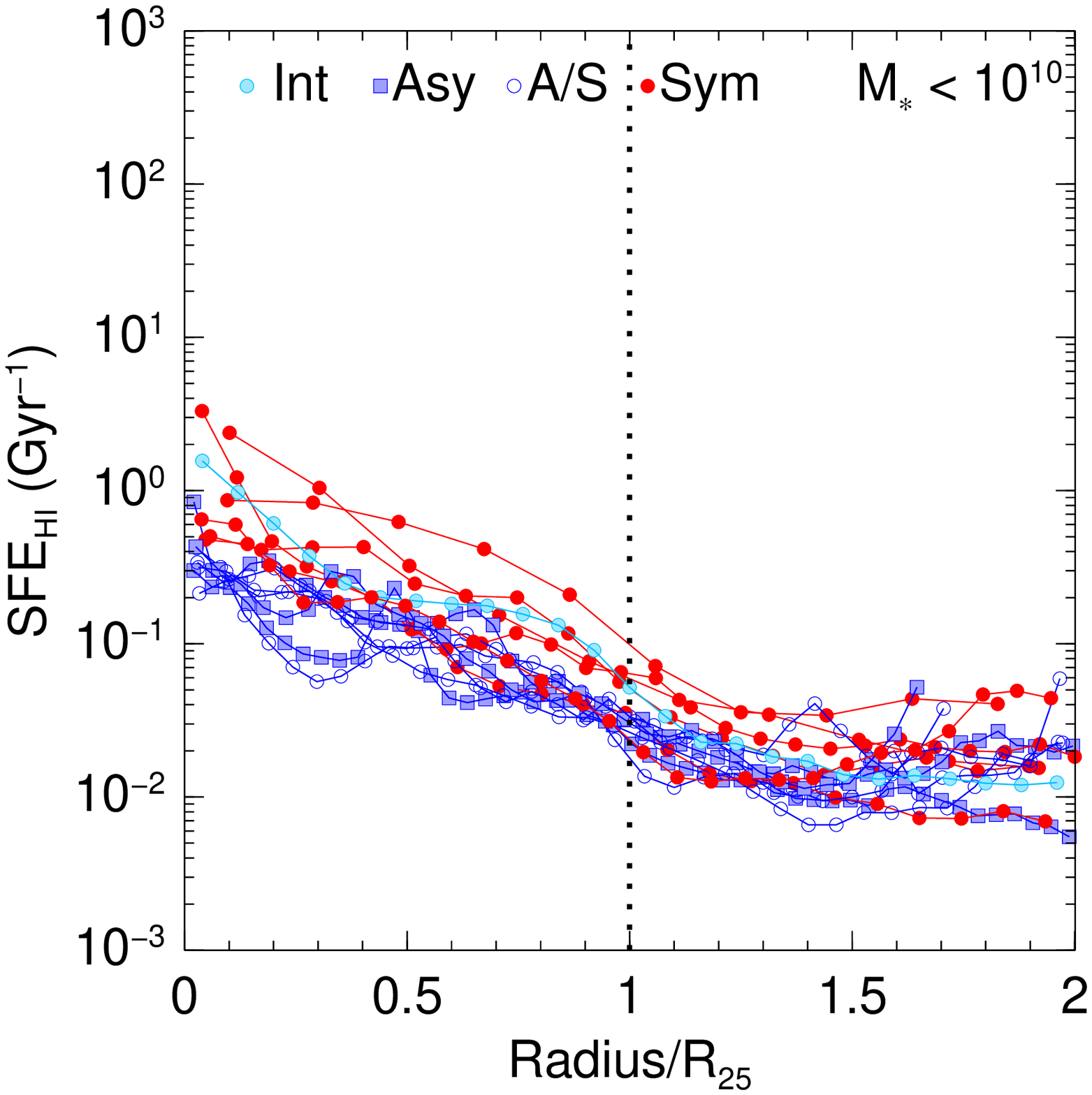}&
\includegraphics[width=0.45\textwidth]{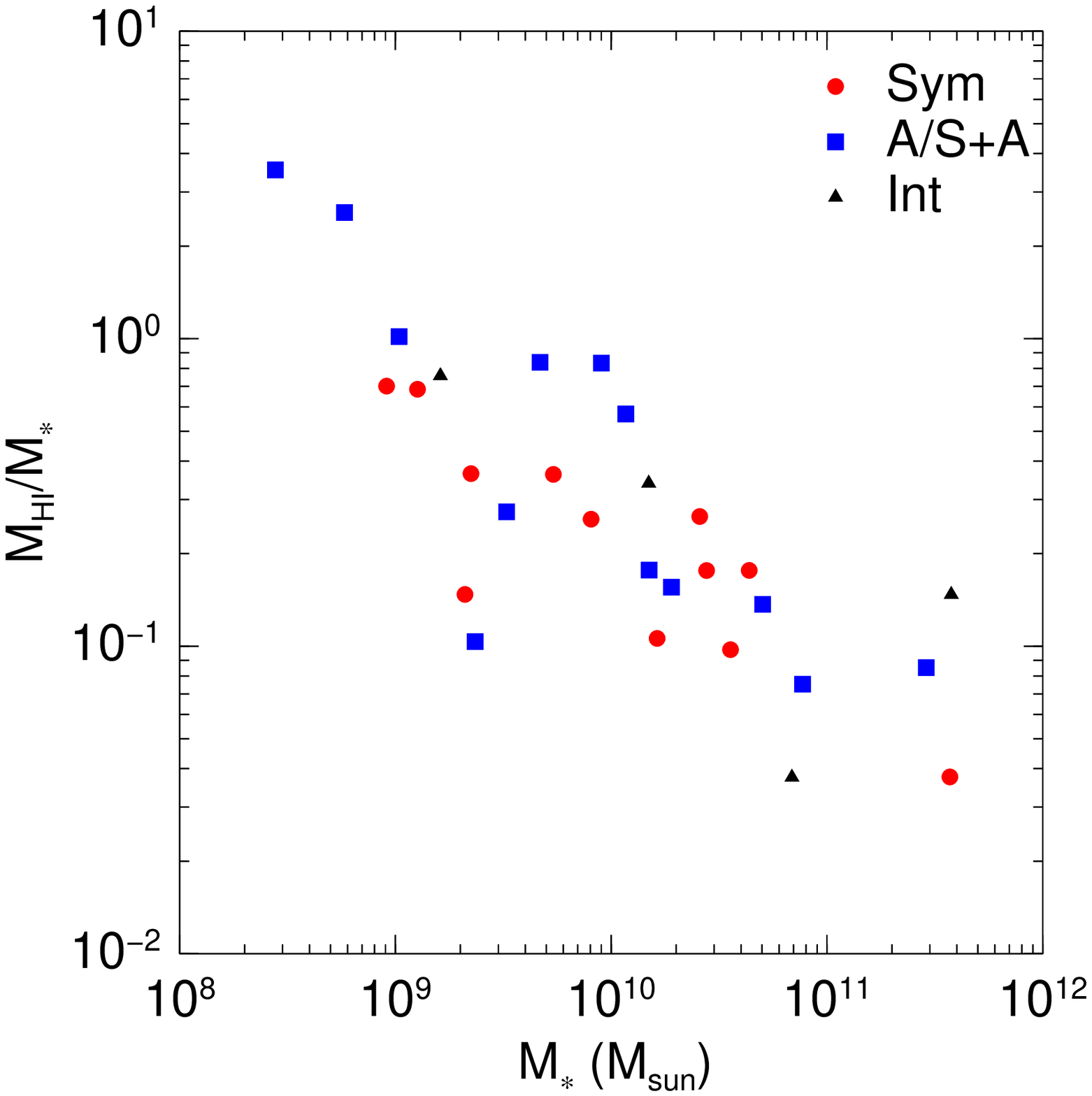}
\end{tabular}
\caption{Left: SFE$_{\rm HI}$ as a function of r/\ropt\ when the total stellar mass $M_{*}$ is less than 10$^{10}$ M$_\odot$. The vertical dotted line represents where $r =$ \ropt.  Right: the ratio of the total \HI\ mass to the total stellar mass (M$_{\rm HI}$/$M_{*}$) versus  $M_{*}$ in units of solar mass. 
\label{fig:Mstar}}
\end{center}
\end{figure*}

We grouped our sample of galaxies  into three bins of the stellar mass: less massive than $10^{10}$ M$_\odot$, intermediate, and more massive than $10^{11}$ M$_\odot$. After comparing the galaxy classes in the same bins for the SFL and SFE, we noticed that there are still no differences between the symmetric, asymmetric, and interacting classes except that SFE$_{\HI}$ of the asymmetric and slightly asymmetric galaxies appears lower than that of the isolated symmetric galaxies when $M_* < 10^{10}$ M$_\odot$ (see the left panel of Figure \ref{fig:Mstar}). The implication is that interactions suppress the SFE or increase the gas reservoir without enhancing the SF.  To verify if the asymmetric galaxies with $M_* < 10^{10}$ are gas rich, we plotted the ratio of the total \HI\ mass to the total stellar mass against $M_*$ in Figure \ref{fig:Mstar} (right). 
It appears that the symmetric galaxies below a stellar mass of 10$^{10}$ M$_\odot$ are systematically less gas rich than the asymmetric galaxies. The reason why the asymmetric galaxies are more gas rich is possibly due either to really lower SFE or the accretion of fresh gas.

\section{Summary and Conclusions}
\label{sum}

In order to find evidence for a positive correlation between gas accretion and star formation, we investigated the K-S law in the inner and outer regions (separately) with symmetric, asymmetric, and interacting groups of galaxies and compared them to each other.  In addition, we compared the scaled radial profiles of \sigsfr, \sightwo, \sighi, and \siggas\ for the groups to see if there is any difference between them. 

1. Among the sub-sample of 16 galaxies with both CO and \HI\ data, several galaxies show  \sightwo\  lower than \sighi\ in the central regions unlike the general trend of CO  and \HI\ radial profiles demonstrating that the molecular gas density is higher than the \HI\ density near the center. 

2. The scaled radial profiles of the total gas are well constrained by the exponential fit regardless of the galaxy types. There is no significant difference between symmetric and asymmetric galaxy groups and the scale lengths are 0.56 (symmetric) and 0.52 (asymmetric). 

3. The examination for the K-S law in the inner regions exhibits no tighter correlation  among the symmetric and asymmetric groups of galaxies and no clear sign for a relationship between gas accretion and star formation. However, the power-law correlation of some symmetric galaxies appears somewhat weaker  than that of the asymmetric and interacting galaxies, especially in the central regions where \sigsfr\ and \siggas\ are higher. The average indices for the molecular K-S law are 0.84 (asymmetric) and 0.88 (symmetric). The indices for the total gas are 2.93 (asymmetric) and 2.15 (symmetric).  
The SFE for the molecular gas is roughly constant with radius while the SFE for the total gas decreases with radius for both the symmetric and asymmetric galaxies. 

4. From the plot of \sigsfr\ vs. \sighi\ for the outer regions beyond \ropt, we noticed that there is a tight correlation between SFR and \HI\ in the outer discs unlike in the inner regions.    The average K-S index for all the galaxies is 1.21. 
The SFE$_{\rm HI}$ decreases (roughly exponentially) until 1.5$\times$\ropt\ and flattens beyond that radius. 
There is no significant  difference between the galaxy groups for the SFL and SFE  except that the isolated symmetric galaxies with small stellar mass ($< 10^{10}$ M$_\odot$) have somewhat higher SFE.

\section*{Acknowledgements}

We thank the anonymous referee for useful suggestions that improved this paper. 
The research leading to these results has received funding from the European Research Council under the European Union's Seventh Framework Programme (FP/2007-2013) / ERC Grant Agreement nr. 291531. 

%%%%%%%%%%%%%%%%%%%%%%%%%%%%%%%%%%%%%%%%%%%%%%%%%%



\bibliographystyle{mnras}
\bibliography{refer} % if your bibtex file is called example.bib

%\begin{thebibliography}{99}
%\end{thebibliography}

%%%%%%%%%%%%%%%%%%%%%%%%%%%%%%%%%%%%%%%%%%%%%%%%%%

%%%%%%%%%%%%%%%%% APPENDICES %%%%%%%%%%%%%%%%%%%%%

\appendix

%\section{Star Formation Law for Each Galaxy}


\renewcommand{\thefigure}{A\arabic{figure}}

\begin{figure*}
\begin{center}
\begin{tabular}{c@{\hspace{0.1in}}c@{\hspace{0.1in}}c@{\hspace{0.1in}}c}
\includegraphics[width=0.35\textwidth]{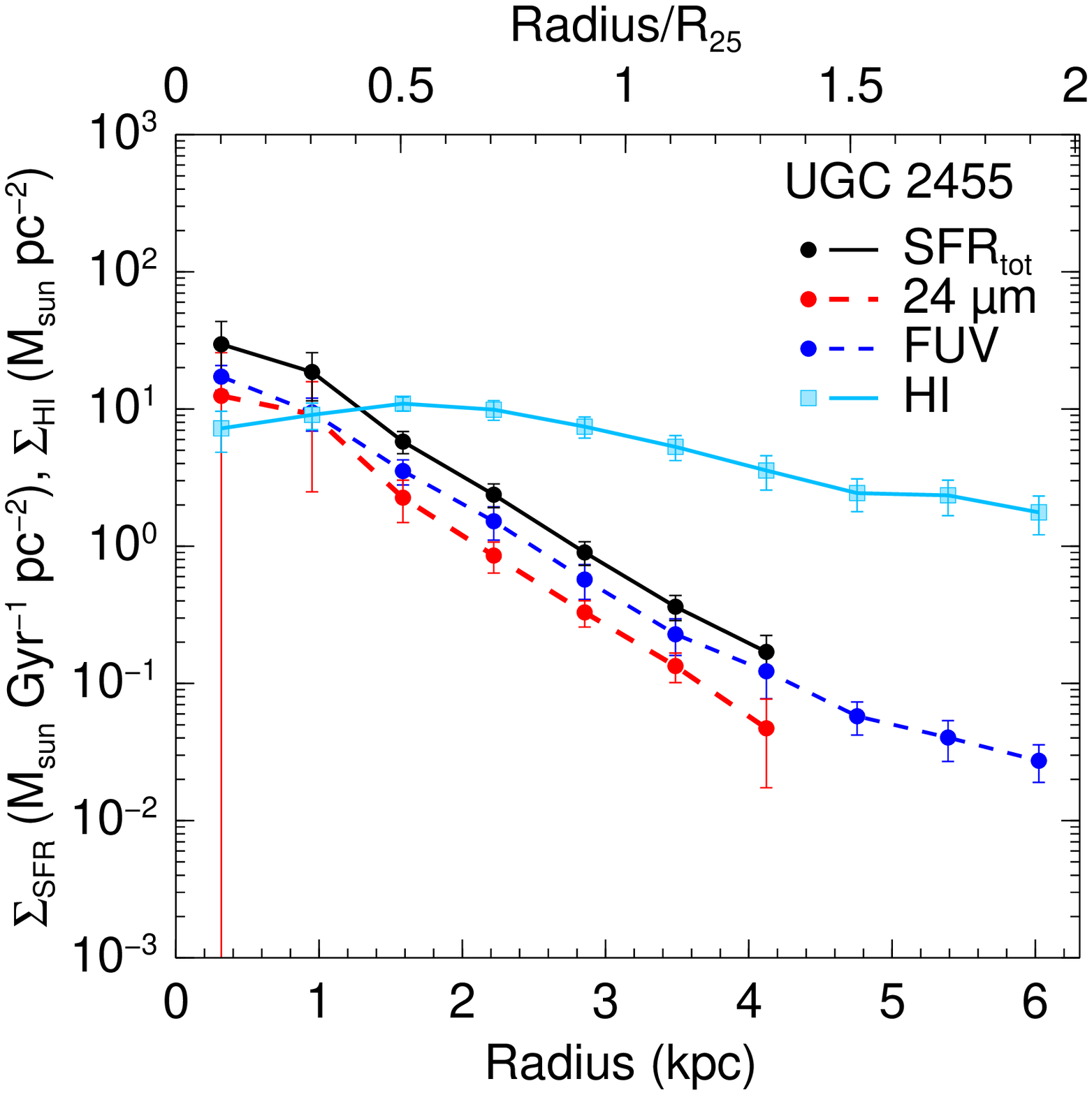}&
\includegraphics[width=0.35\textwidth]{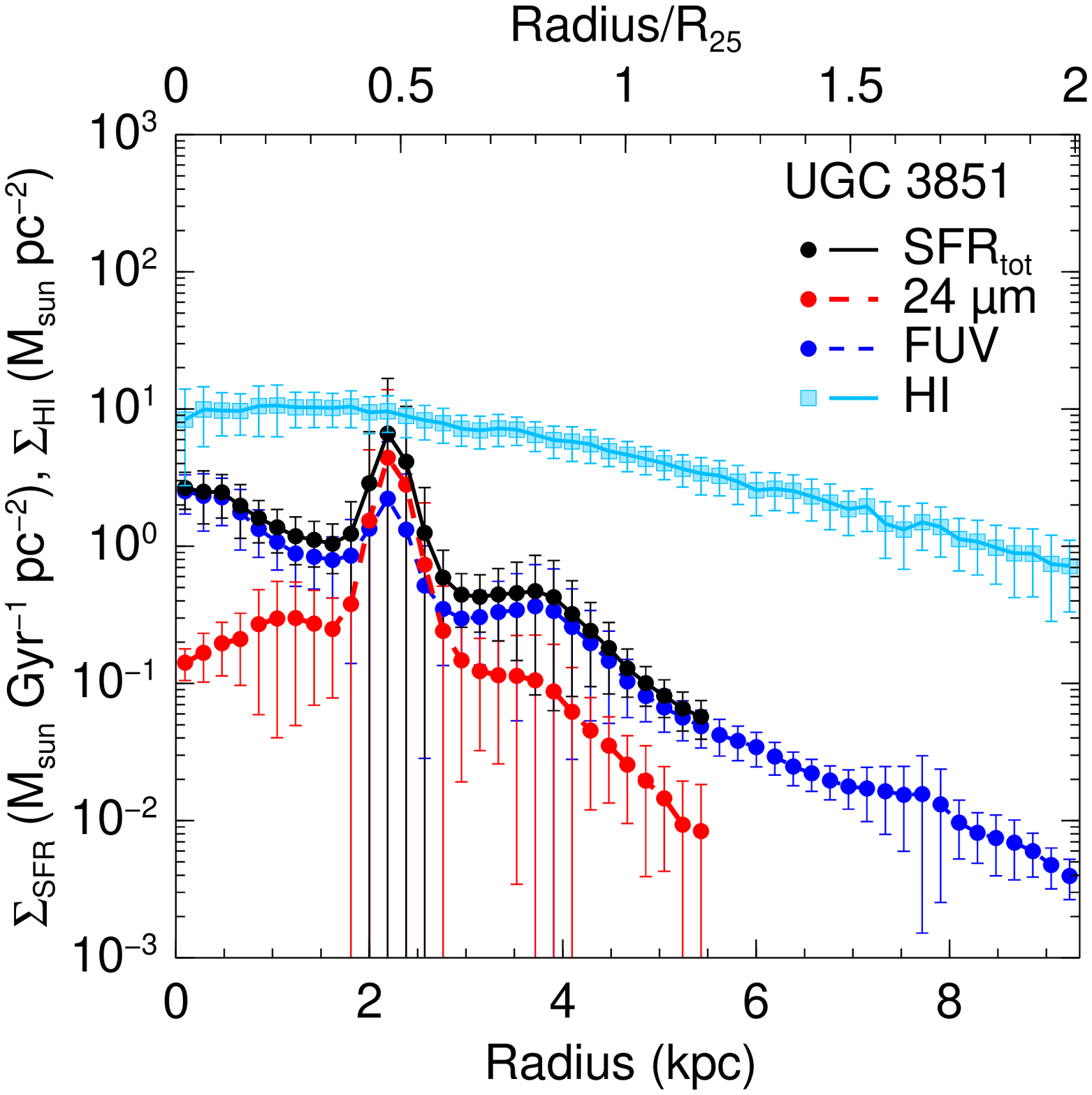}&
\includegraphics[width=0.35\textwidth]{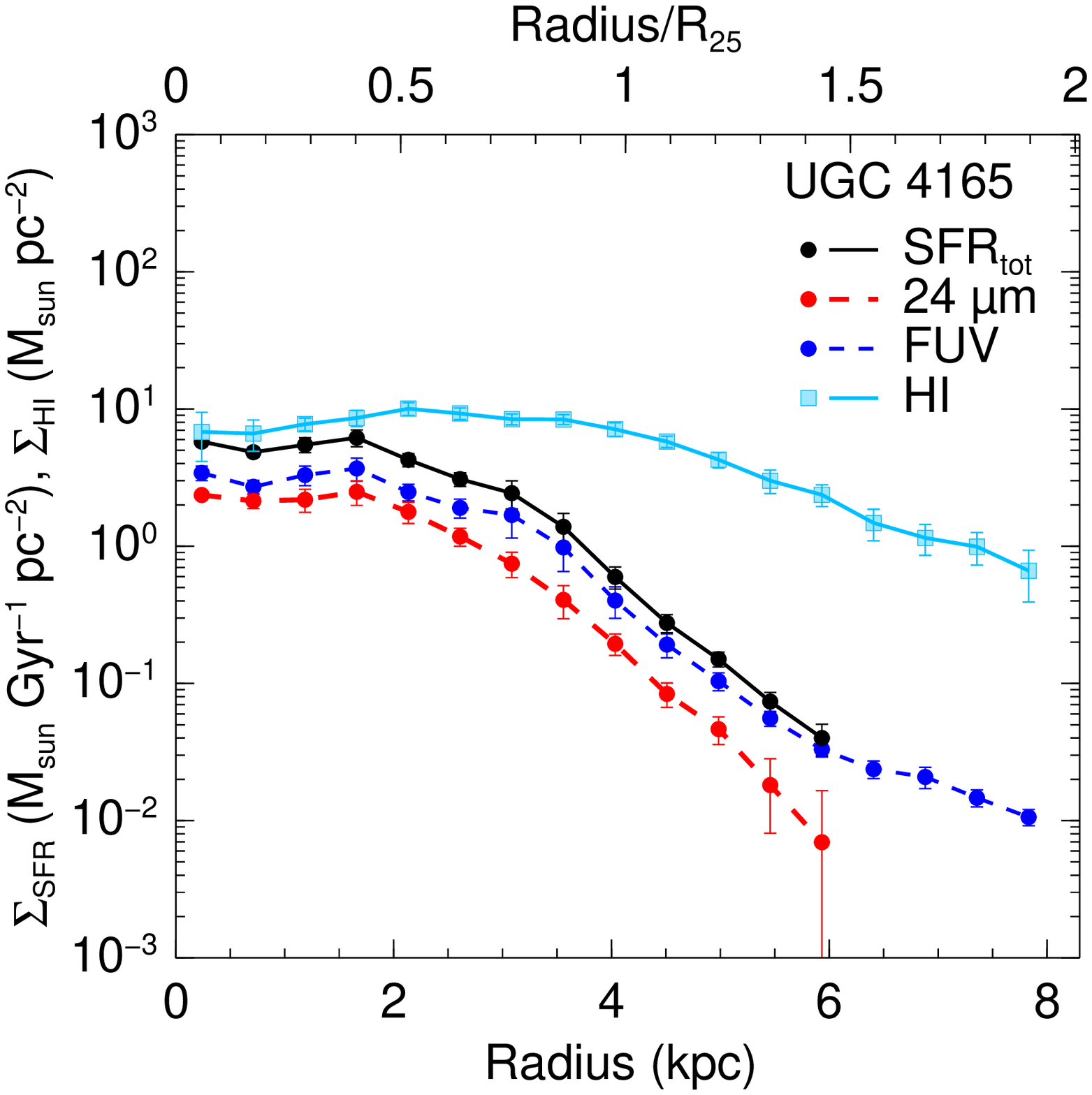}\\
\includegraphics[width=0.35\textwidth]{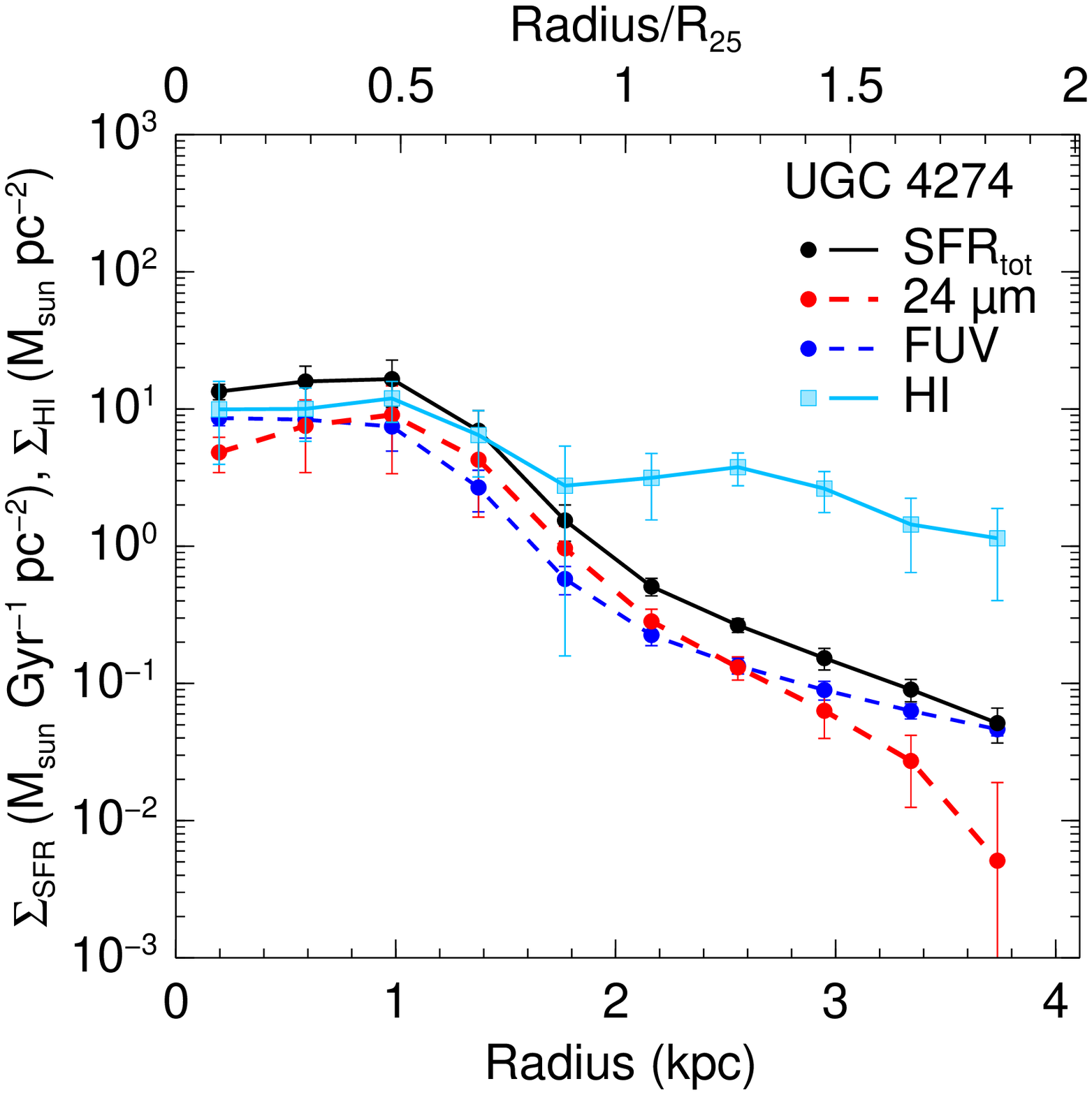}&
\includegraphics[width=0.35\textwidth]{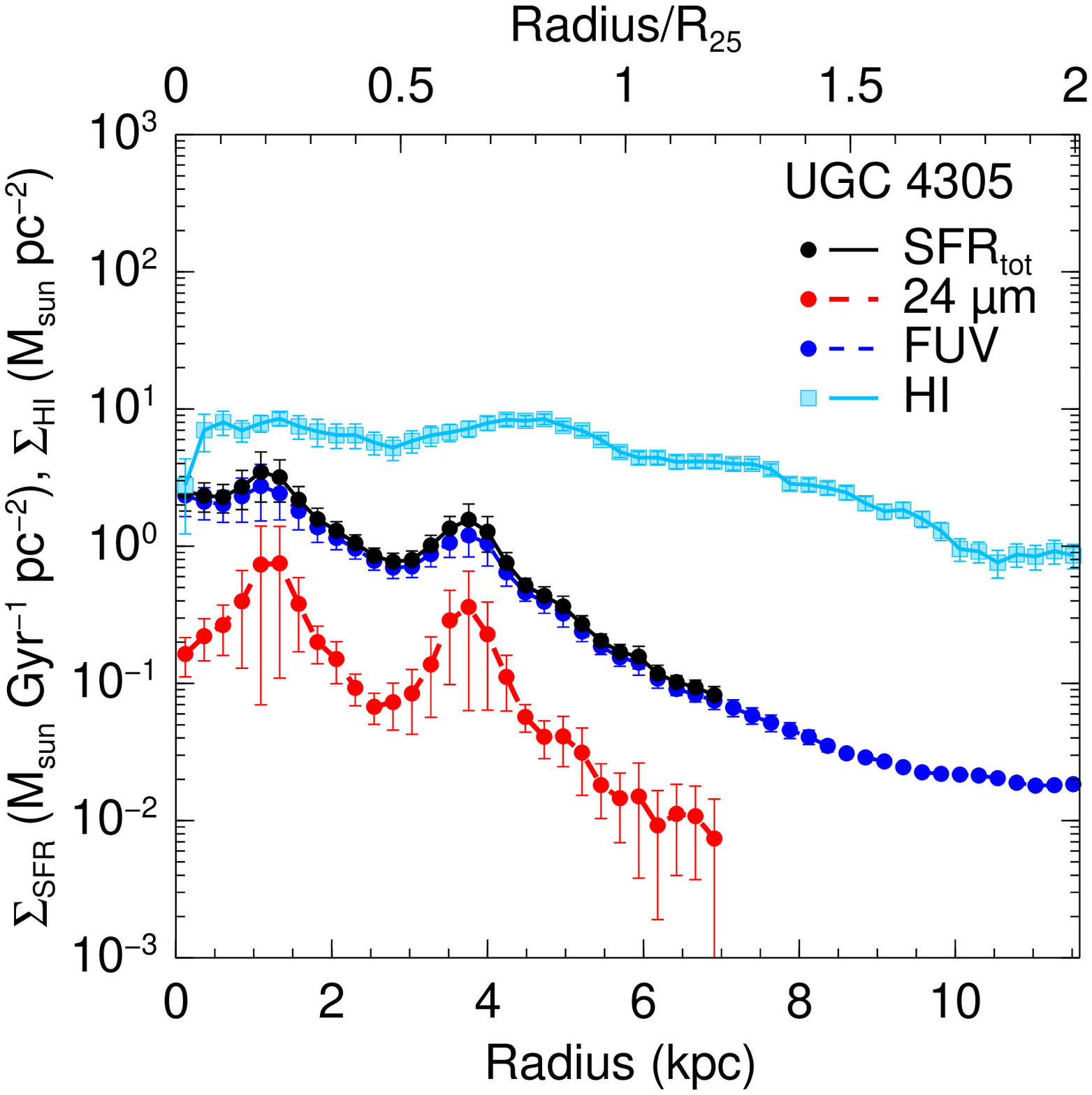}&
\includegraphics[width=0.35\textwidth]{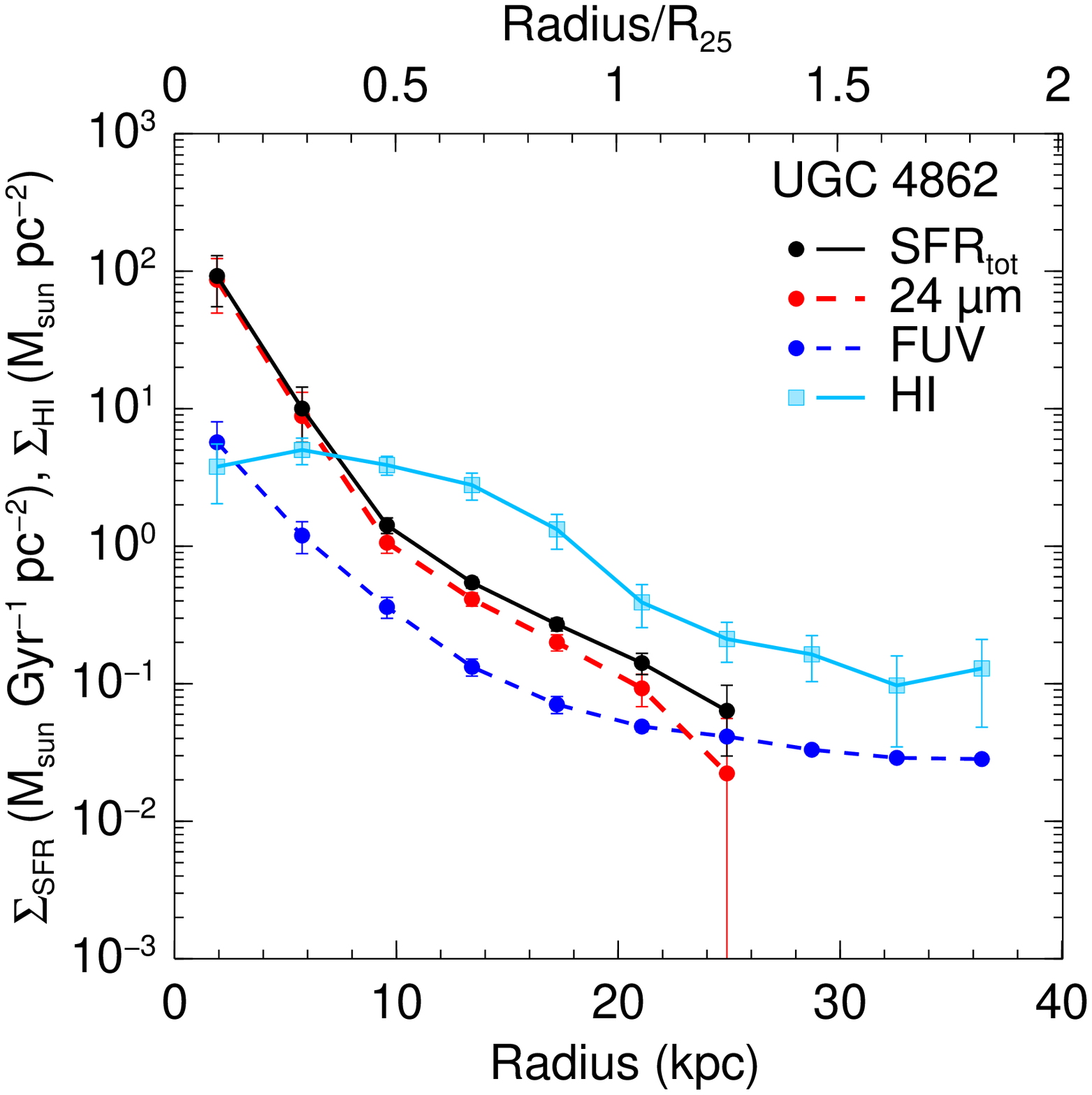}\\
\includegraphics[width=0.35\textwidth]{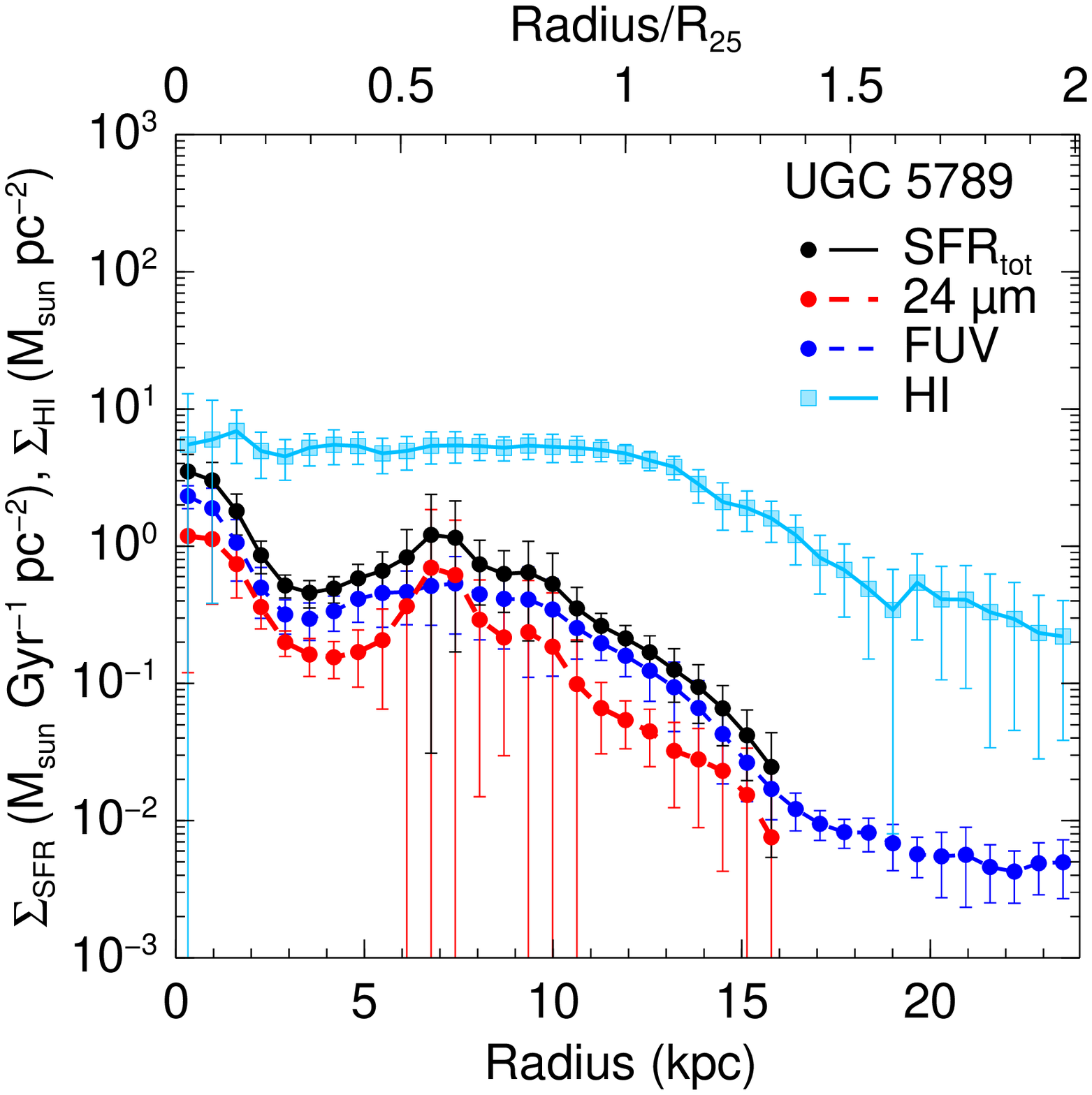}&
\includegraphics[width=0.35\textwidth]{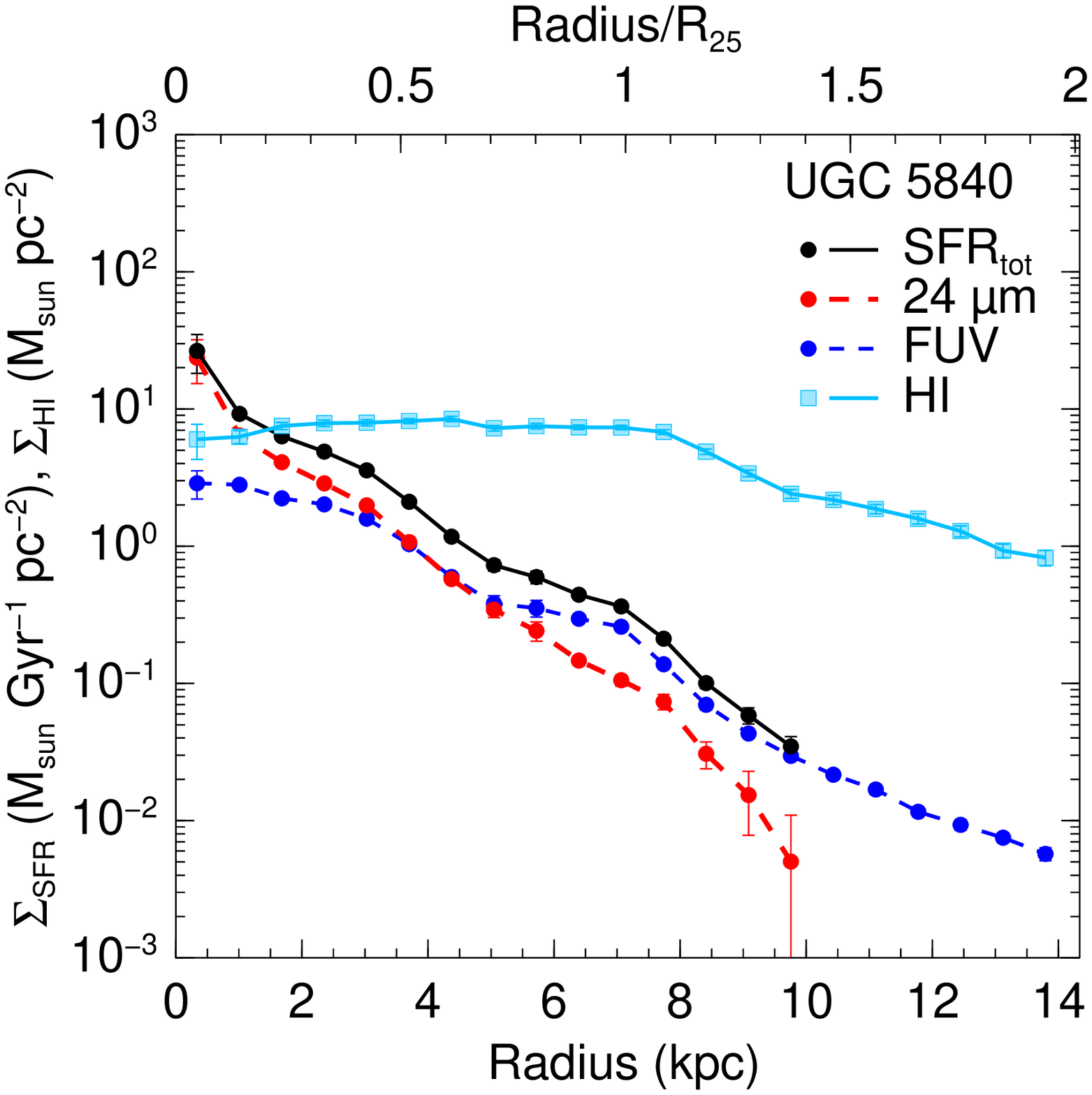}&
\includegraphics[width=0.35\textwidth]{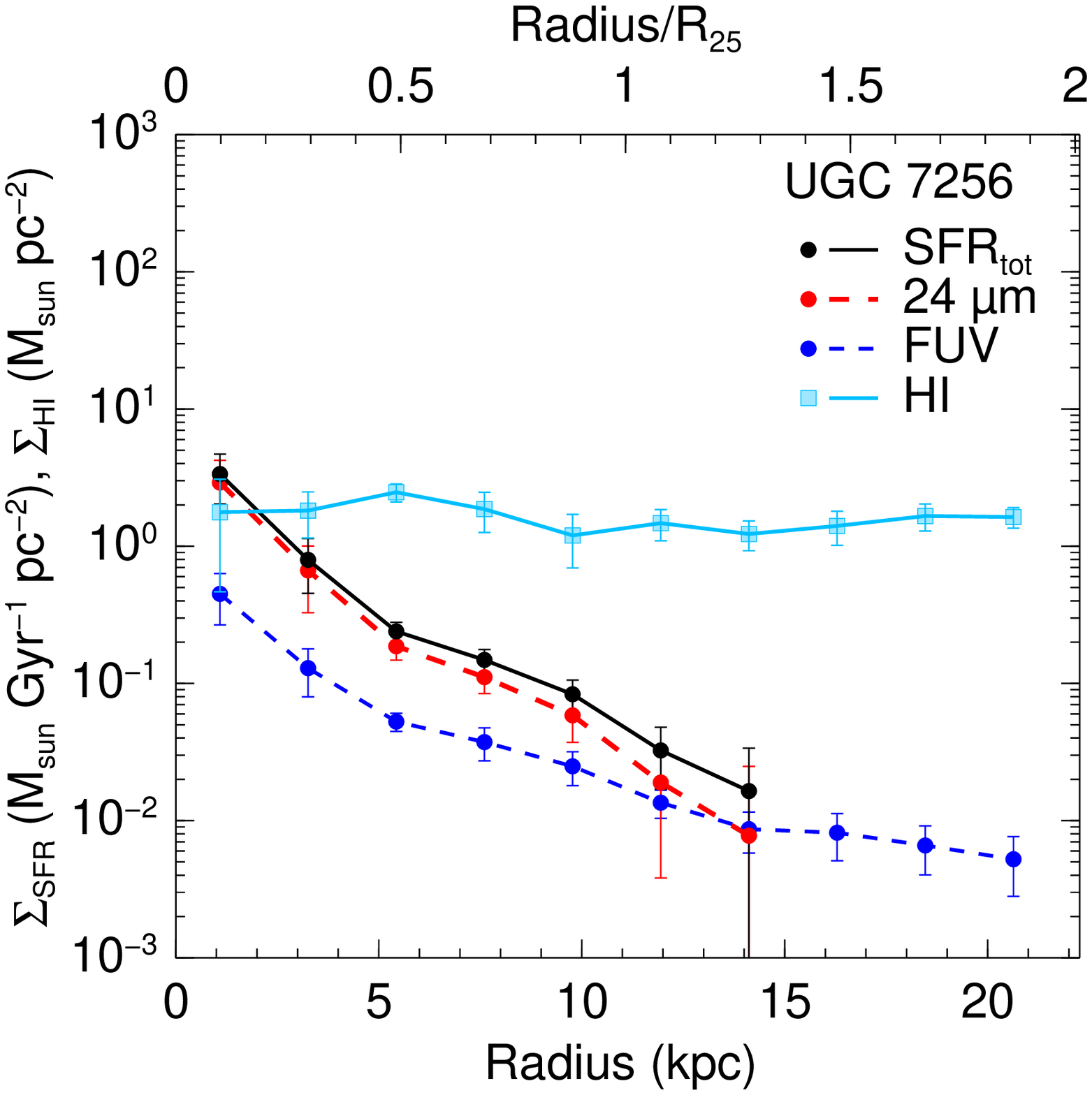}
\end{tabular}
\caption{SFR and \HI\ surface densities as a function of radius for 13 galaxies that have no CO data.} 
\label{restrp}
\end{center}
\end{figure*}

\addtocounter{figure}{-1}

\begin{figure*}
\begin{center}
\begin{tabular}{c@{\hspace{0.1in}}c@{\hspace{0.1in}}c@{\hspace{0.1in}}c}
\includegraphics[width=0.35\textwidth]{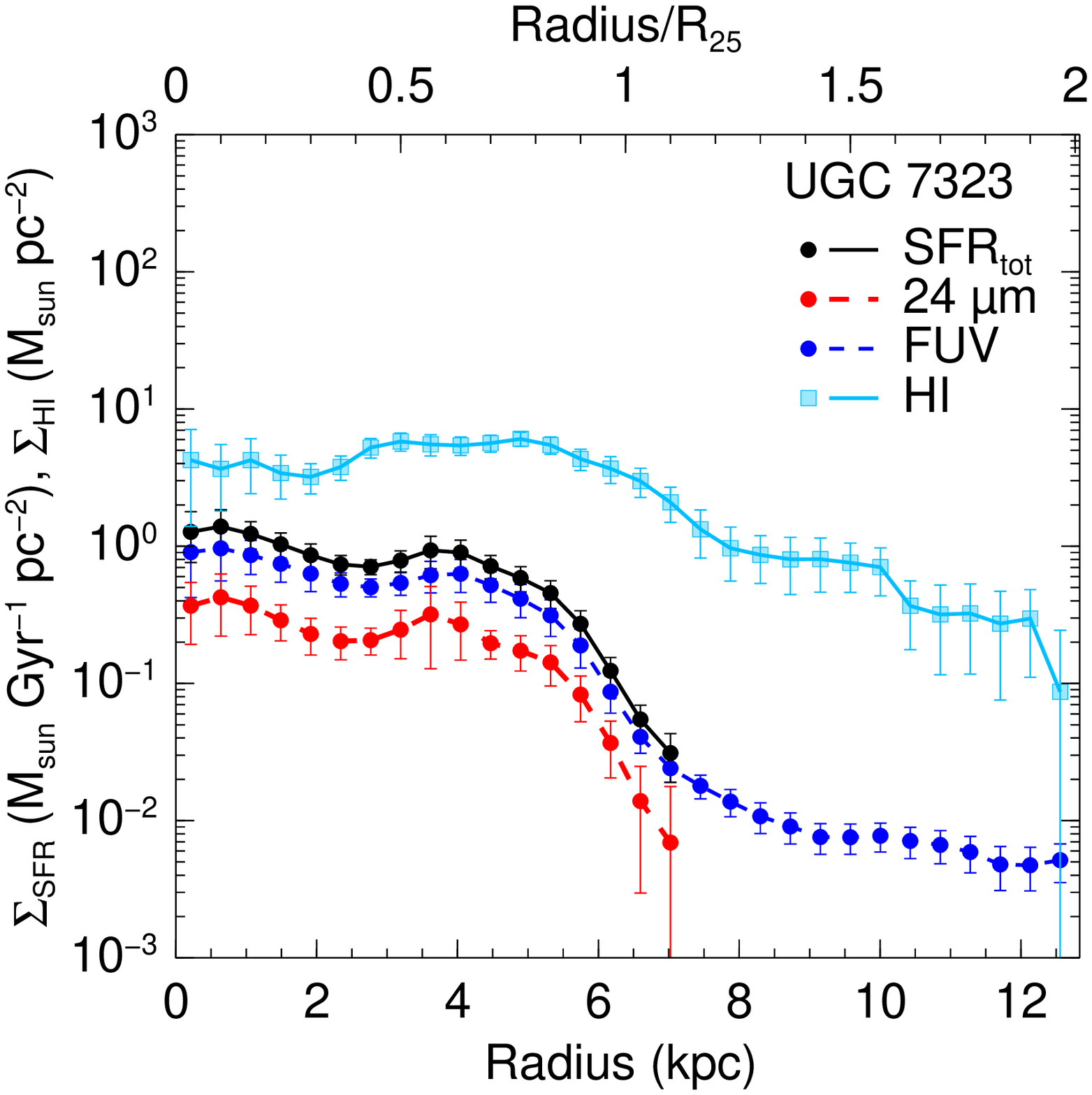}&
\includegraphics[width=0.35\textwidth]{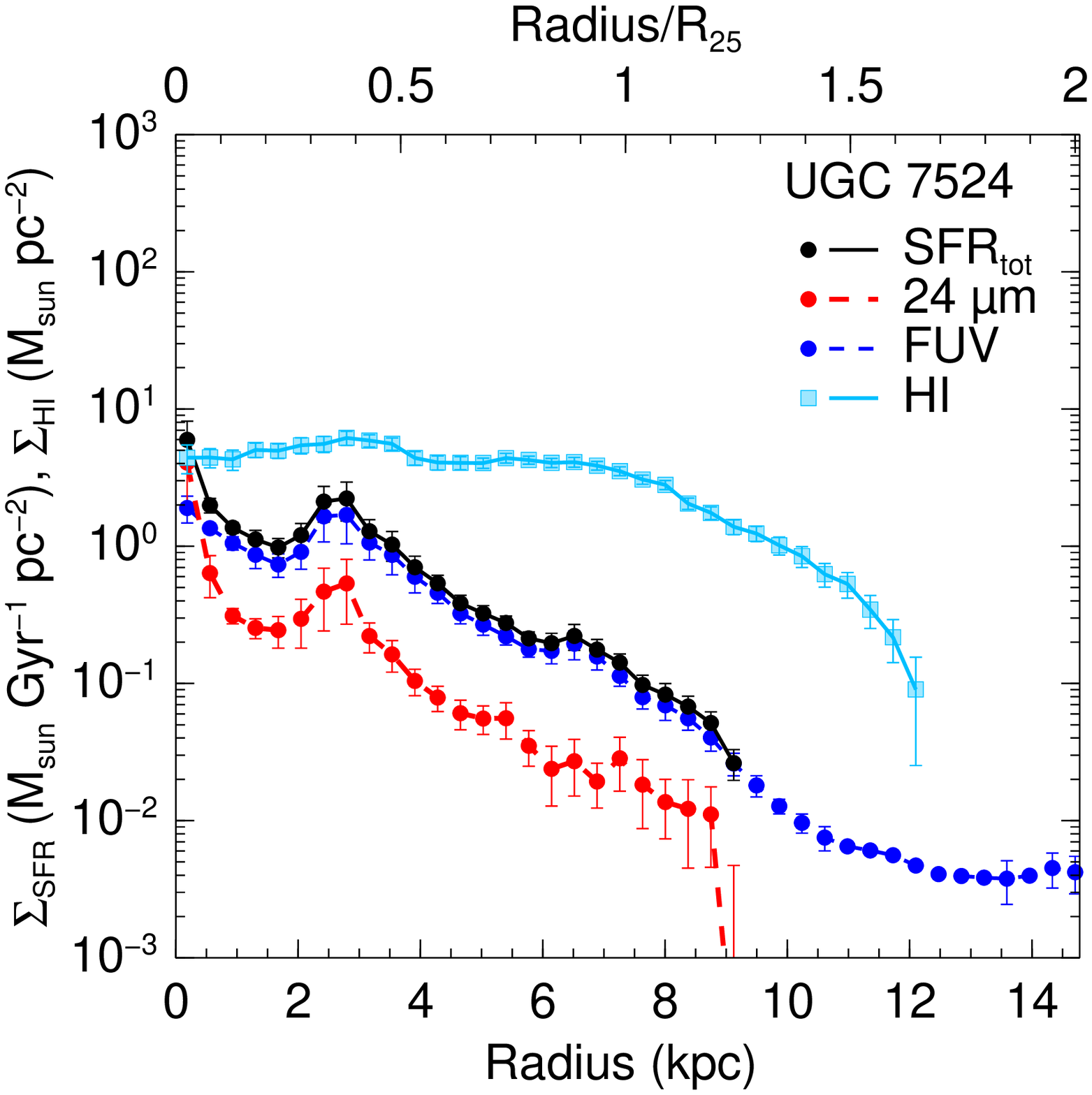}&
\includegraphics[width=0.35\textwidth]{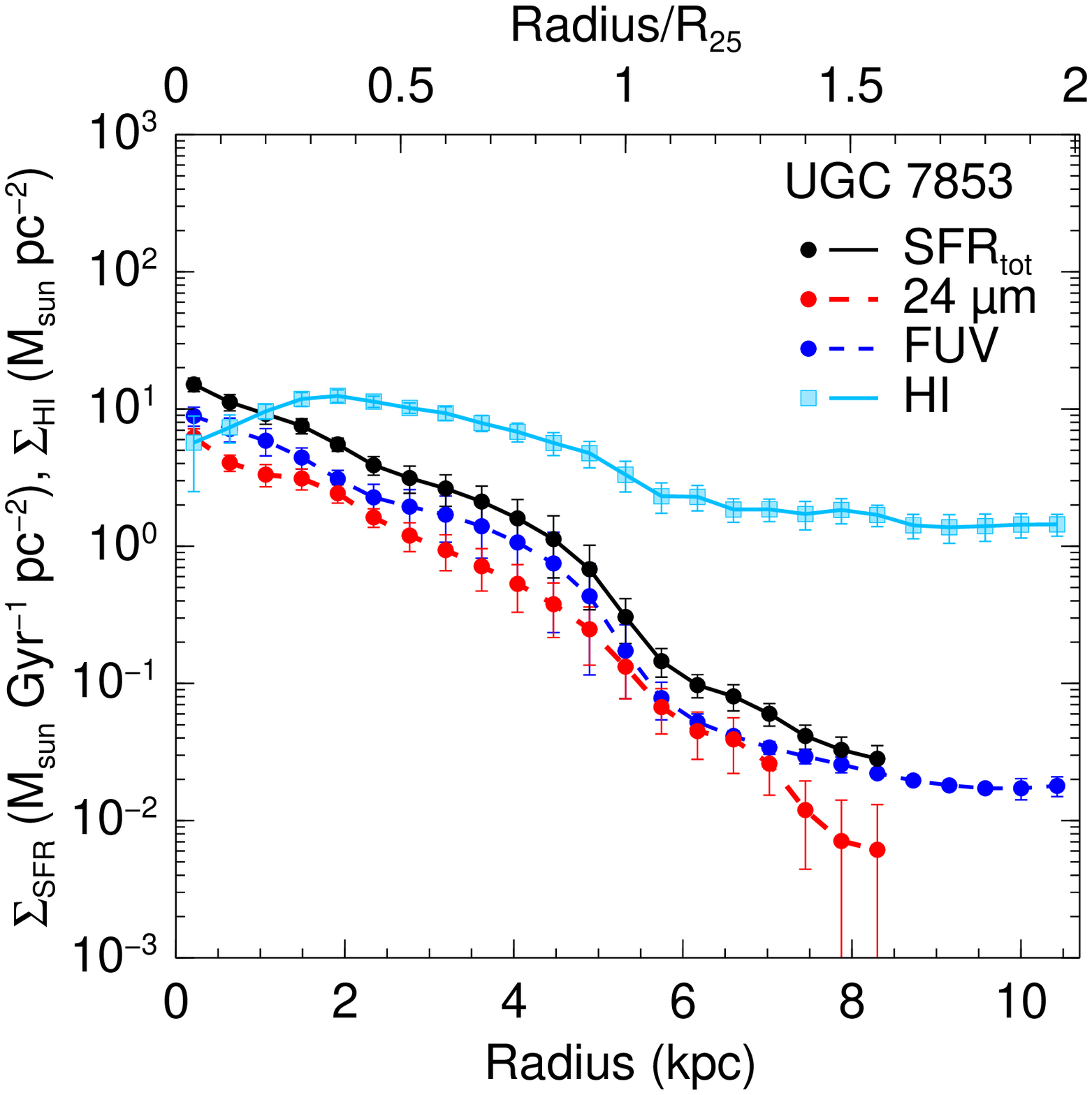}\\
\includegraphics[width=0.35\textwidth]{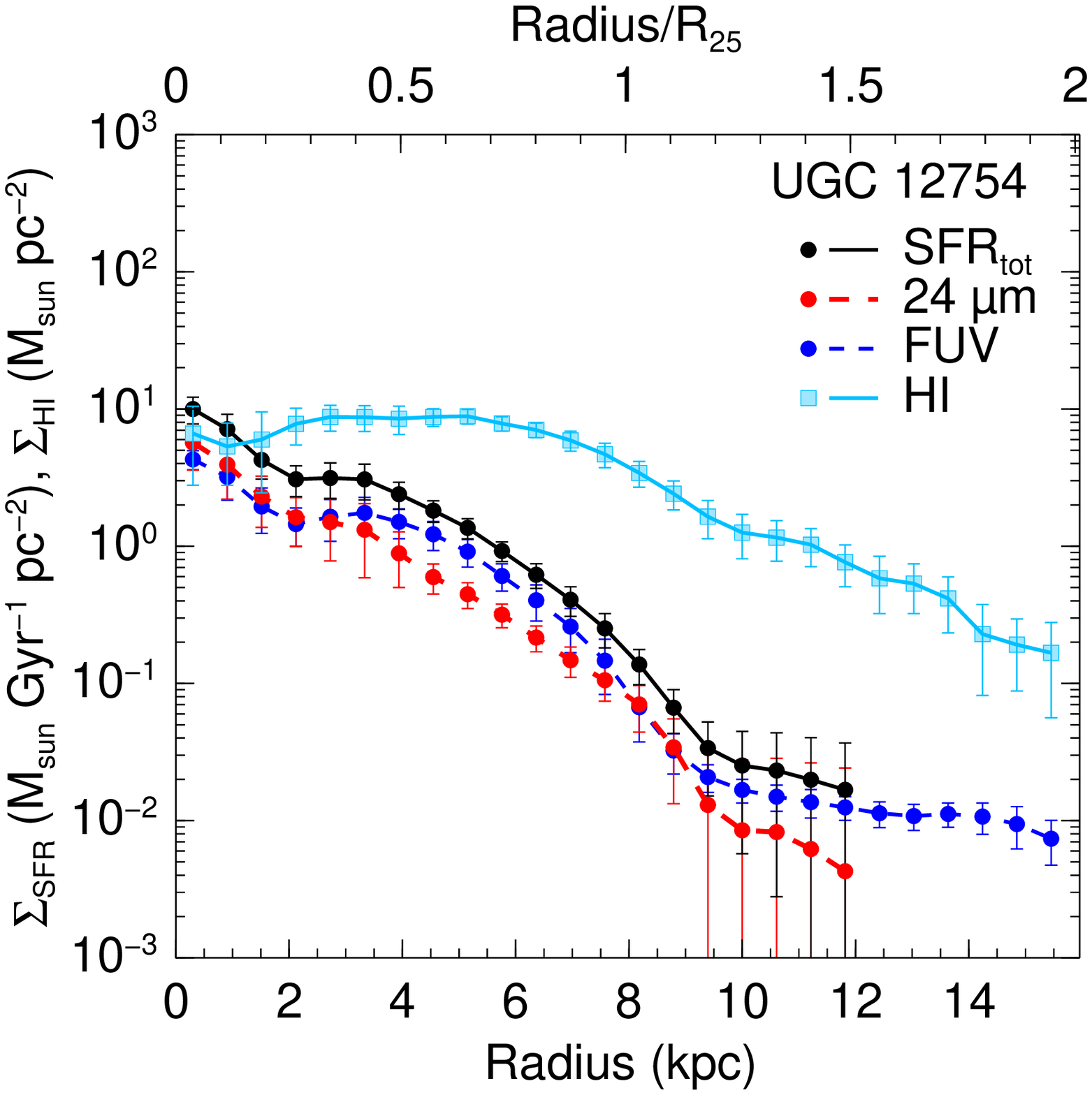}&
\end{tabular}
\caption{continued} 
\label{restrp}
\end{center}
\end{figure*}

\begin{figure*}
\centering
\begin{tabular}{c@{\hspace{0.1in}}c@{\hspace{0.1in}}c@{\hspace{0.1in}}c}
\includegraphics[width=0.35\textwidth]{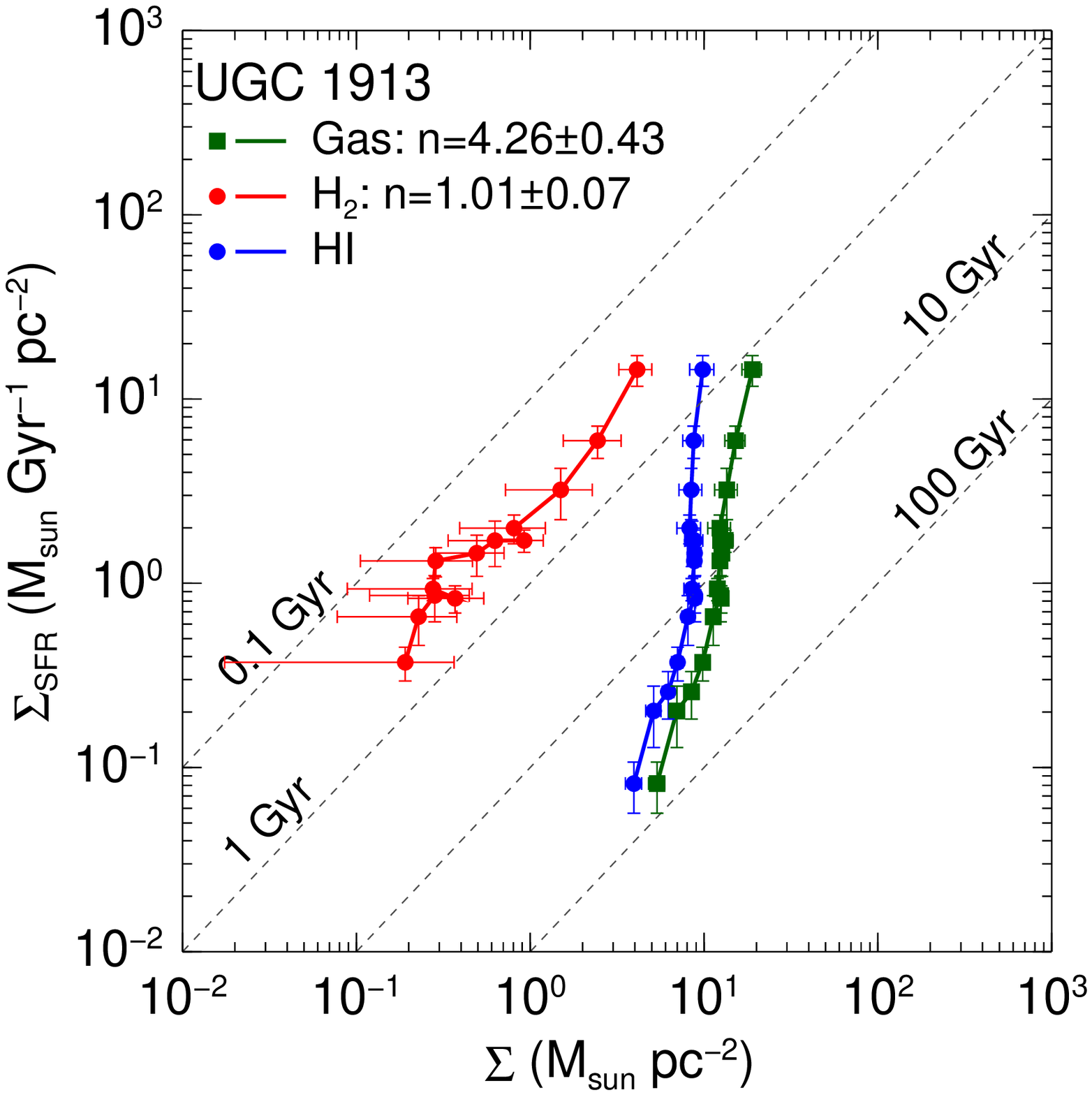}
\includegraphics[width=0.35\textwidth]{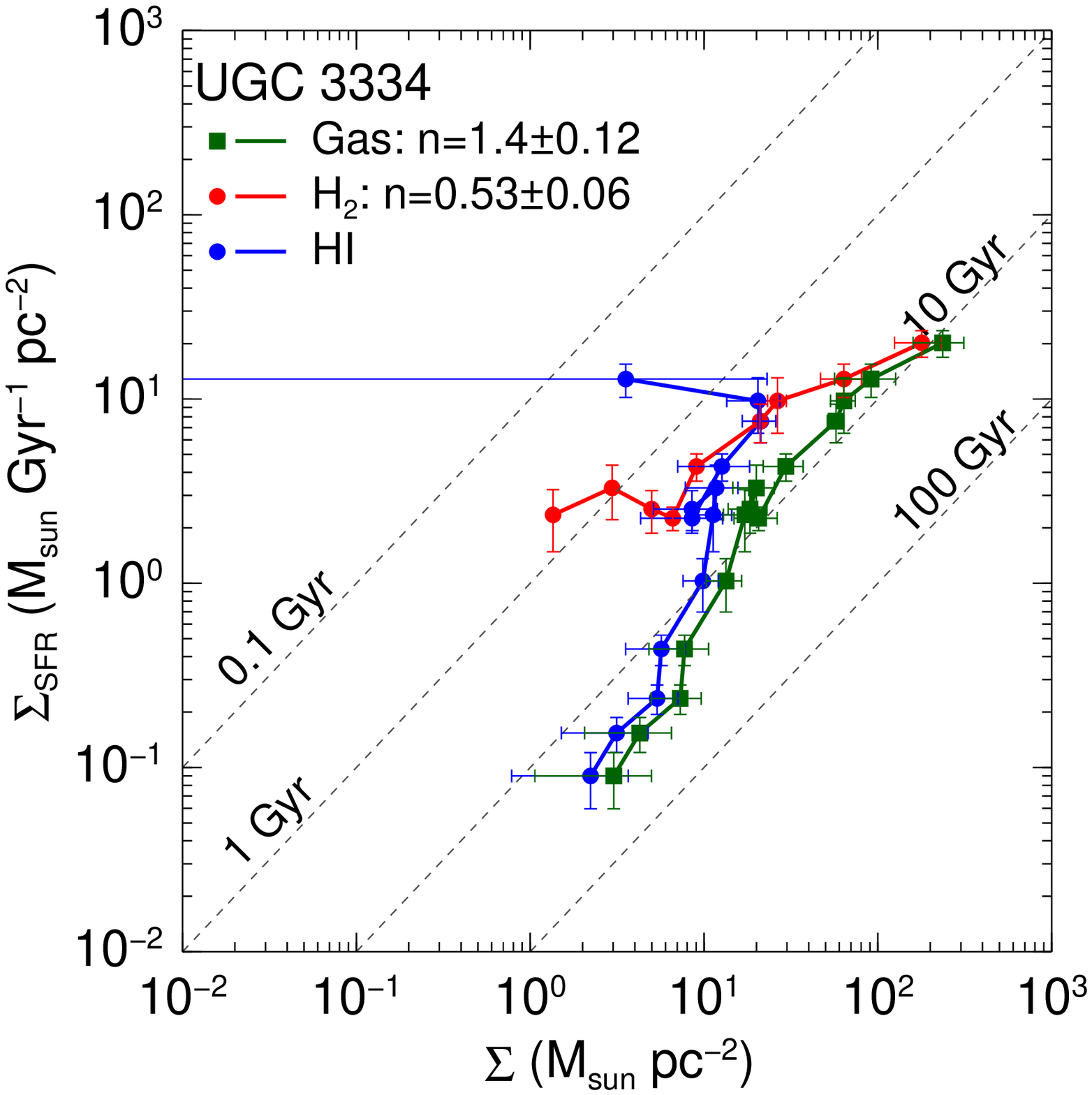}
\includegraphics[width=0.35\textwidth]{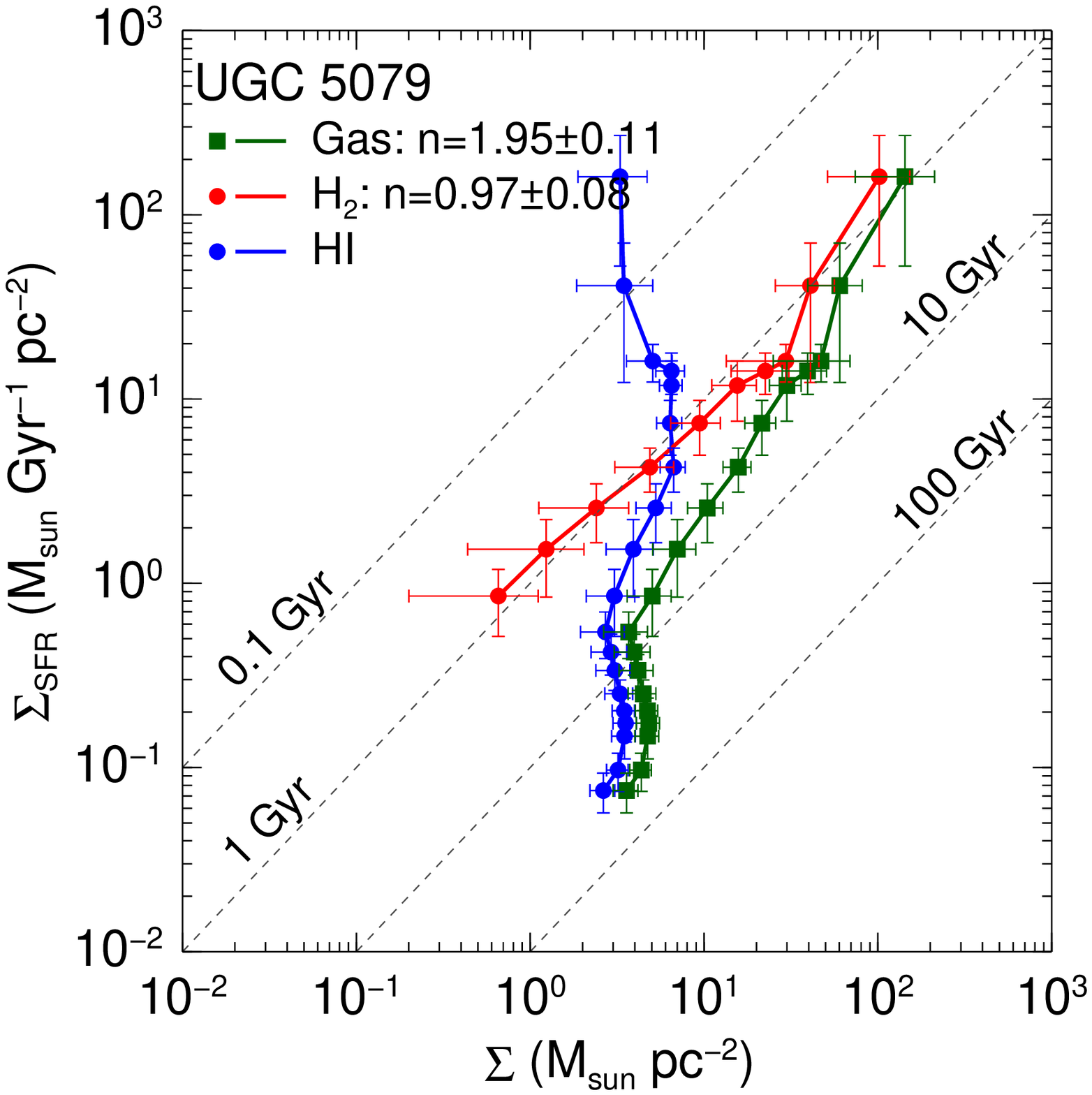}\\
\includegraphics[width=0.35\textwidth]{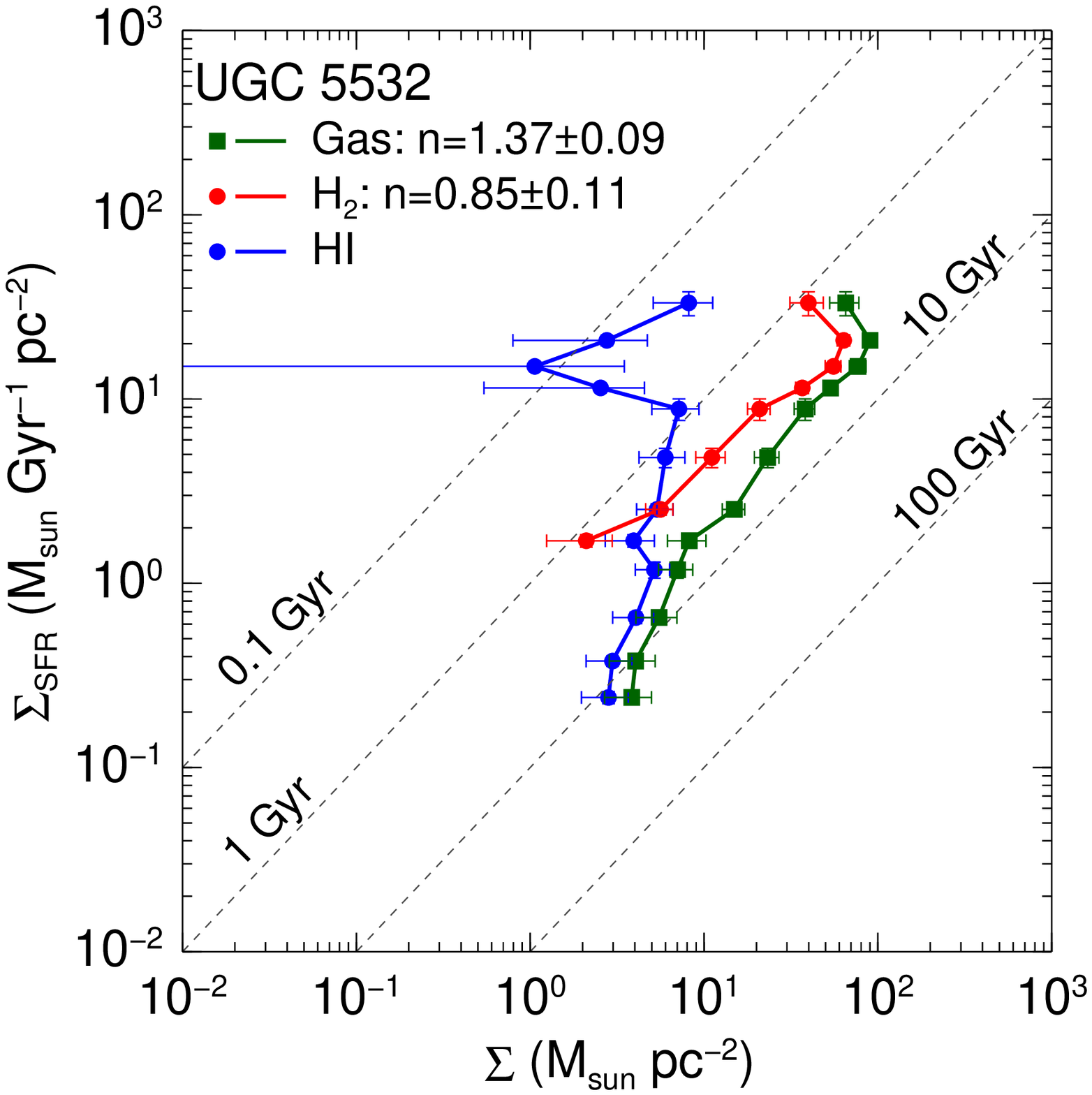}
\includegraphics[width=0.35\textwidth]{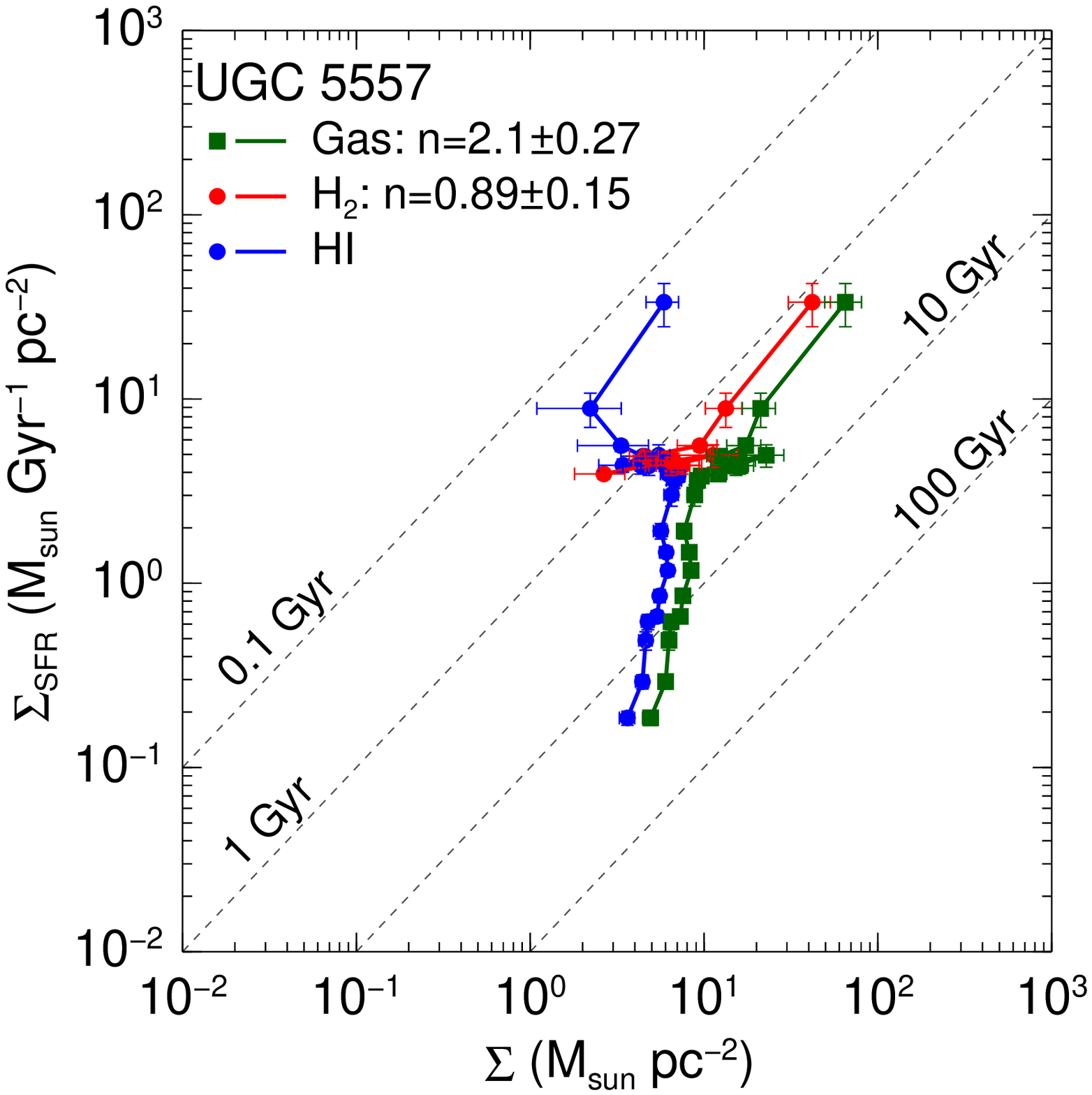}
\includegraphics[width=0.35\textwidth]{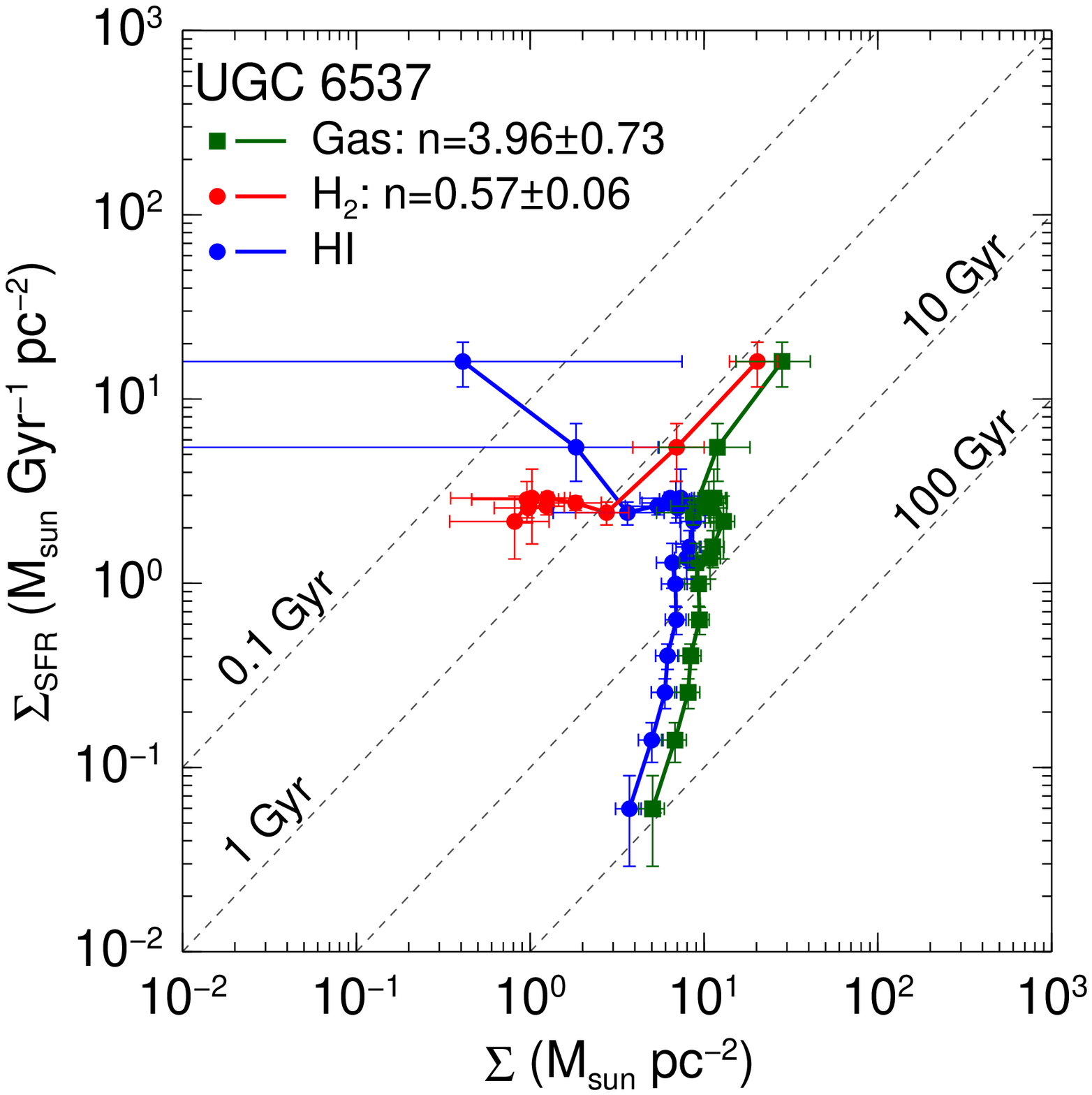}\\
\includegraphics[width=0.35\textwidth]{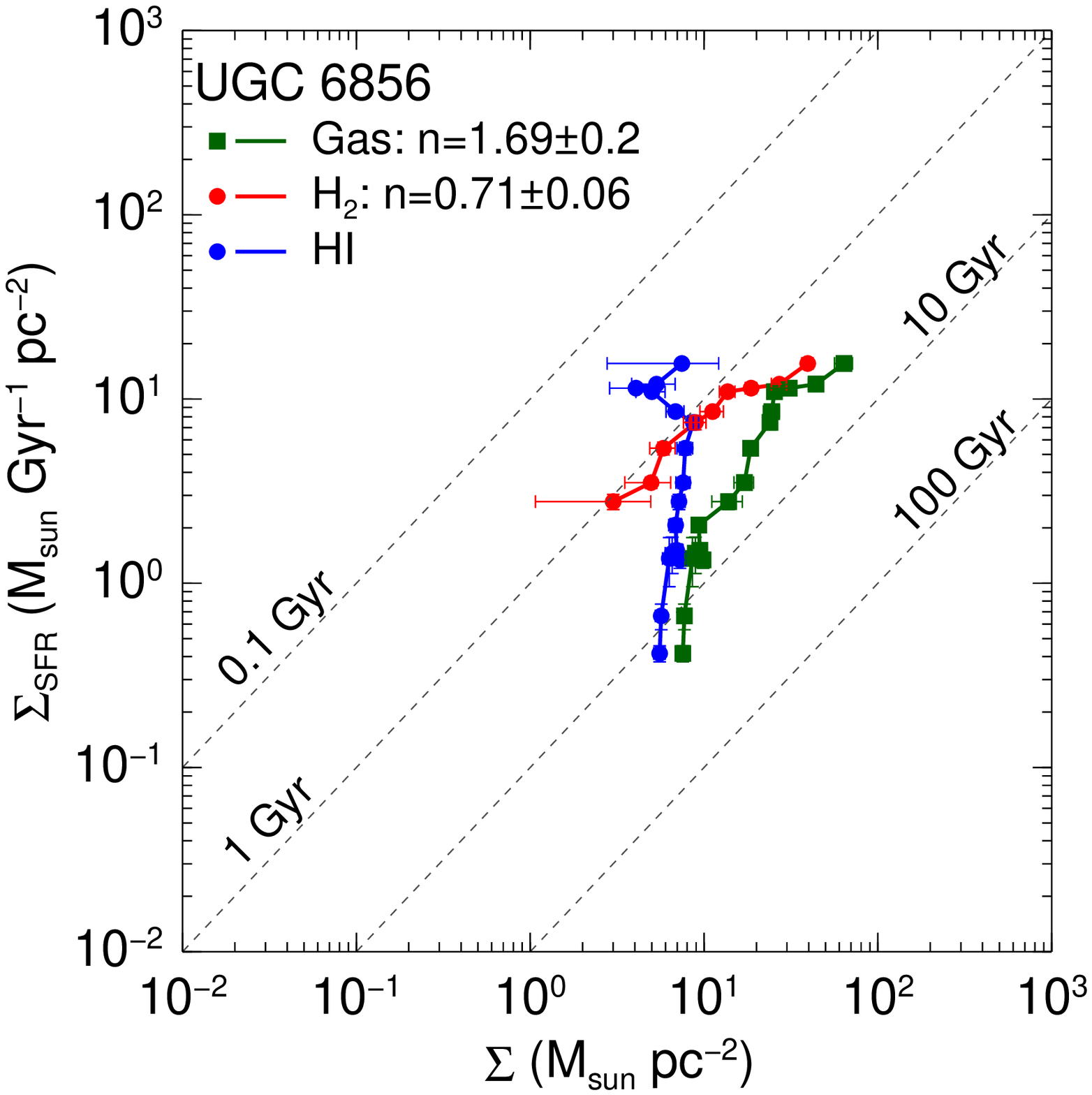}
\includegraphics[width=0.35\textwidth]{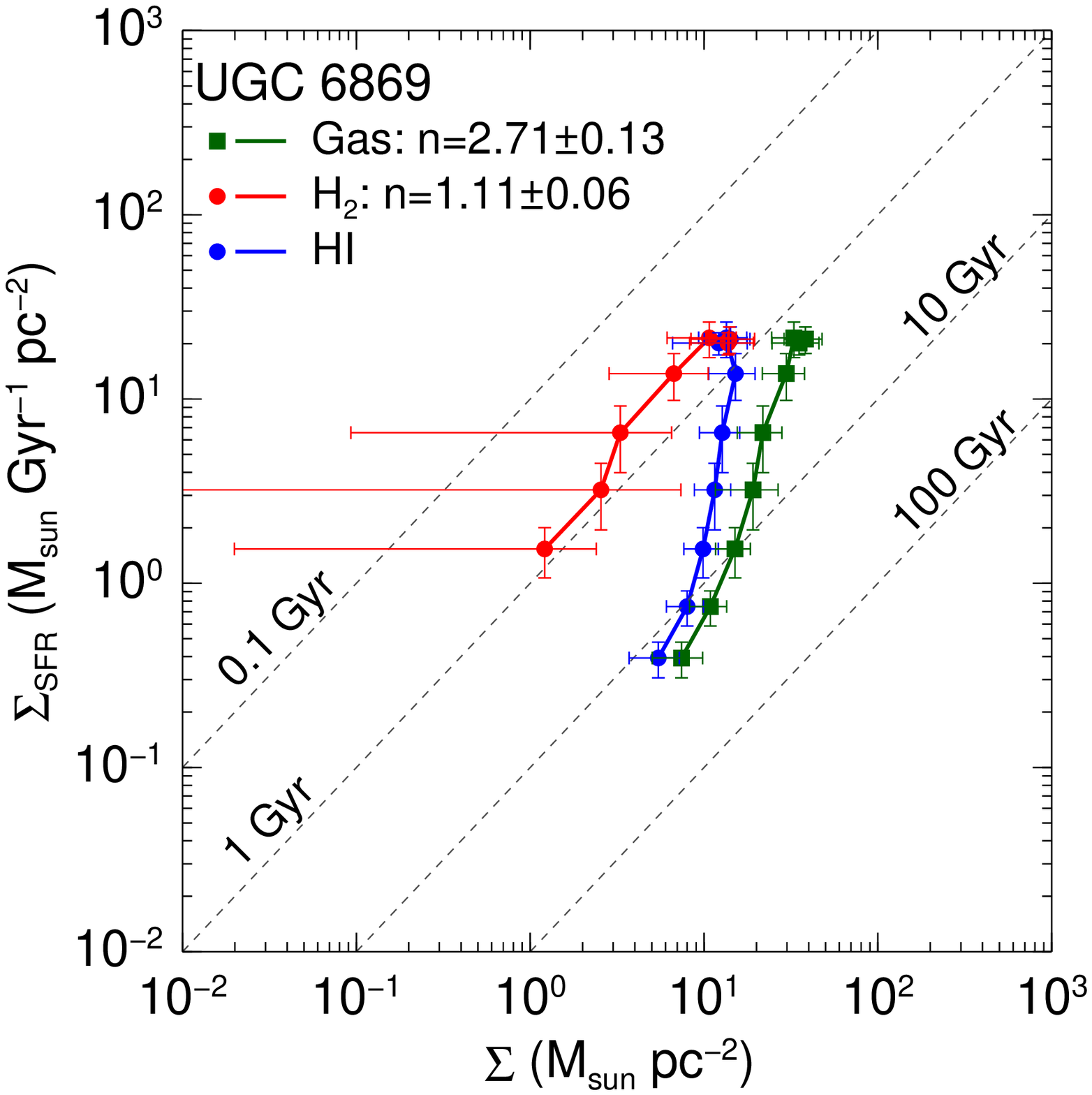}
\includegraphics[width=0.35\textwidth]{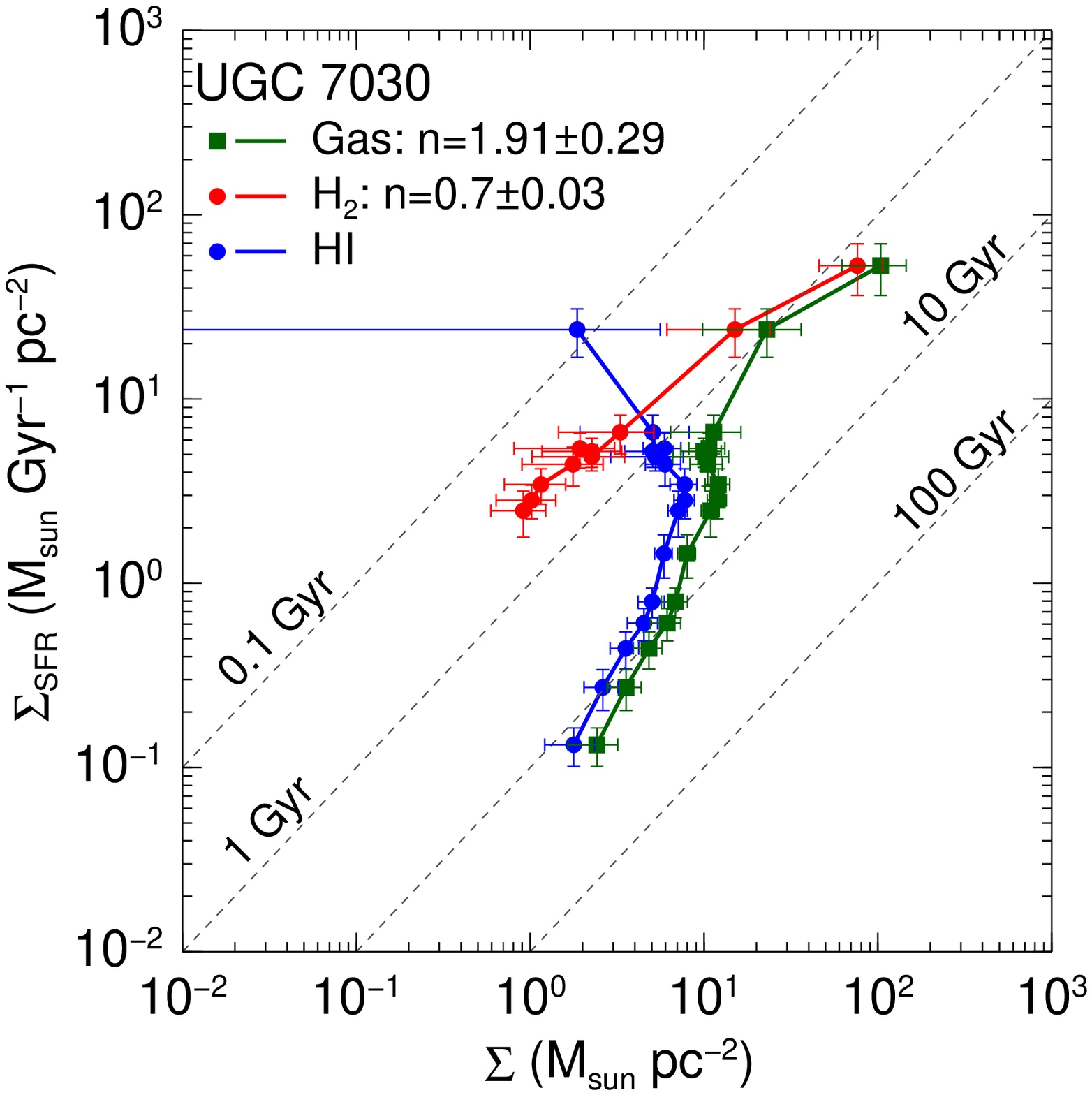}
\end{tabular}
\caption{SFR surface density as a function of \Htwo\ (red), \HI\ (blue), and total gas (green) surface densities for the sub-sample of 16 galaxies with CO data.  
\label{restsfl}}
\end{figure*}

\addtocounter{figure}{-1}

\begin{figure*}
\centering
\begin{tabular}{c@{\hspace{0.1in}}c@{\hspace{0.1in}}c@{\hspace{0.1in}}c}
\includegraphics[width=0.35\textwidth]{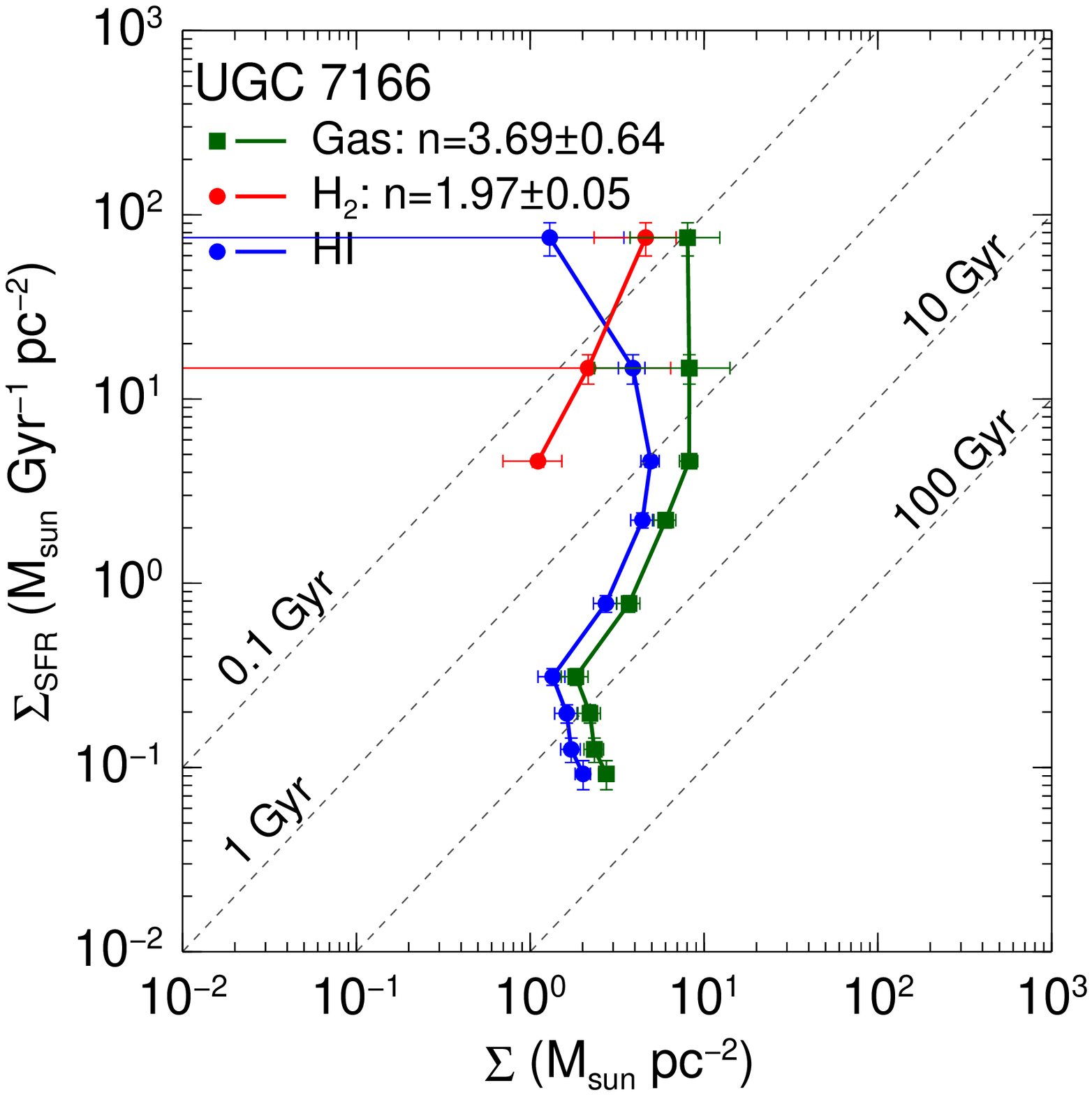}&
\includegraphics[width=0.35\textwidth]{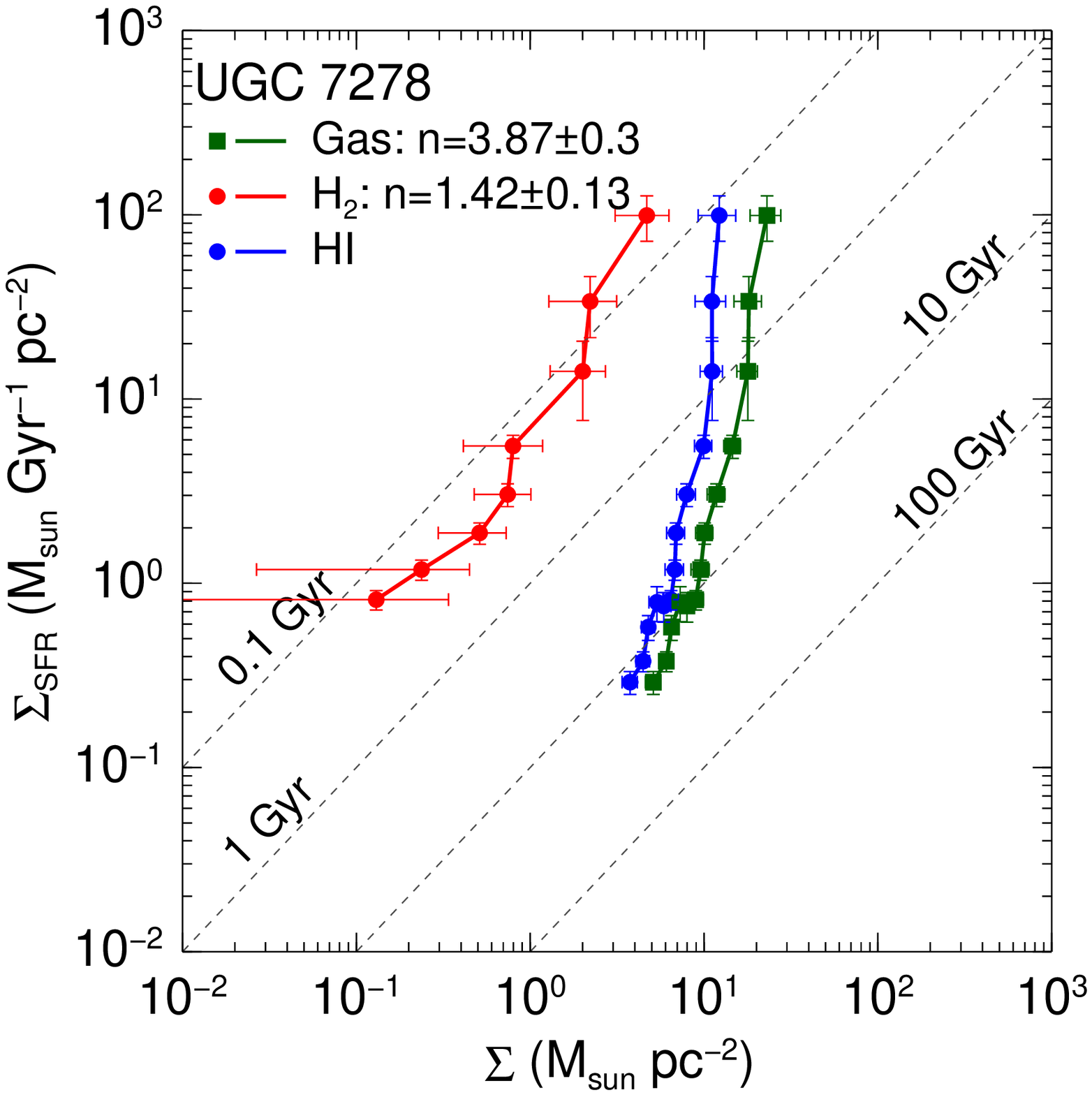}&
\includegraphics[width=0.35\textwidth]{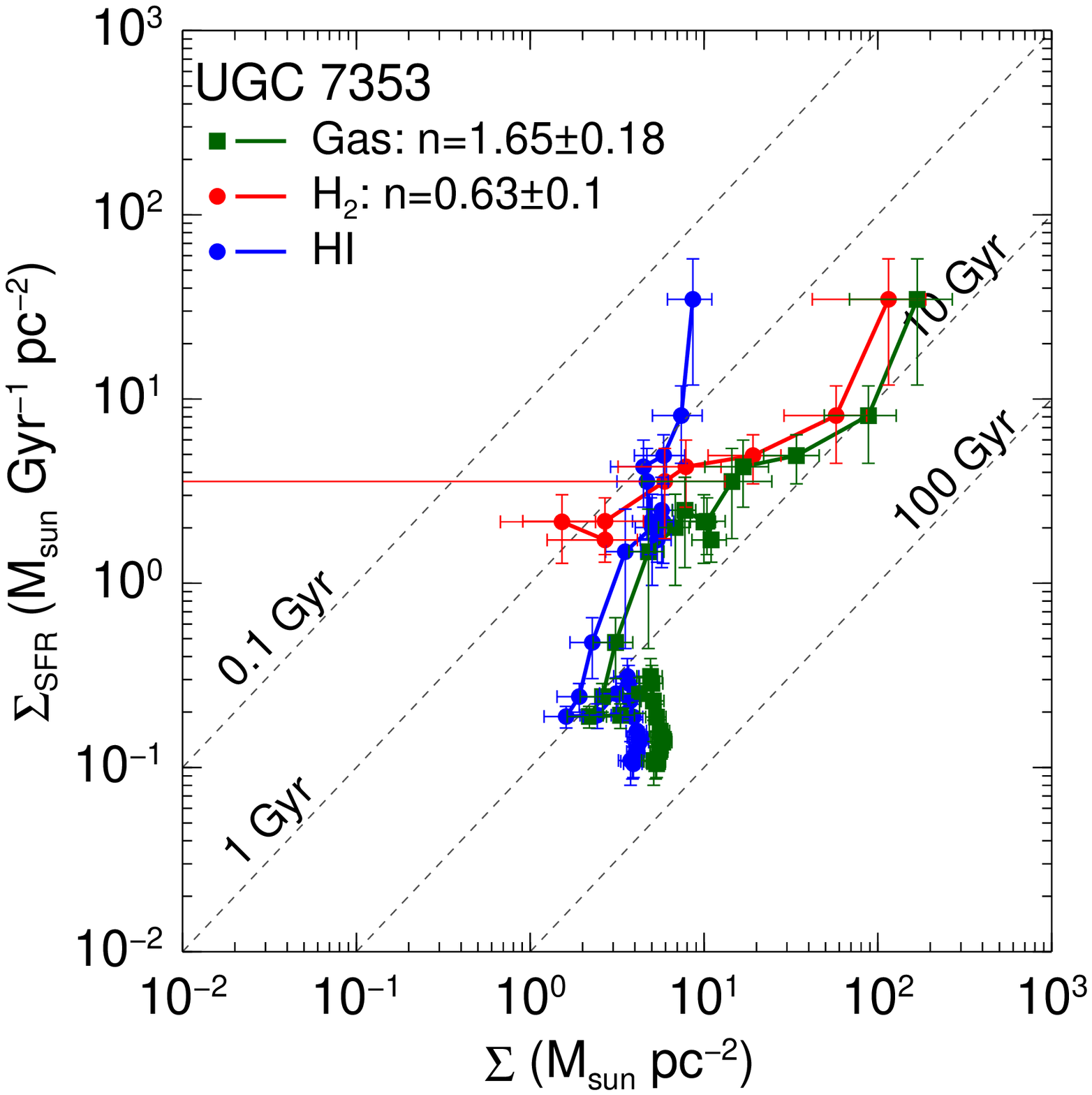}\\
\includegraphics[width=0.35\textwidth]{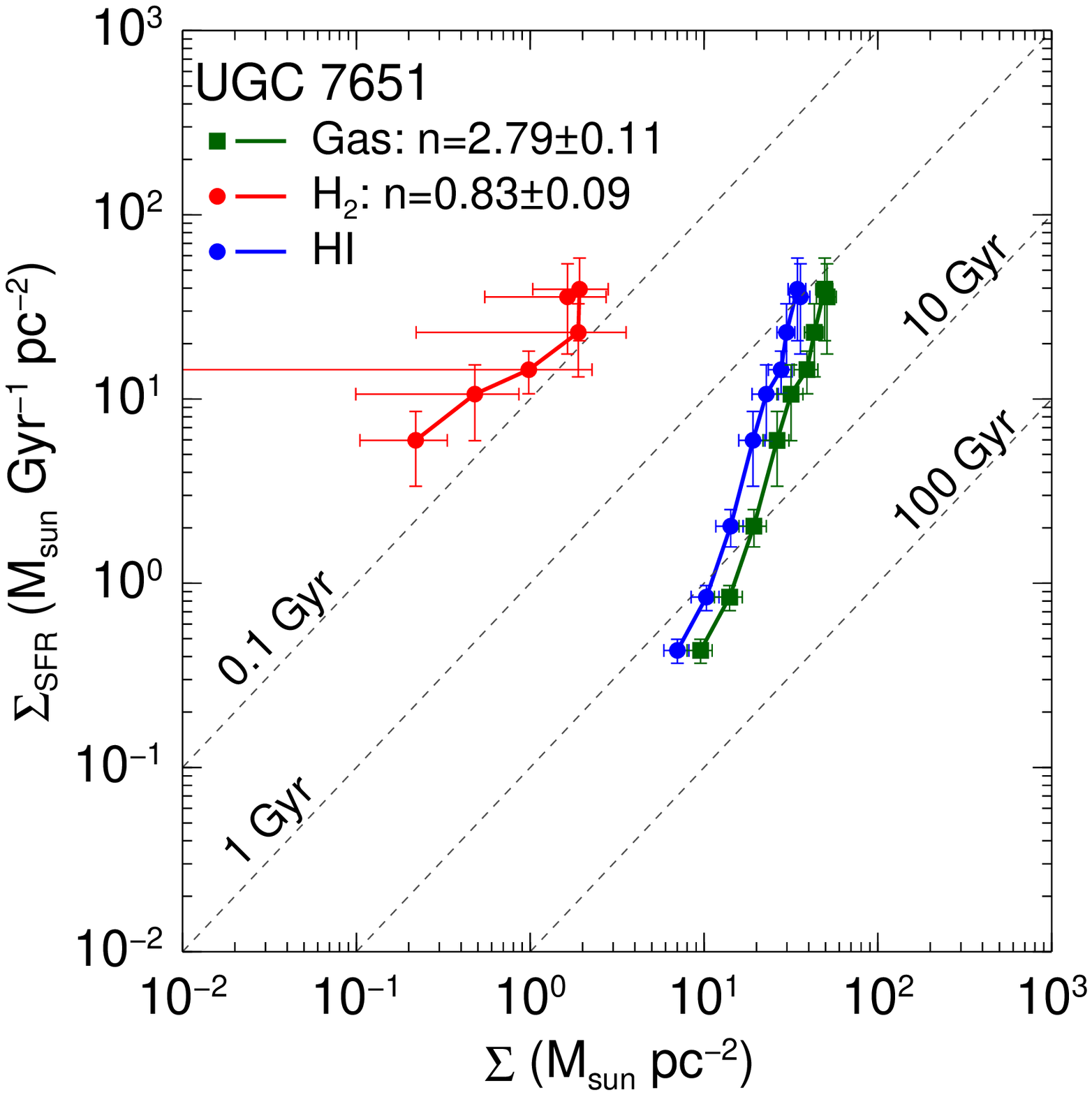}&
\includegraphics[width=0.35\textwidth]{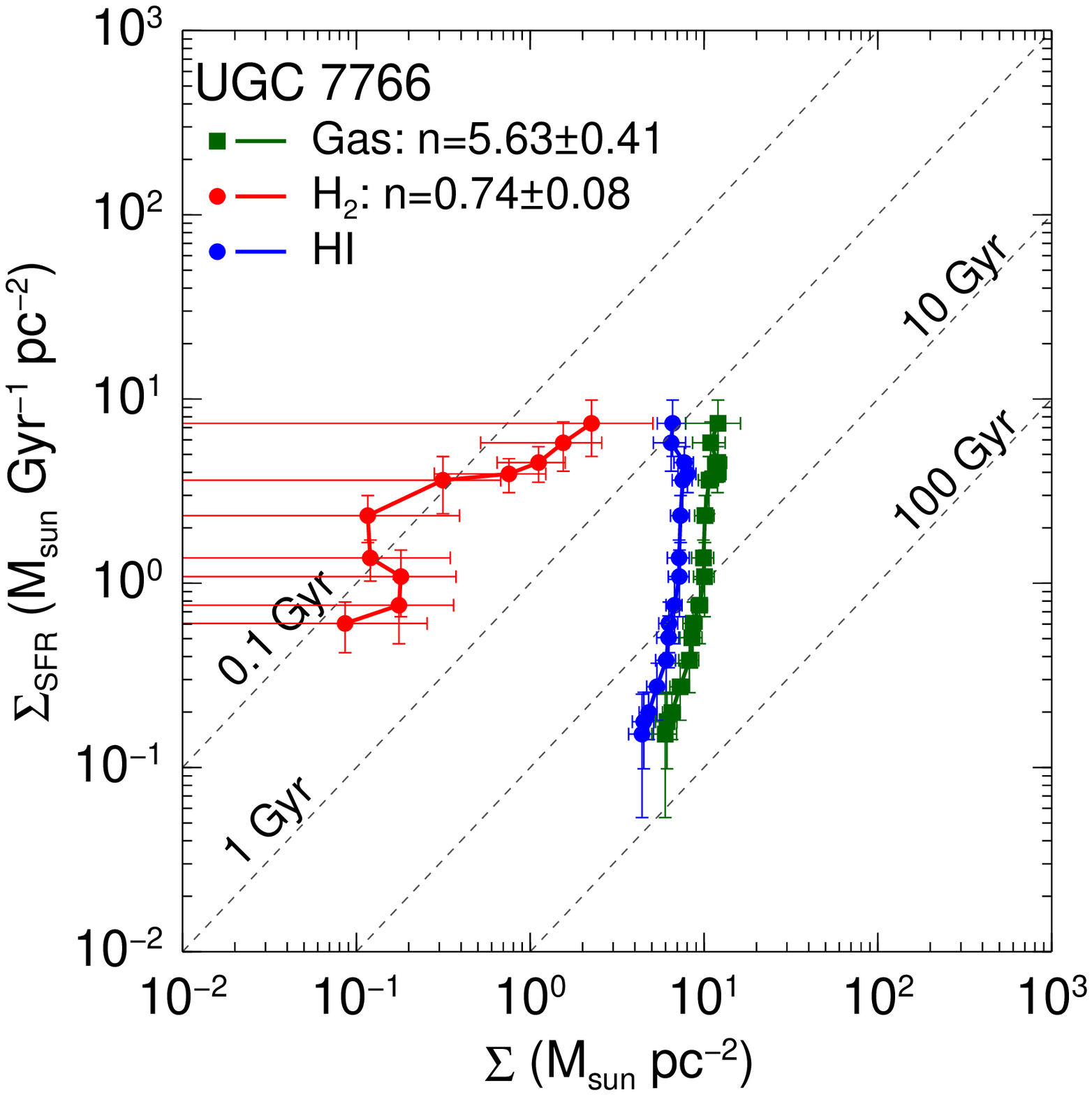}&
\includegraphics[width=0.35\textwidth]{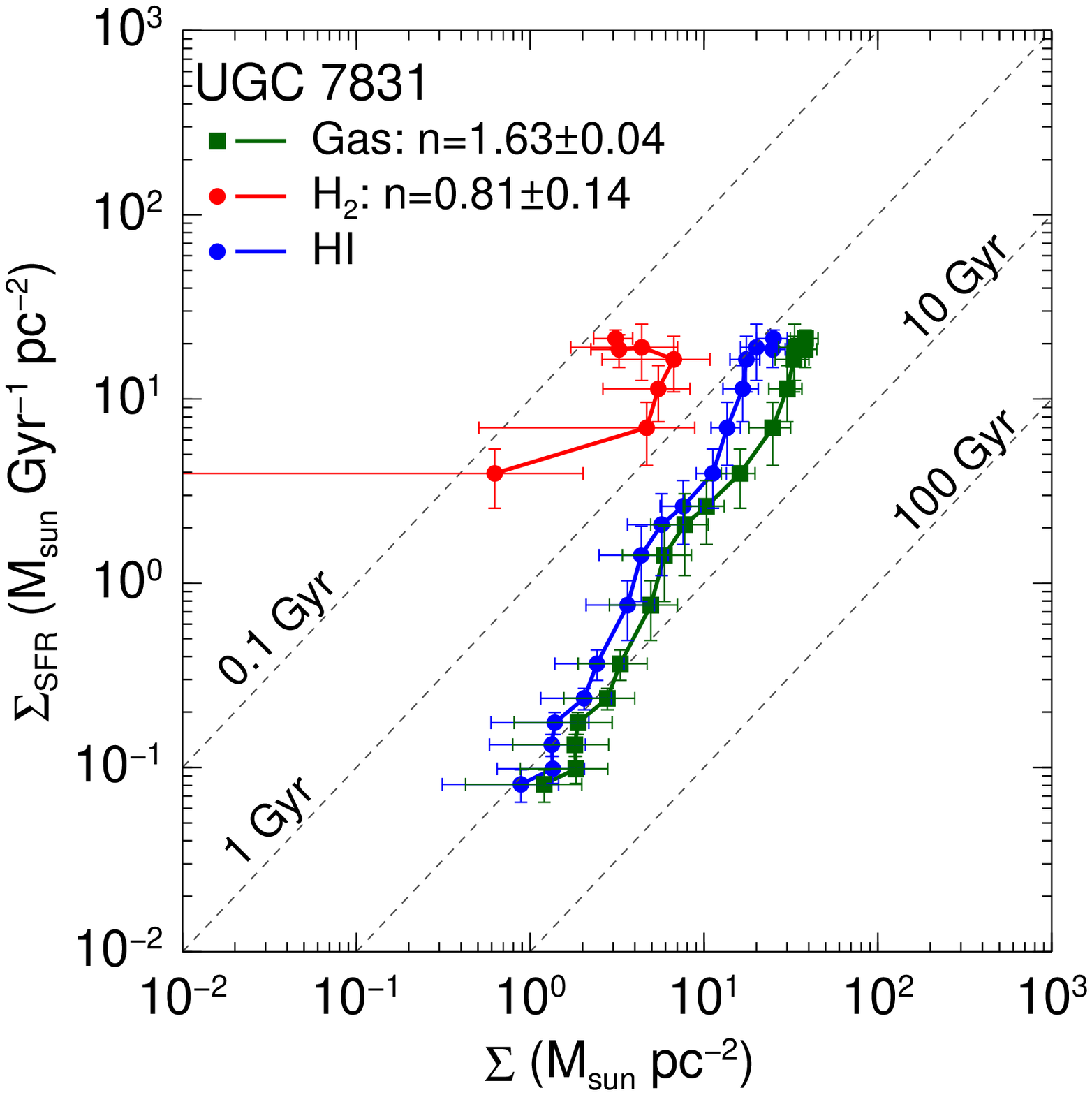}\\
\includegraphics[width=0.35\textwidth]{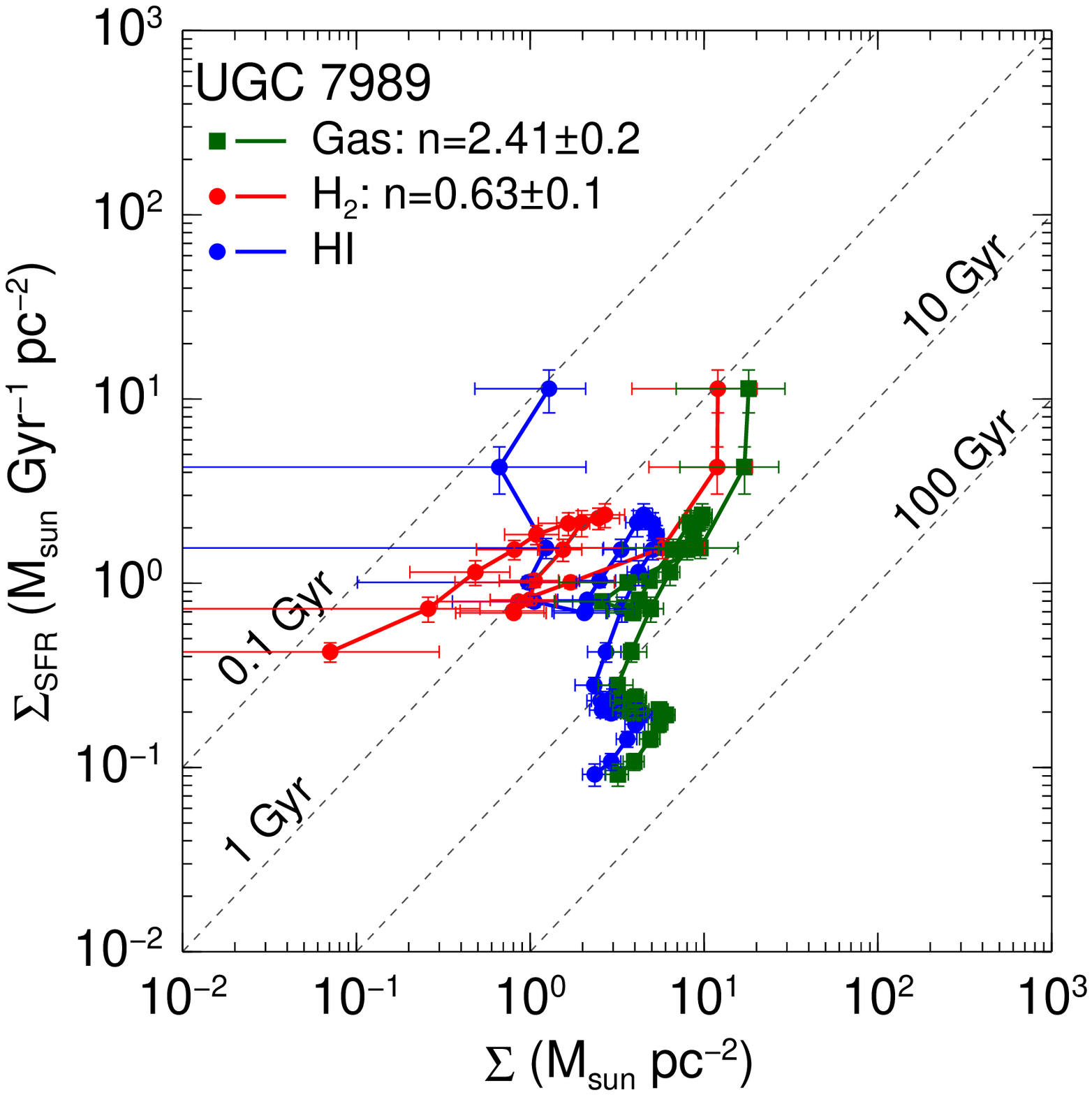}&
\end{tabular}
\caption{continued} 
\label{restsfl}
\end{figure*}
%%%%%%%%%%%%%%%%%%%%%%%%%%%%%%%%%%%%%%%%%%%%%%%%%%

% Don't change these lines
\bsp	% typesetting comment
\label{lastpage}
\end{document}